\documentclass[amsmath,amssymb,nofootinbib,10pt,aps,prd,twocolumn,reprint,floatfix]{revtex4-2}

\usepackage{graphicx}
\usepackage{subfig}
\usepackage{appendix}

\usepackage{csquotes} 

\usepackage[colorlinks=true,linkcolor=blue,
            urlcolor=blue,citecolor=red,
            pdfusetitle]{hyperref}
\usepackage[nameinlink,capitalize]{cleveref}

\graphicspath{
              {./Figures/WaveOpticPlanes}
              {./Figures/Results/KdS}
              {./Figures/Results/SdS}
              {./Figures/Convergence}
              {./Figures/PenroseDiagramInUpMode}
              {./Figures/ObserverSourceLocation}
              {./Figures/Results}
}

\newcommand{\RN}[1]{
  \textup{\uppercase\expandafter{\romannumeral#1}}
}

\renewcommand{\Re}{\text{Re}}
\renewcommand{\Im}{\text{Im}}

\begin{document}
\title{Exact wave-optical imaging of a Kerr-de Sitter black hole using Heun's equation}
\author{Felix Willenborg}\email{felix.willenborg@zarm.uni-bremen.de}
\author{Dennis Philipp}\email{dennis.philipp@zarm.uni-bremen.de}
\author{Claus Lämmerzahl}\email{claus.laemmerzahl@zarm.uni-bremen.de}
\affiliation{
Zentrum für angewandte Raumfahrttechnologie und Mikrogravitation (ZARM), \\
Unversity of Bremen, 28359 Bremen, Germany
}
\affiliation{
Gauss-Olbers Space Technology Transfer Center, \\
c/o ZARM, Unversity of Bremen, 28359 Bremen, Germany
}

\begin{abstract}
\date\today
Spacetime perturbations due to scalar, vector, and tensor fields on a fixed background geometry can be described in the framework of Teukolsky's equation.
In this work, wave scattering is treated analytically, using the Green's function method and solutions to the separated radial and angular differential equations in combination with a partial wave technique for a scalar and monochromatic perturbation. 
The results are applied to analytically describe wave-optical imaging via Kirchhoff-Fresnel diffraction, leading to, e.g., the formation of observable black hole shadows. 
A comparison to the ray-optical description is given, providing new insights into wave-optical effects and properties. 
On a Kerr-de Sitter spacetime, the cosmological constant changes the singularity structure of the Teukolsky equation and allows for an analytical, exact solution via a transformation into Heun's differential equation, which is the most general, second-order differential equation with four regular singularities. 
The scattering of waves originating from a point source involves a solution in terms of the so-called Heun's function $Hf$. 
It is used to find angular solutions that form a complete set of orthonormal functions similar to the spherical harmonics. 
Our approach allows to solve the scattering problem while taking into account the complex interplay of Heun's functions around local singularities. 
\end{abstract}

\maketitle
\tableofcontents

\section{Introduction}
The first-ever images of the black hole (BH) shadows in M87 and Sgr A$^*$ \cite{EHTC2019, EHTC2022} stand as remarkable milestones. 
These images not only visualized strong gravity regimes but also the underlying theory and cutting-edge technology that were developed over several years have provided an innovative tool for testing general relativity (GR) and the nature of our universe.

The sheer magnitude of effort invested in capturing these unprecedented images is unparalleled.
Central to this achievement is the pioneering technology of the Very Long Baseline Interferometry (VLBI) network, called the \enquote{Event Horizon Telescope} (EHT). 
Using the power of multiple strategically positioned radio telescopes aimed at M87 and Sgr A$^*$, this network allows observations in the extended wavelength regime of radio waves. 
The strength of this technique lies in the interference of discrete measurements, resulting in a cohesive visual representation.

One common way to validate and test different theories of gravity is to carefully assess the parameters that characterize black holes at the centers of galaxies.
Great interest, however, lies in studying the shadow of a black hole. 
Theoretical results based on lightlike geodesics in various spacetime geometries are given in, e.g., Refs.\ \cite{Perlick2022,Perlick2018,Grenzebach2015,Bardeen1973}. 

The foundation of theoretical black hole imaging emerges from the perturbation of established spacetime models. 
The literature extensively examines a variety of methods to address this complex task, including analytical techniques, approximations, and numerical approaches \cite{Nambu2016,Nambu2019,Turyshev2020,Turyshev2021,Kanai2013}.  
These investigations involve the use of numerical solutions of differential equations via finite-difference methods, harnessing phase-shift analysis via Prüfer transformations, and utilizing Runge-Kutta algorithms.
The weak gravitational scenario has been explored by Kanai \cite{Kanai2013}, while a discussion of the Schwarzschild and Ellis wormhole spacetimes can be found in the work of Nambu \cite{Nambu2019}. 
In addition, the case of Kerr black holes has been addressed by Glampedakis \cite{Glampedakis2001}.
For approximate solutions, Andersson \cite{Andersson1995} introduced the phase-integral method applied to the Schwarzschild spacetime. 
Furthermore, the widely used Wentzel–Kramers–Brillouin (WKB) approximation has been used by Nambu \cite{Nambu2016} to treat high-frequency scalar wave scattering in a Schwarzschild spacetime.
In a comprehensive review, Andersson \cite{Andersson2000} provided an introduction to the field of perturbations, approaching it from various angles.
In the present work, the focus is on exact solutions for scalar perturbations within various spacetime contexts.

A significant part of the aforementioned studies focuses on  differential cross-sections, as well as the emergence of back- and rainbow scattering \cite{Dolan2017,Stratton2019,Nambu2019}. 
The formalism employed in these works also provides a means of exploring wave-optical imaging on an observer plane \cite{Kanai2013,Nambu2016}.

Within the weak-field regime, imaging concepts can be extended to cover various celestial objects. 
Turyshev \cite{Turyshev2020,Turyshev2021} examined imaging possibilities for stars, while Feldbrugge \cite{Feldbrugge2020} applied these ideas to binary systems. 
The use of imaging techniques also extends to gravitational waves in the microlensing regime, as explored by, e.g.,  Cheung \cite{Cheung2021}. 
In this context, rays that carry phase information, which undergo curvature-induced alterations \cite{Schneider1999}, introduce an additional observable for wave scattering. 
The wave-optical approach complements the ray-optical approach, which is based on tracing lightlike geodesics, and provides theoretical and analytical support of prospective observations.

However, it is important to note that despite the various advances, currently no exact and analytical derivation is available in the field of wave-optical imaging.
And, thus, the purpose of this paper is to remedy this fact.

There are several approaches that discuss linear perturbations of a black hole induced by exterior sources. 
One key method involves the Teukolsky partial differential equation and the associated separated radial and angular ordinary differential equations. 
However, the boundary conditions for the radial equation present a challenge, especially at spatial infinity. 
Thus, there has not been an explicit analytical solution found for most black hole spacetimes yet, leading to the necessity of approximations mentioned above.

Interestingly, when a cosmological constant is present, the radial equation can be supplemented by a well-defined boundary condition \cite{Motohashi2021}. 
This insight is made possible by an involved discussion of Heun's equation \cite{Suzuki_1998,Suzuki_2000,Hatsuda2020,Batic_2007,Kamran1987,Hortacsu2018,Hortacsu2020,Nambu2022,Schmidt2023}. 
It is the general, second-order, linear differential equation with four regular singularities.
This framework with its singularity structure and known solutions provides an analytical tool to solve the separated Teukolsky equations on BH spacetimes in the presence of a cosmological constant.  

The primary goal of this work is to study the exact wave-optical imaging of a point source emitting scalar waves in a Kerr-de Sitter spacetime. 
The results allow us to reproduce and validate the established shadow formulae for black holes within the realm of wave optics. 

In \cref{KdS} the metric of interest, the Kerr-de Sitter metric, is introduced as a special case of the Plebanski-Demianski metric, the most general metric of Petrov type D spacetimes in GR. 
The presence of a (positive) cosmological constant $\Lambda$ introduces some modification of the horizon structure compared to the Kerr spacetime around rotating black holes.

The Teukolsky master equation, a second-order linear differential equation that describes linear perturbations in the Newman-Penrose formalism, is introduced in \cref{TME}. 
Necessary expressions for the derivation are given, resulting in separated radial and angular equations.
After a transformation, they can be solved by (i) the so-called local Heun functions $Hl$ and (ii) the Heun functions $Hf$ covered in \cref{Heun}. 
However, we focus on the extension by the Heun function $Hf$ because of its importance for the angular Teukolsky equation.

Of paramount significance, an orthogonality relation for the solutions can be constructed, closely related to the Sturm-Liouville eigenvalue problem. 
It plays a crucial role in the normalization of the angular solutions and leads to a complete set of orthonormal functions, similar to the normalization of associated Legendre functions in the context of spherical harmonics.
On the basis of this, in \cref{TMESol} solutions of the separated angular and radial equations are given in terms of solutions to Heun's equation. 
For the angular case, the focus lies on the eigenvalue problem. 
In terms of the Heun function, the non-Kerr limit is discussed, which leads to spin-weighted spherical harmonics. 
On the other hand, the radial solution requires an involved discussion of the boundary condition.
Its solution in terms of the Heun function is derived mainly from \cite{Motohashi2021}, which is why it will only be briefly revised here.

The Green's function method is used with corresponding solutions that follow naturally from the solution to Heun's equation using the physical boundary conditions. 
It enables a description of wave scattering and the resulting interference at arbitrary locations around a black hole.
The aforementioned normalization becomes particularly crucial in the final stages. 

Finally, \cref{Results} treats wave-optical imaging for scalar waves and an arbitrarily placed observer plane. 
Point sources around a Kerr-de Sitter and Schwarzschild-de Sitter black hole are considered, respectively.
Images are constructed for different Kerr parameters $a$ and wave frequencies $\omega$.
The results are compared to known results and properties of black holes, e.g., the formation of the Einstein ring in an appropriate setup, frame-dragging in the presence of spinning black hole, ray-optical shadows of black holes requiring multiple point sources, and finally the appearance of splitting images in a particular alignment of source and observer.

\section{Kerr-de Sitter spacetime (KdS)}
\label{KdS}
The Plebanski-Demianski (PB) spacetime is the most general axial-symmetric and stationary spacetime in general relativity and describes the geometry around a rotating black hole in the presence of a cosmological constant $\Lambda$, a spin parameter $a$, the Taub-NUT charge $\ell$, an acceleration parameter $\alpha$, a charge $\beta = q_e^2 + q_m^2$ including the electric and magnetic monopol charges $q_e$, $q_m$, respectively, and finally a mass parameter $M$.  
The spacetime admits two Killing vector fields $\partial_t$ and $\partial_\phi$ and can be classified as a Petrov type D geometry \cite{Griffiths2012}. 
In Boyer-Lindquist coordinates and ${G = c = 1}$ it is represented by
\begin{align}
ds^2 =~ &\frac{1}{\Omega^2} \bigg(\frac{1}{\Sigma} \left(\Delta_r - a^2 \Delta_\theta \sin^2\theta \right) dt^2 \label{KdS:Eq:PBmetric} \\
&+ \frac{2}{\Sigma} \left(\Delta_r \chi - a (\Sigma + a \chi) \Delta_\theta \sin^2 \theta\right) dt d\phi \notag \\
& - \frac{1}{\Sigma} \left[(\Sigma + a \chi)^2 \Delta_\theta \sin^2 \theta - \Delta_r \chi^2 \right] d\phi^2 \notag \\
& - \frac{\Sigma}{\Delta_r} dr^2 - \frac{\Sigma}{\Delta_\theta} d\theta^2\bigg) \notag
\end{align}
The metric functions of the PB metric are 
\begin{subequations}
\begin{align}
    \Omega &=: \Omega(r, \theta) &&= 1 - \frac{\alpha_P}{\omega_P} (\ell + a \cos\theta) r \, , \\
    \Sigma &=: \Sigma(r, \theta) &&= r^2 + (\ell + a \cos\theta)^2 \, , \\
    \chi &=: \chi(\theta) &&= a \sin^2\theta - 2 \ell (\cos\theta + C) \, , \\
    \Delta_\theta &=: \Delta_\theta(\theta) &&= 1 - a_3 \cos\theta - a_4 \cos^2\theta \, , \\
    \Delta_r &=: \Delta_r(r) &&= b_0 + b_1 r + b_2 r^2 + b_3 r^3 + b_4 r^4 \, ,
\end{align}
\end{subequations}
where the coefficients of the $\Delta_\theta$ and $\Delta_r$ polynomials are  
\begin{subequations}
\begin{align}
    \label{KdS:Eq:DeltaThetaPBCoeff}
    a_3 =& 2 \frac{\alpha_P}{\omega_P} a M - 4 a \ell \left(\frac{\alpha_P^2}{\omega_P^2} (k + \beta) + \frac{\Lambda}{3}\right) \, , \\
    a_4 =& -a^2 \left(\frac{\alpha_P^2}{\omega_P^2} (k + \beta) + \frac{\Lambda}{3}\right) \, ,
\end{align}
\end{subequations}
and
\begin{widetext}
\label{KdS:Eq:DeltarPBCoeff}
\begin{subequations}
\begin{align}
    b_0 =& k + \beta \, , \\
    b_1 =& -2 M \, , \\
    b_2 =& \frac{k}{a^2 - \ell^2} + 4 \frac{\alpha_P}{\omega_P} \ell M - \left(a^2 + 3 \ell^2\right) \left(\frac{\alpha_P^2}{\omega_P^2} (k + \beta) + \frac{\Lambda}{3}\right) \, , \\
    b_3 =& -2 \frac{\alpha_P}{\omega_P} \left[ \frac{k \ell}{a^2 - \ell^2} - \left(a^2 - \ell^2\right) \left(\frac{\alpha_P}{\omega_P} M - \ell\left(\frac{\alpha_P^2}{\omega_P^2} (k + \beta) + \frac{\Lambda}{3}\right)\right) \right] \, , \\
    b_4 =& -\left(\frac{\alpha_P^2}{\omega_P^2} k + \frac{\Lambda}{3}\right) \, .
\end{align}
\end{subequations}
\end{widetext}

The abbreviated definitions $k$, $\omega_P$, and $\beta$, appearing in \cref{KdS:Eq:DeltaThetaPBCoeff,KdS:Eq:DeltarPBCoeff}, are
\begin{subequations}
\begin{align}
k &= \frac{1 + 2 \frac{\alpha_P}{\omega_P} \ell M - 3 \ell^2 \left(\frac{\alpha_P^2}{\omega_P^2} \beta + \frac{\Lambda}{3}\right)}{1 + 3 \frac{\alpha_P^2}{\omega_P^2} \ell^2 \left(a^2 - \ell^2\right)} \, , \\
\omega_P &= \sqrt{a^2 + \ell^2} \, ,
\end{align}
\end{subequations}

In this work we reduce to the Kerr-de Sitter (KdS) case, in which the spacetime is described by the black holes angular moment, its mass and the (positive) cosmological constant, or rather the (positive) constant global curvature. 
Thus $\alpha_P = \beta = \ell = 0$. 
Written in terms of the PB representation, the KdS metric is
\begin{align}
ds^2 =~ 
    &\frac{1}{\Sigma \Xi^2} \left(\Delta_r - a^2 \Delta_\theta \sin^2\theta \right) dt^2 \label{KdS:Eq:metric} \\
    &+ \frac{2}{\Sigma \Xi^2} \left(\Delta_r \chi - a (\Sigma + a \chi) \Delta_\theta \sin^2 \theta\right) dt d\phi \notag \\
    & - \frac{1}{\Sigma \Xi^2} \left[(\Sigma + a \chi)^2 \Delta_\theta \sin^2 \theta - \Delta_r \chi^2 \right] d\phi^2 \notag \\
    & - \frac{\Sigma}{\Delta_r} dr^2 - \frac{\Sigma}{\Delta_\theta} d\theta^2 \, . \notag
\end{align}
An additional rescaling of $dt \rightarrow \frac{dt}{\Xi}$ and $d\phi \rightarrow \frac{d\phi}{\Xi}$ is considered here, with $\Xi := 1 + \alpha$ and $\alpha := \frac{\Lambda}{3} a^2$, see Ref.\ \cite{Griffiths2012}. 
As a result, the metric near the axis is well behaved, and conical singularities are resolved. 
A convenient explanation can be found in \cite{Akcay_2011}.
The metric functions are given by
\begin{subequations}
\begin{align}
    \rho &= -\frac{1}{r - i a \cos\theta} \, , \\
    \Sigma &= \frac{1}{\rho \rho^*} = r^2 + a^2 \cos^2\theta \, , \\
    \chi &= a \sin^2\theta \, , \\
    \Delta_r &= a^2 - 2 M r + \left(1 - \frac{\Lambda a^2}{3}\right) r^2 - \frac{\Lambda}{3} r^4 \label{KdS:Eq:Deltar} \, , \\
    \Delta_\theta &= 1 + \alpha \cos^2 \theta \,.
\end{align}
\end{subequations}
Here, $\Delta_r$ encodes an important physical property of black holes. 
Its zeros give the radii of possible horizons. 
In the case of KdS, ${\Delta_r = 0}$ is a fourth-order polynomial.
Thus, it can also be written as 
\begin{align}
\Delta_r(r) = -\frac{3}{\Lambda} (r - r'_-) (r - r_-) (r - r_+) (r - r'_+) \label{KdS:Eq:DeltarZeroDecomp} \, .
\end{align}
Comparing \cref{KdS:Eq:Deltar,KdS:Eq:DeltarZeroDecomp} results in the identity
\begin{align}
r'_- + r_- + r_+ + r'_+ = 0 \, .
\end{align}
Since the polynomial is of fourth-order, the expressions for its zeros in the full analytic representation are quite lengthy. 
However, an expansion up leading order in $\Lambda$ yields
\begin{subequations}
\begin{align}
r_\pm 
    &= M \pm \sqrt{M^2 - a^2} + \mathcal{O}(\Lambda^1) \, , \\
r'_\pm 
    &= \pm \sqrt{\frac{3}{\Lambda}} + \mathcal{O}(\Lambda^0) \, ,
\end{align}
\end{subequations}
revealing more physical context.
$r_+$ is the event horizon, $r_-$ the inner Cauchy horizon, $r'_+$ the (positive) cosmological horizon, and $r'_-$ its negative counterpart. 

For the following discussion, it is assumed that the four possible zeros of $\Delta_r$ are all of real and distinct nature, and hence $r'_- < 0 < r_- < r_+ < r'_+$. 
This assumption guarantees the existence of the mandatory horizon structure required in \cref{TMESol:Rad:BC}. 
By examining the discriminant of $\Delta_r$, an inequality defines the parameter ranges of $M$, $a$, and $\Lambda$ that satisfy this assumption \cite{Belgiorno2009},
\begin{subequations}
\label{KdS:Eq:aRange}
\begin{align}
    \alpha &< 7 - 4 \sqrt{3} \, ,
\label{KdS:Eq:alphaLim}\\
    M_{c,-} &< M < M_{c,+} \, ,    
\label{KdS:Eq:ExistenceCond}
\end{align}
\end{subequations}
where 
\begin{subequations}
\begin{align}
    M_{c,\pm} 
    &= \frac{(1 - \alpha)^{3/2}}{3 \sqrt{2 \Lambda}} \sqrt{1 \pm \gamma} (2 \mp \gamma), \\
    \gamma 
    &= \sqrt{1 - \frac{12 \alpha}{(1 - \alpha)^2}} \, .
\end{align}
\end{subequations}
The interested reader is referred to \cite{Akcay_2011} for a discussion of this range in the KdS spacetime. 
One remarkable consequence is the possible exceeding of the critical Kerr parameter $a = M$, for which the horizon structure is still preserved in a de Sitter spacetime due to the cosmological drift. 
The new upper limit is derived from \cref{KdS:Eq:aRange}, where the inequality is replaced by an equality. 
This implies, for example, that for $a \rightarrow 0$ there are no nonrotating black holes with the required horizon structure for $\Lambda M^2 > 1/9 $.
Increasing $\Lambda$ causes the upper and lower bounds of the Kerr parameter $a$ to converge until the horizon structure collapses for larger parameter choices, exposing a naked curvature singularity.

\section{KdS Teukolsky equations}
\label{TME}
Scattering involves a perturbation of the underlying spacetime and is conveniently approached in a first step by examining linear perturbations. 
Thus, these perturbations can be described by $g_{\mu\nu} = \tilde{g}_{\mu\nu} + h_{\mu\nu}$, where $\tilde{g}_{\mu\nu}$ is the background spacetime to be considered, e.g. the KdS spacetime, and $h_{\mu\nu}$ is the linear perturbation term. 
This approach led, for example, to the derivation of differential equations whose solution yields quasi-normal modes for the Schwarzschild spacetime as background \cite{Zerilli1970}. 
The problem can also be considered in the Newman-Penrose (NP) formalism, introducing the spinor formalism to GR\cite{Newman1962}. 
All NP formalism-dependent expressions are linearly perturbed, leading to an analog differential equation, as shown by Teukolsky \cite{Teukolsky_1973}. 
This so-called Teukolsky Master Equation (TME) is a second-order linear partial differential equation.
Examples of the Kerr TME can be found in \cite{Bini2002} and \cite{Teukolsky_1973}. 
\cite{Bini_2003} shows a TME for the Kerr-Taub-NUT spacetime.
The TME for KdS, expressed in a representation similar to Teukolsky's, is
\begin{widetext}
\begin{align}
& \Delta_r^{-s} \frac{\partial}{\partial r}\left( \Delta_r^{s + 1} \frac{\partial \Psi}{\partial r} \right) + \frac{1}{\sin\theta } \frac{\partial}{\partial \theta} \left(\Delta_\theta \sin\theta \frac{\partial \Psi}{\partial \theta} \right) 
 - (1 + \alpha)^2 \left(\frac{a^2}{\Delta_r} - \frac{\csc^2 \theta}{\Delta_\theta} \right) \frac{\partial^2 \Psi}{\partial \phi^2} \notag \\
& + (1 + \alpha)^2 \left(\frac{2 a (\Sigma + a \chi)^2}{\Delta_r} - \frac{\csc^2 \theta ~\chi^2}{\Delta_\theta} \right)\frac{\partial^2 \Psi}{\partial t^2} 
+ 2 (1 + \alpha)^2 \left(\frac{a (\Sigma + a \chi)}{\Delta_r} - \frac{\csc^2\theta~ \chi}{\Delta_\theta} \right) \frac{\partial^2 \Psi}{\partial t \partial \phi} \label{TME:Eq:TME} \\
& + s ( 1 + \alpha) \left(a \frac{\Delta_r'}{\Delta_r} + i \csc\theta \left(2 \cot\theta + \frac{\Delta_\theta'}{\Delta_\theta}\right)\right) \frac{\partial \Psi}{\partial \phi} + s (1 + \alpha) \bigg(\frac{4}{\rho^*} + (\Sigma + a \chi)\frac{\Delta_r'}{\Delta_r} + i \chi \csc\theta \left(2 \cot\theta + \frac{\Delta_\theta'}{\Delta_\theta}\right)\bigg) \frac{\partial \Psi}{\partial t} \notag \\
& - \bigg(\frac{2}{3} \Lambda \Sigma + \frac{1}{2} s \Delta_r'' - s^2 \left(\frac{4}{3} \Lambda r^2 + (1 + \alpha^2) \frac{\cot^2\theta}{\Delta_\theta}\right)\bigg) \Psi = 4 \pi \Sigma T \notag \, , 
\end{align}
\end{widetext}
where $s$ is the spin weight, $T$ describes the source terms, and $\Psi := \Psi(t, r, \phi, \theta)$ is the corresponding field quantity. 
The actual expressions of $\Psi$ and $T$ depend on the choice of $s$, see \cite{Teukolsky_1973} for more details. 
The apostrophes on $\Delta_r$ and $\Delta_\theta$ denote derivatives w.r.t.\ the respective independent variable.
The final differential equation for the scalar case $s = 0$ coincides with the Klein-Gordon equation in de Sitter spacetimes $\left(\Box - \frac{R}{6}\right) \Phi = 0$.
Moreover, all bosonic and fermionic perturbations, e.g., solutions to Maxwell's equations on this background, can be constructed based upon solutions of the TME.
However, in the following, the vacuum case ($T = 0$) will be considered. 

The TME can be solved by a separation of variables, resulting in ordinary rather than partial differential equations. 
The separability of Petrov type D geometries and their TME is shown for $s \in \{0, \pm 1/2, \pm 1, \pm 2\}$ in \cite{Kamran1987}. 
With
\begin{align}
    {}_s\Psi_{lm}(t,r,\phi,\theta) 
    = {}_sR_{lm}(r) ~{}_sS_{lm}(\theta) e^{-i \omega t} e^{i m \phi} \, ,
\label{TME:Sep:Eq:Sep}
\end{align}
where $l,m$ are the multipole expansion indices, the TME is separated into radial and angular equations,
\begin{align}
    \Delta_r^{-s} \frac{d}{dr}\left(\Delta^{s+1}_r \frac{d}{dr} {}_sR_{lm}(r) \right) + {}_sV^\text{(rad)}_{lm}(r) \,{}_sR_{lm}(r) &= 0   
\label{TME:Sep:Eq:Rad}
\end{align}
and
\begin{align}
    \frac{1}{\sin\theta} \frac{d}{d\theta} \left(\sin\theta ~\Delta_\theta \frac{d}{d\theta} {}_sS_{lm}(\theta)\right) + {}_sV^\text{(ang)}_{lm}(\theta) \,{}_sS_{lm}(\theta) &= 0 \, .
\label{TME:Sep:Eq:AngTheta}
\end{align}

The radial Teukolsky equation for KdS is characterized by the potential term \cite{Motohashi2021}
\begin{align}
    {}_sV^\text{(rad)}_{lm}(r) = &\frac{K_m^2(r) - i s K_m(r) \Delta'_r(r)}{\Delta_r(r)} + 2 i s K'_m(r)     
\label{TME:Sep:Eq:RadV}\\
    & - \frac{2 \alpha}{a^2} (s + 1) (2s + 1) r^2 + s (1 - \alpha) - {}_s\lambda_{lm} \notag 
\end{align}
with
\begin{align}
    K_m(r) = (1 + \alpha) \left((a^2 + r^2) \omega - a m\right) \, .
\label{TME:Sep:Eq:RadK}
\end{align}
This differential equation has five regular singularities $\{r'_-, r_-, r_+, r'_+, \infty\}$, of which the first four coincide with the radii of the horizons and ${}_s\lambda_{lm}$ is the separation constant related to an eigenvalue problem in the context of the Sturm-Liouville theory. 

In the case of the angular equation \cref{TME:Sep:Eq:AngTheta} it is more convenient to introduce a new variable $x := \cos \theta$. 
Consequently, its domain transforms to $x \in [-1, 1]$ and the differential equation becomes
\begin{align}
    \frac{d}{dx} \left(\Delta_x \frac{d}{dx} {}_sS_{lm}(x)\right) + {}_sV^\text{(ang)}_{lm}(x) \,{}_sS_{lm}(x) &= 0 \, ,
\label{TME:Sep:Eq:Ang}
\end{align}
where ${ \Delta_x(x) = (1 - x^2) \Delta_\theta(\arccos x) }$. 
The characterizing term of \cref{TME:Sep:Eq:Ang} is now
\begin{align}
    {}_sV^\text{(and)}_{lm}(x) = 
    &-\frac{{}_sG_m^2(x)}{\Delta_x(x)} - 2 \alpha x^2 + {}_s\lambda_{lm} \\
    & + s \frac{4 x (1 + \alpha) (m \alpha - c (1 + \alpha))}{\Delta_\theta(x)} \, , \notag
\end{align}
with
\begin{align}
{}_sG_m(x) = (1 + \alpha) \left(m + s x - a \omega (1 - x^2)\right) \, .
\end{align}
The regular singularities of the angular Teukolsky equation are 
\begin{align}
x_1 = -i / \sqrt{\alpha} \, , \quad x_2 = i / \sqrt{\alpha} \, , \quad x_3 = -1 \, , \quad x_4 = 1 \, .
\end{align}

The separated radial and angular differential equations each have five regular singularities.
In both cases, the singularities at infinity are removable and can be eliminated by a suitable transformation.
The solution of these equations can be conveniently derived by transforming the differential equations into the form of Heun's equation.
In fact, the separated Teukolsky equations are Heun's equations in disguise \cite{Suzuki_1998,Batic_2007}, which is why a thorough discussion of this class is important. 

\section{Heun's differential equation}
\label{Heun}
The canonical representation of Heun's differential equation is \cite{Ronveaux1995}
\begin{align}
&\frac{d^2 y(z)}{dz^2} + \left(\frac{\gamma}{z} + \frac{\delta}{z - 1} + \frac{\epsilon}{z - a_H}\right) \frac{dy(z)}{dz} \notag \\
& + \frac{\alpha \beta z - q}{z (z - 1) (z - a_H)} y(z) = 0 \, ,
\label{Heun:Eq:DEQ}
\end{align}
where $a_H$ is the singularity parameter and $q$ is called accessory (or auxiliary) parameter, which is closely related to a Sturm-Liouville eigenvalue problem discussed in \cref{TMESol:Ang}. 
The exponents $\gamma$, $\delta$, $\epsilon$, $\alpha$, $\beta$ are related to the Frobenius method applied to the Heun's equation, where two indicial exponents $\{0, 1-\gamma\}, \{0, 1-\delta\}, \{0, 1-\epsilon\}, \{\alpha, \beta\}$ exist at its regular singularities $\{0, 1, a_H, \infty\}$, respectively.
The sum of all exponents must be equal to two, thus the identity 
\begin{align}
    \gamma + \delta + \epsilon = \alpha + \beta + 1
\label{Heun:Eq:Identity}
\end{align}
holds. 

Around each singularity there exists a solution in the form of a local Heun function $Hl$ with a convergence radius extending to the next neighboring singularity, as discussed in \cref{Heun:Hl}. 
By means of an analytical extension, it is possible to go beyond the convergence radius, e.g., establish a common domain of convergence. 
However, the result generally does not show the same behavior as the analytical solutions obtained at other singularities. 
The nature of the Heun equation, nevertheless, allows for solutions of which the convergence domain contains two or three singularities by imposing a certain condition on $q$. 
These are called Heun functions $Hf$ and Heun polynomials $Hp$, respectively. 
Of these, only $Hf$ will be of interest in this work; see \cref{Heun:Hf}. 

\subsection{Local Heun function \texorpdfstring{$Hl$}{Hl}}
\label{Heun:Hl}
\cref{Heun:Eq:DEQ} can be solved by two different series expansions: A simple power series or a function series expansion. 
While the power series is usually used for $Hl$, the function series is applied for $Hf$, as discussed in \cref{Heun:Hf}.
The power series expansion around $z = 0$ and the choice of the first respective indicial exponent
\begin{align}
y(z) = \sum\limits^\infty_{r=0} c_r z^r
\label{Heun:Hl:Eq:PowSeries}
\end{align}
yields a three-term recurrence relation for the series coefficients $c_r$,
\begin{align}
R_r \, c_{r+1} - (Q_r + q) \, c_r + P_r \, c_{r-1} = 0 \, ,
\end{align}
where
\begin{subequations}
\begin{align}
    R_r &= (r + 1) (r + \gamma) a_H, \\
    Q_r &= r \left[(r - 1 + \gamma) (1 + a_H) + a_H \delta + \epsilon\right], \\
    P_r &= (r - 1 + \alpha) (r - 1 + \beta) \, ,
\end{align}
\end{subequations}
and $c_n = 0$ for $n < 0$. 
Due to the normalization ${c_0 = 1}$. 
Note that $\gamma \in \mathbb{N}$, otherwise the so-called logarithmic case has to be considered. 
In \cref{TMESol:Ang:Final} this coincides with solutions that are irregular at the poles and are not considered further. 

By construction, the solution converges inside a circle around $z = 0$ with radius $z < \min(1, |a_H|)$. 
The notation convention for the solution to \cref{Heun:Hl:Eq:PowSeries} is given in \cref{Heun:Hl:Eq:y01} below, where the first subscript denotes the singularity and the second denotes the solution index. 
$\epsilon$ is omitted and implicitly defined by the identity \cref{Heun:Eq:Identity}.
\begin{widetext}
\begin{subequations}
\label{Heun:Hl:Eq:y}
\begin{align}
    y_{01}(z) &= Hl(a_H, q; \alpha, \beta, \gamma, \delta; z)
\label{Heun:Hl:Eq:y01} \, , \\
    y_{02}(z) &= z^{1 - \gamma} Hl(a_H, (a_H \delta + \epsilon) (1 - \gamma) + q; \alpha + 1 - \gamma, \beta + 1 - \gamma, 2 - \gamma, \delta; z)
\label{Heun:Hl:Eq:y02} \, , \\
    y_{11}(z) &= Hl(1 - a_H, \alpha \beta - q; \alpha, \beta, \delta, \gamma; 1 - z) \label{Heun:Hl:Eq:y11} \, , \\
    y_{12}(z) &= (1 - z)^{1 - \delta} Hl(1 - a_H, ((1 - a_H) \gamma + \epsilon) (1 - \delta) + \alpha \beta - q; \alpha + 1 - \delta, \beta + 1 - \delta, 2 - \delta, \gamma; 1 - z) 
\label{Heun:Hl:Eq:y12} \, .
\end{align}
\end{subequations}
\end{widetext}
In total, eight solutions can be formulated with their respective three-term recurrence relations. 
However, by means of automorphisms, the solution \cref{Heun:Hl:Eq:y01} can be used to construct all other solutions. 
In general, all other solutions can be derived by applying an appropriate transformation of the independent variable $z \mapsto \zeta(z)$. 
It may be necessary to transform the dependent variable as well so that after the transformation the Frobenius exponents are $\{0, \rho_1\}, \{0, \rho_2\}, \{0, \rho_3\}, \{\rho_4, \rho_5\}$. 
Then the Heun differential equation can be obtained again with new coefficients leading to $Hl(a_H^*, q^*\; \alpha^*, \beta^*, \gamma^*, \delta^*, \zeta)$. 
For example, the solution $y_{02}$ with the second exponent at $z = 0$ can be expressed by $y_{01}$, as shown in \cref{Heun:Hl:Eq:y02}. 
The procedure can then be repeated for solutions at other singularities, e.g., at $z = 1$, leading to \cref{Heun:Hl:Eq:y11,Heun:Hl:Eq:y12} for indicial exponents $\{0, 1-\delta\}$, respectively. 
The remaining solutions are not of interest in this paper and are omitted here. 
With \cref{Heun:Hl:Eq:PowSeries,Heun:Hl:Eq:y01,Heun:Hl:Eq:y02,Heun:Hl:Eq:y11,Heun:Hl:Eq:y12}, the behavior of the respective solutions near their singularities can be studied. 
For solutions around $z = 0$ the leading-order terms are\footnote{These equations clarify the role of the so-called indicial exponents as exponents of the leading order terms.}
\begin{subequations}
\label{Heun:Hl:Eq:Ay0}
\begin{align}
    y_{0 1}(z) &= 1 + \mathcal{O}(z)
\label{Heun:Hl:Eq:Ay01}, \\
    y_{0 2}(z) &= z^{1 - \gamma} \left(1 + \mathcal{O}(z)\right) \, ,
\label{Heun:Hl:Eq:Ay02}
\end{align}
\end{subequations}
and for solutions around $z = 1$ 
\begin{subequations}
\label{Heun:Hl:Eq:Ay1}
\begin{align}
    y_{1 1}(z) &= 1 + \mathcal{O}(1 - z) 
\label{Heun:Hl:Eq:Ay11}, \\
    y_{1 2}(z) &= (1 - z)^{1 - \delta} \left(1 + \mathcal{O}(1 - z)\right) 
\label{Heun:Hl:Eq:Ay12}.
\end{align}
\end{subequations}
These functions are implemented in many commonly used CAS programs, such as Mathematica \cite{WolframMathematicaHeunG2020} or Maple \cite{MaplesoftHeunG2022}, which also use the same power series implementation. 
There are also open source implementations, e.g., in Octave \cite{Motygin2015}. 
A promising result for a Python implementation is described in \cite{Birkandan2021a}, where the integral series representation of the solutions to the Heun equation is implemented. 
In this work, however, we use the Mathematica implementation for reliability reasons.

\subsection{Connection coefficients}
\label{Heun:CC}
Local solutions around different singularities are not proportional in overlapping convergence domains.
The (two-point) connection problem is an approach to express a solution by other linearly independent solutions in a domain of mutual convergence. 
Motohashi et al.\ \cite{Motohashi2021} and Hatsuda et al.\ \cite{Hatsuda2020} dealt extensively with this problem in the context of local Heun functions, using the results of Dekar et al.\ \cite{Dekar1998}. 
An alternative approach to the connection problem is offered by Fiziev \cite{Fiziev2016}, who solved it by transforming the Heun equation to another domain, which allowed to respect the occurring branch cuts more carefully. 
However, we will stick to the notation of Motohashi and Hatsuda. 

Local Heun functions formulated at $z = 0$ are expressed by those at $z = 1$ through the linear combinations
\begin{subequations}
\begin{align}
    y_{0 1}(z) &= C_{11} y_{11}(z) + C_{12} y_{12}(z) 
\label{Heun:CC:Eq:y01} \, ,\\
    y_{0 2}(z) &= C_{21} y_{11}(z) + C_{22} y_{12}(z) 
\label{Heun:CC:Eq:y02} \, ,
\end{align}
\end{subequations}
and vice versa,
\begin{subequations}
\begin{align}
    y_{1 1}(z) &= D_{11} y_{01}(z) + D_{12} y_{02}(z) 
\label{Heun:CC:Eq:y11} \, ,\\
    y_{1 2}(z) &= D_{21} y_{01}(z) + D_{22} y_{02}(z)
\label{Heun:CC:Eq:y12} \, .
\end{align}
\end{subequations}
The exact form of the coefficients is given by \cite{Dekar1998}. 
As discussed in \cite{Motohashi2021}, a more computationally efficient and convergent expression is given in terms of Wronskians of local Heun functions. 
The coefficients for the first case are 
\begin{subequations}
\label{Heun:CC:Eq:C}
\begin{align}
    C_{11} &= \frac{W_z\left[y_{01}, y_{12}\right]}{W_z\left[y_{11}, y_{12}\right]} \, , 
\label{Heun:CC:Eq:C11} \\
    C_{12} &= \frac{W_z\left[y_{01}, y_{11}\right]}{W_z\left[y_{12}, y_{11}\right]} \, ,
\label{Heun:CC:Eq:C12} \\
    C_{21} &= \frac{W_z\left[y_{02}, y_{12}\right]}{W_z\left[y_{11}, y_{12}\right]} \, ,
\label{Heun:CC:Eq:C21} \\
    C_{22} &= \frac{W_z\left[y_{02}, y_{11}\right]}{W_z\left[y_{12}, y_{11}\right]} \, , 
\label{Heun:CC:Eq:C22}
\end{align}
\end{subequations}
and for the second case
\begin{subequations}
\label{Heun:CC:Eq:D}
\begin{align}
    D_{11} &= \frac{W_z\left[y_{11}, y_{02}\right]}{W_z\left[y_{01}, y_{02}\right]} \, , 
\label{Heun:CC:Eq:D11} \\
    D_{12} &= \frac{W_z\left[y_{11}, y_{01}\right]}{W_z\left[y_{02}, y_{01}\right]} \, , 
\label{Heun:CC:Eq:D12} \\
    D_{21} &= \frac{W_z\left[y_{12}, y_{02}\right]}{W_z\left[y_{01}, y_{02}\right]} \, , 
\label{Heun:CC:Eq:D21} \\
    D_{22} &= \frac{W_z\left[y_{12}, y_{01}\right]}{W_z\left[y_{02}, y_{01}\right]} \, . 
\label{Heun:CC:Eq:D22} 
\end{align}
\end{subequations}
The evaluation is performed at any point $z$ within the mutual convergence domain.

\subsection{Heun Function \texorpdfstring{$Hf$}{Hf}}
\label{Heun:Hf}
For the local Heun function $Hl$, a power series is an appropriate choice. 
On the other hand, a suitable choice for the Heun function $Hf$ is a series of functions, which ensure more efficient convergence with the mathematical properties of special cases appearing more naturally. 
Here, the hypergeometric function ${}_2F_1\left(\alpha, \beta; \gamma; z\right)$ is a good choice, which seems appropriate since Heun's equation is historically the result of a generalization of the hypergeometric equation. 
It is a solution of a second-order linear differential equation with only two regular singularities and one irregular singularity. 
This approach is similar to expanding the spheroidal wave function in a series of Bessel functions. 
Despite more efficient convergence, another very important property motivates the use of the hypergeometric function, as will be shown below for a particular choice of parameters.

The solution of \cref{Heun:Eq:DEQ} at $z = 0$ with indicial exponent $\gamma = 0$ is constructed as
\begin{subequations}
\label{Heun:Hf:Eq:FunSeries}
\begin{align}
    y(z) &= \sum\limits^\infty_{n=0} c_n y_{\nu_0 + n}(z) \, , \\
    y_\nu(z) &= ~{}_2F_1(-\nu, \nu + \varpi; \gamma; z) \, ,
\end{align}
\end{subequations}
where $\varpi = \gamma + \delta - 1 = \alpha + \beta - \epsilon$.
The corresponding three-term recurrence relation for the coefficients is
\begin{align}
    P^*_n \, c_{n+1} + S^*_n \, c_n + R^*_n \, c_{n-1} = 0 \, ,
\label{Heun:Hf:Eq:threeterm}
\end{align}
with
\begin{align}
    P^*_n &= F_{\nu_0 + n - 1} \, , \quad S^*_n &= J_{\nu_0 + n} \, , \quad R^*_n &= D_{\nu_0 + n + 1} \, ,
\end{align}
where
\begin{subequations}
\begin{align}
    F_\nu 
    &= - \frac{(\nu + \alpha) (\nu + \beta) (\nu + \gamma) ( \nu + \varpi)}{(2 \nu + \varpi) (2 \nu + \varpi - 1)} \, , \\
    D_\nu 
    &= -\frac{\nu (\nu + \varpi - \alpha) (\nu + \varpi - \beta) (\nu + \delta - 1)}{(2 \nu + \varpi) (2 \nu + \varpi - 1)} \, , \\
    J_\nu 
    &= -E_\nu - q, \notag \\
    &= \frac{Z_\nu}{(2 \nu + \varpi + 1) (2 \nu + \varpi - 1)} - a_H \nu (\nu + \varpi) - q \, ,
\end{align}
\end{subequations}
and
\begin{multline}
Z_\nu = \epsilon \nu (\nu + \varpi)(\gamma - \delta) + (\nu(\nu + \varpi) + \alpha \beta) (2 \nu (\nu + \varpi) \\
+ \gamma(\varpi - 1)) \, .
\end{multline}
Again, $c_{n} = 0, \, \forall n < 0$ and $c_{0} = 1$. 
The choice of $\nu_0$ is important and affects the resulting type of hypergeometric function and its convergence behavior. 
In the literature, two types are discussed: the so-called Erdélyi type (E-type) \RN{1} and \RN{2} solutions, which differ significantly in their convergence behavior.
While E-type \RN{1} solutions have a "lima\c{c}on" as convergence space and can contain one or two singularities (the second one will be at the edge of the convergence domain but is included), E-type \RN{2} solutions have an ellipse as convergence domain with singularities at the foci.
E-type \RN{2} essentially uses degenerate hypergeometric functions as expansion functions in \cref{Heun:Hf:Eq:FunSeries}, which predetermines the solution to have two singularities in its convergence space. 
By choosing ${\nu_0 \in \{0, -\beta-\varpi\}}$, E-type \RN{2} solutions are obtained for class\footnote{See \cref{Heun:Hf:Eq:Classes}} \RN{1} or \RN{3} Heun functions, respectively. 
For the sake of simplicity, the discussion here is restricted to the first case. 
Solutions of other classes can be derived by automorphisms similar to those used for solutions of $Hl$. 

The condition imposed on $q$ is important to build a Heun function $Hf$. 
By rewriting the functions of the three-term recurrence relation \cref{Heun:Hf:Eq:threeterm} 
\begin{subequations}
\begin{align}
    M_n &= \frac{c_n}{c_{n-1}} := - \frac{R^*_n}{S^*_n + P^*_n M_{n+1}} \, , \\
    L_n &= \frac{c_n}{c_{n+1}} := - \frac{P^*_n}{S^*_n + R^*_n L_{n-1}} \, ,
    \end{align}
\end{subequations}
the identity 
\begin{align}
M_n L_{n-1} = 1
\label{Heun:Hf:Eq:infcontfracshort}
\end{align}
can be derived \cite{Suzuki_1998}.
Replacing $S^*_n = -(E_n + q)$ in \cref{Heun:Hf:Eq:infcontfracshort}, inserting the recurrence relation and rearranging for $q$, the necessary condition for the transformation of $Hl$ into $Hf$ follows: 
\begin{align}
    q_k = 
    &-E_{n-1} - \frac{R^*_n P^*_{n-1}}{-(E_n + q_k) + P^*_{n-1} L_{n-2}} + P^*_n M_{n+1}, \notag \\
    = &-E_{n-1} - \frac{R^*_n P^*_{n-1}}{-(E_n + q_k) -} \frac{R^*_{n-1} P^*_{n-2}}{-(E_{n-2} + q_k) - ...} \label{Heun:Hf:Eq:infcontfrac} \\
    & - \frac{P^*_n R^*_{n+1}}{-(E_{n+1} + q_k) -} \frac{ P^*_{n+1} R^*_{n+2}}{-(E_{n+2} + q_k) - ...} \, . \notag
\end{align}
Note that the parameter $q$ now has an index $k \in \mathbb{N}$ and $q_k$ is an infinitely countable set of possible auxiliary parameters that provide solutions to the problem. 
\cref{Heun:Hf:Eq:infcontfrac} involves a finite continued fraction\footnote{Continued fractions can be written compactly using a notation where, for example, $\frac{1}{a - } \frac{1}{b -} \frac{1}{c} = \frac{1}{a - \frac{1}{b - \frac{1}{c}}}$.} in the second term and an infinite one in the third term. 
The second term is finite due to \cref{Heun:Hf:Eq:infcontfracshort} and the conditions on the coefficients of \cref{Heun:Hf:Eq:threeterm}. 
Consequently, $L_n = 0$ for $n < 0$ and $M_n = 0$ for $n < 1$. 
The continued fraction is centered on $n$ and can be chosen arbitrarily, e.g., $n = 0$ reduces to a single infinite continued fraction. 
To solve for $q_k$, a successive approximation is performed, similar to the eigenvalue derivation in \cite{Meixner1954}. 
Despite the infinite continued fraction in \cref{Heun:Hf:Eq:infcontfrac}, another property leads to the same result.
Two local solutions $Hl$ developed around two different singularities share a mutual convergence domain and a particular choice of $q_k$ will make them linearly dependent \cite{Becker1997}, i.e., the Wronskian vanishes,
\begin{multline}
    W_z\left[y_{0i}(q; z), y_{1j}(q; z)\right] = y_{0i}(q; z) y'_{1j}(q; z) \\ - y'_{0i}(q; z) y_{1j}(q; z) 
\overset{!}{=} 0 \,,
\label{Heun:Hf:Eq:Wronskian}
\end{multline}
where the indices $i, j \in \{1, 2\}$ refer to the respective exponents of the local solutions \cref{Heun:Hl:Eq:y01,Heun:Hl:Eq:y02,Heun:Hl:Eq:y11,Heun:Hl:Eq:y12}. 
Despite its analytical property, this equation will be evaluated numerically by a root-finding algorithm and will complement the infinite continued fraction \cref{Heun:Hf:Eq:infcontfrac} approach in later calculations. 

The construction of the auxiliary parameter $q_k$ as in \cref{Heun:Hf:Eq:infcontfrac,Heun:Hf:Eq:Wronskian} transforms the involved local Heun functions $Hl$ into a Heun function $Hf$ for which the convergence domain contains two singularities instead of one. 
Despite the different notation, the evaluation of the resulting $Hf$ can still be performed by local solutions \cref{Heun:Hl:Eq:y01,Heun:Hl:Eq:y02,Heun:Hl:Eq:y11,Heun:Hl:Eq:y12}. 
In the literature, the technical description of an $Hf$ solution is
\begin{align}
    y_{0i}(q_k; z) &:= (s_1, s_2)Hf_k^\text{(X)} (a_H, q_k; \alpha, \beta, \gamma, \delta; z) \, .
\label{Heun:Hf:Eq:HfNotation}
\end{align}
The proportionality of local solutions is defined by
\begin{align}
    \Theta^{i \rightarrow j} := \frac{y_{0i}(q_k; z)}{y_{1j}(q_k; z)} \, ,
\end{align}
which is a constant and independent of $z$ in the region of mutual convergence.

A connection problem as in \cref{Heun:CC} has become obsolete, as can be seen in \cref{Heun:Hf:Eq:Wronskian}. 
In the notation above of $Hf$, $s_1$ and $s_2$ are the singularities at which the solutions are simultaneously regular. 
Their respective Frobenius exponents are $\{\rho_1, \rho_2\}$ and $\{\sigma_1, \sigma_2\}$. 
$X \in \{\RN{1}, \RN{2}, \RN{3}, \RN{4}\}$ refers to the class of the exponent combination, $\{\rho_1, \sigma_1\}$, $\{\rho_2, \sigma_1\}$, $\{\rho_1, \sigma_2\}$, and $\{\rho_2, \sigma_2\}$, respectively. 
For further discussion, the singularities will be considered to be $s_1 = 0$ and $s_1 = 1$, thus $\rho_1 = 0, \rho_2 = 1 - \gamma, \sigma_1 = 0$ and $\sigma_2 = 1 - \delta$. 
This, together with the fact that for \cref{Heun:Hl:Eq:y01} the parameter $\gamma$ must be a positive integer, establishes the existence conditions for the class definitions just introduced \cite{Becker1997},
\begin{subequations}
\label{Heun:Hf:Eq:Classes}
\begin{align}
    \text{class I:}~~~ &\Re (\gamma) > 0,~~~ \Re (\delta) > 0 \label{Heun:Hf:Eq:I} \\
    \text{class II:}~~~ &\Re (\gamma) < 2,~~~ \Re (\delta) > 0 \label{Heun:Hf:Eq:II} \\
    \text{class III:}~~~ &\Re (\gamma) > 0,~~~ \Re (\delta) < 2 \label{Heun:Hf:Eq:III} \\
    \text{class IV:}~~~ &\Re (\gamma) < 2,~~~ \Re (\delta) < 2 \,. \label{Heun:Hf:Eq:IV}
\end{align}
\end{subequations}
As a result, only certain classes are relevant for the Heun function.

\subsubsection{Orthogonality and normalization}
\label{Heun:Normal}
The orthogonality of spherical harmonics and the resulting normalization play a crucial role in creating a complete set of orthonormal functions, useful to expand any square-integrable function.
This particular property will be important later in evaluating the scattering of waves by a black hole.

Although $Hl$ has no orthogonality relation and therefore does not provide normalization, $Hf$ has an orthogonality relation \cite{Ronveaux1995}. 
This is
\begin{align}
    \left(q_k - q_n\right) \int_C w(z) y_k(z) y_n(z) dz = \left[p(z) W_z\left[y_k, y_n\right]\right]_C \, ,
\label{Heun:Hf:Normal:Eq:Ortho}
\end{align}
where 
\begin{subequations}
\begin{align}
    w(z) 
    &= z^{\gamma - 1} (z - 1)^{\delta - 1} (z - a_H)^{\epsilon - 1} \, , \\
    p(z) 
    &= z^\gamma (z - 1)^\delta (z - a_H)^\epsilon \, .
\end{align}
\end{subequations}
$y_k$ and $y_n$ are Heun functions $Hf$ of the same class with different auxiliary parameters $q_k \neq q_n$, and $C$ is a contour along which the integral is evaluated. 
It is assumed that the singularity parameter $a_H \notin [0, 1]$. 
The right-hand side vanishes for the class \RN{1} Heun functions when $C$ is a real line from $z = 0$ to $z = 1$. 
Thus, as long as $k \neq n$, the integral is equal to zero and the orthogonality of $y_k,y_n$ holds. 
When $k = n$ \cref{Heun:Hf:Normal:Eq:Ortho} reveals a normalization constant 
\begin{align}
    \int_C w(z) y_{0i}^2(q_k; z) dz = \zeta_{ij} \, .
\end{align}

In this short excerpt, only the class \RN{1} Heun functions are discussed. 
This restriction can be lifted if $C$ is a closed Pochhammer double-loop contour \cite{Ronveaux1995}. 
However, Becker \cite{Becker1997} carried out an approach for each class that is still feasible along the line $z \in [0, 1]$. 
The normalization constant $\zeta_{ij}$ has the same expression for all classes \cref{Heun:Hf:Eq:I,Heun:Hf:Eq:II,Heun:Hf:Eq:III,Heun:Hf:Eq:IV},
\begin{align}
    \zeta_{ij} = -\Theta^{i \rightarrow j} p(z) \left.\frac{\partial W_z^{ij}}{\partial q}\right\rvert_{q=q_k} \, .
\label{Heun:Hf:Normal:Eq:NormConst}
\end{align}
Here, $W_z^{ij} := W_z\left[y_{0i}(q, z), y_{1j}(q, z)\right]$. 
An important property is that $\zeta_{ij}$ is independent of $z$, since $\Theta^{i \rightarrow j}$ and $p(z) \frac{\partial W_{ij}}{\partial q}\rvert_{q=q_k}$ are independent of $z$ in the mutual convergence domain!

\section{Solving the separated Equations}
\label{TMESol}
Insights into Heun's differential equation and its solutions are now applied to the TMEs \cref{TME:Sep:Eq:Ang} and \cref{TME:Sep:Eq:Rad}.
A general instruction on how to transform a differential equation into the Heun form is given in Ref.\ \cite{Ronveaux1995}.
The independent variable undergoes a Moebius transformation
\begin{align}
    z(u) = \frac{u_2 - u_4}{u_2 - u_1} \frac{u - u_1}{u - u_4} \, ,
\end{align}
which modifies the regular singularities positions' ${\{u_1, u_2, u_3, u_4, \infty\} \rightarrow \{0, 1, a_H, \infty, z_\infty\}}$. 
An f-homotopic transformation of the dependent variable follows,
\begin{align}
    y(z) = z^{\rho_1} (z - 1)^{\rho_2} (z - a_H)^{\rho_3} f(z) \, .
\end{align}
Reading off the exponents of each singularity gives the Heun form \cref{Heun:Eq:DEQ}. 

\subsection{Angular solution}
\label{TMESol:Ang}
Starting with the solution of the angular Teukolsky equation \cref{TME:Sep:Eq:Ang}, 
\begin{align}
    z(x) = \frac{x_4 - x_2}{x_4 - x_3} ~\frac{x - x_3}{x - x_2} \, ,
\label{TMESol:Ang:Eq:IndepTrans}
\end{align}
transforms  $\{x_1, x_2, x_3, x_4, \infty\}$ to $\{z_a, \infty, 0, 1, z_{a\infty}\}$, where 
\begin{subequations}
\begin{align}
    z   _{a\infty} &= z(x)\rvert_{x \rightarrow \infty} = \frac{x_4 - x_2}{x_4 - x_3}, \\
    z_a &= z(x)\rvert_{x \rightarrow x_1} = \frac{x_4 - x_2}{x_4 - x_3} \frac{ x_1 - x_3}{x_1 - x_2} \, .
\end{align}
\end{subequations}
The physically interesting region between the two poles is now located in the region $z \in [0, 1]$. 
It should be noted that $z(x)$ describes a complex path for $x \in [-1, 1]$ and not a straight line connecting both singularities on the real axis. 

The dependent variable is transformed using the f-homotopic transformation 
\begin{multline}
    {}_sS^\text{(ij)}_{lm}(z) = z^{A_1} (z - 1)^{A_2} (z - z_a)^{A_3} (z - z_{a\infty})^{A_5} \\
\times y^{(a)}_{ij}({}_s\lambda_{lm}; z) \, .
\end{multline}
Instead of using $q_k$ in the notation of the local Heun function, the corresponding eigenvalue ${}_s\lambda_{lm}$ appears. 
Since the differential equation \cref{TME:Sep:Eq:Ang} has five regular singularities, $A_5$ is also included here as an additional exponent for the regular singularity at $z = z_{a\infty}$. 
This singularity is removable and does not obstruct the formalism.
Its exponent turns out to be $A_5 = 1$. 

To get to the Heun form, the exponents $A_i$ will have to take a particular form, which is
\begin{align}
    A_i &= \pm \left|A(x_i)\right|
\end{align}
with
\begin{align}
    A(x) &= \frac{{}_sG_m(x)}{\Delta'_x (x)}
\end{align}
where $i \in \{1, 2, 3, 4\}$. 
Note that $A_4 = A^*_3$. 
Explicitly written out, the exponents are
\begin{subequations}
\begin{align}
    A_1 &= \pm \left|\frac{m - s}{2}\right| \, , \\
    A_2 &= \pm \left|\frac{m + s}{2}\right| \, , \\
    A_3 &= \mp \frac{1}{2} \left|\left(s - i \left(a \omega \frac{1 + \alpha}{\sqrt{\alpha}} - m \sqrt{\alpha}\right)\right)\right| \, , \\
    A_4 &= \pm \frac{1}{2} \left|\left(s + i \left(a \omega \frac{1 + \alpha}{\sqrt{\alpha}} - m \sqrt{\alpha}\right)\right)\right| \, .
\end{align}
\end{subequations}
The choice of sign for $A_1$, $A_2$ depends on the boundary condition and it is arbitrary for $A_3$. 
Finally, the transformed differential equation \cref{TME:Sep:Eq:Ang} takes the form
\begin{align}
    &\frac{d^2 y_a}{d z^2} + \left(\frac{2 A_1 + 1}{z} + \frac{2 A_2 + 1}{z -1} + \frac{2 A _3 + 1}{z - z_a}\right)\frac{d y_a}{d z} \notag \\
    &+ \frac{\rho_+ \rho_- z + u}{z (z -1) (z - z_a)}y_a = 0,
\label{TMESol:Ang:Eq:AngHeunEq}
\end{align}
where
\begin{subequations}
\begin{align}
    \rho_\pm &= (1 - A_4) \pm A_4, \\
    u &= -\left[\frac{i \lambda}{4 \sqrt{\alpha}} + \frac{1}{2} + A_1 + \left(m + \frac{1}{2}\right) \left(A_3 - A_4\right)\right] \, .
\end{align}
\end{subequations}
A coefficient comparison with \cref{Heun:Eq:DEQ} leads to the identification of the Heun parameters,
\begin{subequations}
\begin{align}
    \gamma &= 2 A_1 + 1 \, , \\
    \delta &= 2 A_2 + 1 \, , \\
    \epsilon &= 2 A_3 + 1 \, , \\
    \alpha &= \rho_+ \, , \\
    \beta &= \rho_- \, , \\
    a_H &= z_a \, , \\
    q &= -u \, .
\end{align}
\end{subequations}
Inserting the Heun parameters into \cref{Heun:Eq:Identity}, the identity for the exponents yields
\begin{align}
    A_1 + A_2 + A_3 + A_4 = 0.
\end{align}
The arbitrary choice of signs for $A_3$, $A_4$ must respect this identity.

A remarkable fact is that for $a \rightarrow 0$, in the absence of frame dragging effects, the angular Teukolsky equation reduces to a spin-weighted Legendre equation. 
With normalization of the solution due to the orthogonality property, the polar part of the spin-weighted spherical harmonics is obtained, \cite{Goldberg1967}
\begin{widetext}
\begin{align}
    {}_sY_{lm}(\theta) = &(-1)^m \sqrt{\frac{(l + m)! (l - m)! (2l + 1)}{2 (l + s)! (l - s)!}} \sin^{2l} \left(\frac{\theta}{2}\right) e^{i m \phi} \sum\limits^{l -s}_{p = 0} \binom{l - s}{p} \binom{l + s}{p + s - m} (-1)^{l - p - s} \cot^{2p + s - m} \left(\frac{\theta}{2}\right) \, ,
\label{TMESol:Ang:Eq:spinSpherHarmonic}
\end{align}
\end{widetext}
which becomes the usual spherical harmonics for $s = 0$.

This fact is also reflected in \cref{TMESol:Ang:Eq:AngHeunEq} - or, more evidently, in the Heun's equation. 
The limit of the Kerr parameter $a \rightarrow 0$ also leads to the limit of $a_H \rightarrow 0$. 
Dividing \cref{Heun:Eq:DEQ} by $a_H$, considering the aforementioned limit, and letting $\beta, \epsilon, q \rightarrow \infty$ simultaneously, such that
\begin{align}
    \frac{\beta}{a} \rightarrow -\nu,~~~ \frac{\epsilon}{a} \rightarrow -\nu,~~~ \frac{q}{a} \rightarrow -\sigma \, ,
\end{align}
the general Heun differential equation reduces to the confluent Heun equation
\begin{align}
    \frac{d^2 y}{dz^2} + \left[\nu + \frac{\gamma}{z} + \frac{\delta}{z - 1}\right] \frac{d y}{d z} + \left[\frac{\alpha \nu z - \sigma}{z (z - 1)}\right] y = 0 \, ,
\label{TMESol:Ang:Eq:ConfHeun}
\end{align}
which has only two regular singularities at $z \in \{0, 1\}$ and one irregular singularity at $z = \infty$ due to the merging of $z = a_H, z=\infty$. 

\cref{TMESol:Ang:Eq:ConfHeun} can also be derived starting from the Legendre equation and transforming it into the confluent Heun equation, but bearing in mind that now there are only two regular singularities. 
Another equivalent approach is shown in \cite{Suzuki_1998}, where the general Heun equation is examined and various limits are considered in the three-term recurrence relations.
 
So far, the solution of the angular Teukolsky equation has been derived in terms of the Heun function. 
What is still missing is the resolution of the boundary condition and the derivation of the separation constant, or rather the eigenvalue ${}_s\lambda_{lm}$. 

\subsubsection{Eigenvalue problem}
\label{TMESol:Ang:Eigen}
The eigenvalues of spherical harmonics in an axially symmetric spacetime have been extensively discussed in the past \cite{Fackerell1977,Berti2006,Meixner1954}. 
An extension of this are the spin-weighted spherical harmonics, where the spin-weight modifies the polar component \cite{Goldberg1967} (cf. \cref{TMESol:Ang:Eq:spinSpherHarmonic}). 
In \cite{Suzuki_1998} the authors performed a successive approximation of the eigenvalue for KdS spacetimes by Heun functions. 
We will briefly present the results in the notation used here. 

To apply the successive approximation formula for \cref{Heun:Hf:Eq:infcontfrac}, the desired eigenvalue ${}_s\lambda_{lm}$ is extracted from the auxiliary parameter ${}_sq_{lm} = {}_s\tilde{q}_{lm} - i \frac{{}_s\lambda_{lm}}{4 \sqrt{\alpha}}$ and rearranged for ${}_s\lambda_{lm}$. 
An expansion in terms of the Kerr parameter $a$ is 
\begin{align}
    {}_s\lambda_{lm} = \sum\limits^\infty_{i=0} b_i a^i \, .
\label{TMESol:Ang:Eigen:Eq:SuccApprox}
\end{align}
It is important for comparability to emphasize the relation of this eigenvalue to Teukolsky's definition of the eigenvalue, which is ${}_s\lambda_{lm} := {}_sA_{lm} + 2s - 2 a \omega m + a^2 \omega^2$ \cite{Hatsuda2020,Teukolsky_1973}. 

Relabeling the index $n = l - (A_1 + A_2)$ in \cref{Heun:Hf:Eq:threeterm} introduces the new index $l$, which gives the zeroth-order coefficient \cite{Fackerell1977}.
\begin{align}
    b_0 = (l + A_1 + A_2 + 1) (l - (A_1 + A_2)) \, .
\label{TMESol:Ang:Eigen:Eq:0thOrder}
\end{align}
As mentioned above, the solution of the angular Teukolsky equation for spherically symmetric cases, i.e., Schwarzschild(-de Sitter), turns out to be the spin-weighted spherical harmonics ${}_sY_{lm}$, whose defining property is the regularity at the poles $x \in \{-1, 1\}$. 
Since the differential equation and its eigenvalue still qualify as a Sturm-Liouville problem, for which a Dirichlet-type boundary condition imposes the said regularity, the exact analytic eigenvalue in this case is
\begin{align}
    \lambda = l(l + 1) - s(s - 1) \, .
\label{TMESol:Ang:Eigen:Eq:SpinWghtSphHarmLambda}
\end{align}
This case must be reproduced in the limit $a \rightarrow 0$ of the more general case of the KdS spacetime treated here.
Thus, comparing \cref{TMESol:Ang:Eigen:Eq:SpinWghtSphHarmLambda,TMESol:Ang:Eigen:Eq:0thOrder},
\begin{align}
    A_1 + A_2 \overset{!}{=} -s ~\forall m,s. 
\end{align}
This can only be achieved if the signs of $A_1$, $A_2$ are chosen such that
\begin{subequations}
\begin{align}
A_1 &= \begin{cases}
+\left|\frac{m - s}{2}\right|, &m - s \geqslant 0 \\
-\left|\frac{m - s}{2}\right|, &m - s < 0
\end{cases} = \frac{m - s}{2} \, , \\
A_2 &= \begin{cases}
-\left|\frac{m + s}{2}\right|, &m + s \geqslant 0 \\
+\left|\frac{m + s}{2}\right|, &m + s < 0
\end{cases} = -\frac{m + s}{2} \, .
\end{align}
\end{subequations}
After the sign uncertainty is resolved by the boundary condition, the remaining coefficients of the successive approximation can be determined in \cref{TMESol:Ang:Eigen:Eq:SuccApprox}. 
The coefficients up to the fifth order are given in \cite{Suzuki_1998}.

In principle, the eigenvalues of $a \neq 0$ can be determined analytically by successive approximations. 
Due to its approximate nature, the resulting eigenvalues will not be accurate enough for applications. 
However, they serve as seed values for a numerical root-finding algorithm. 
Recalling that the eigenvalue can be determined by two approaches, the infinite continued fraction \cref{Heun:Hf:Eq:infcontfrac} or the Wronskian method \cref{Heun:Hf:Eq:Wronskian}, the second approach qualifies for a root-finding algorithm.

\subsubsection{Final solution}
\label{TMESol:Ang:Final}
The final solution is a combination of all possible solutions in the region of interest, i.e., the region between the north pole ($z = 1$) and the south pole ($z = 0$).
Thus,
\begin{align}
    {}_sS_{lm}(z) = \sum\limits_{i=0}^1 \sum\limits_{j=1}^2 X_{ij} \,{}_sS_{lm}^{(ij)}(z) \, .
\end{align}
However, depending on $m, s$, not all solutions are regular on the whole domain including the boundaries at the same time. 
We are only interested in the solutions that are regular on the poles for the given parameter set. 
Observing that the expansion of each solution at its respective regular singularity is \cref{Heun:Hl:Eq:Ay0,Heun:Hl:Eq:Ay1}, the coefficients $X_{ij}$ must be
\begin{subequations}
\begin{align}
X_{01}(z) = \begin{cases}
1, &A_1 \geqslant 0 \notag \\
0, &A_1 < 0
\end{cases} \, , ~&
X_{02}(z) = \begin{cases}
0, &A_1 \geqslant 0 \notag \\
1, &A_1 < 0
\end{cases} \, , \\
X_{11}(z) = \begin{cases}
1, &A_2 \geqslant 0 \notag \\
0, &A_2 < 0
\end{cases} \, , ~&
X_{12}(z) = \begin{cases}
0, &A_2 \geqslant 0 \notag \\
1, &A_2 < 0
\end{cases}
\end{align}
\end{subequations}
in order to reduce the sum to its regular solutions. 
The sum, combined with the aforementioned conditions, can also be rewritten as
\begin{align}
{}_sS_{lm}(z) = {}_sS_{lm}^{(0i)}(z) + {}_sS_{lm}^{(1j)}(z) \, ,
\end{align}
where the indices $i,j$ refer to the angular solutions
\begin{align}
i = \begin{cases}
1, & A_1 \geqslant 0 \\
2, & A_1 < 0
\end{cases} \, ,~ 
j = \begin{cases}
1, & A_2 \geqslant 0 \\
2, & A_2 < 0
\end{cases} \, .
\label{TMESol:Ang:Final:Eq:IndexCond}
\end{align}
The general solution now consists of two solutions in total. 
Recalling that the eigenvalue discussion transforms local Heun functions $Hl$ into Heun functions $Hf$, and the remaining two solutions of each regular singularity become linearly dependent, this property, expressed as ${}_sS_{lm}^{(0i)}(z) = {}_s\Theta^{(i \rightarrow j)}_{lm} {}_sS_{lm}^{(1j)}(z)$, simplifies the general solution even more to
\begin{align}
{}_sS_{lm}(z) = \left(1 + \frac{1}{{}_s\Theta^{(i \rightarrow j)}_{lm}}\right) {}_sS_{lm}^{(0i)}(z). \label{TMESol:Ang:Final:Eq:Final}
\end{align}
${}_sS_{lm}(z)$ can be reformulated in a last step by dividing the prefactor of the left-hand side, so that the definition
\begin{align}
{}_s\tilde{S}_{lm}(z) := {}_sS_{lm}^{(0i)}(z) \notag
\end{align}
finally expresses the general solution by only one of the remaining solutions, which still depends on $m,s$ via the $i$-index.

Since the Heun function is evaluated in Mathematica using the series representation \cref{Heun:Hf:Eq:FunSeries}, the greater the distance between the evaluation point and the respective regular singularity, the more terms must be calculated, which may take a substantial amount of time. 
In particular, this means that, e.g., computing a point near $z = 1$ by a solution of $z = 0$ takes considerably longer. 
Using the linear dependence of the Heun functions reduces the computational cost at this point. 
The final general solution is therefore calculated by
\begin{align}
    {}_s\tilde{S}_{lm}(z) = \begin{cases}
{}_sS_{lm}^{(0i)}(z), &|z| \leqslant \frac{1}{2} \\
    {}_s\Theta^{(i \rightarrow j)}_{lm} {}_sS_{lm}^{(1j)}(z), &|z| > \frac{1}{2}
\end{cases}\, ,
\label{TMESol:Ang:Final:Eq:FinalComp}
\end{align}
where the proportionality constant of the linear dependence is ${}_s\Theta^{(i \rightarrow j)}_{lm} = {}_sS_{lm}^{(0i)}(1/2) / {}_sS_{lm}^{(1j)}(1/2)$. 
Here, $z = 1/2$ is chosen to switch from one definition to another because it lies between the two regular singularities within the mutual convergence region, thus minimizing computational cost. 
The modulus of $z$ in the case of conditions of \cref{TMESol:Ang:Final:Eq:FinalComp} is necessary since $z(x)$ for $x \in [-1, 1]$ is a path in the complex plane. 

\subsection{Radial solution}
\label{TMESol:Rad}
The radial Teukolsky equation \cref{TME:Sep:Eq:Rad} is solved similarly to the angular case. 
The transformation of the independent variable uses the regular singularities, which are equal to all radial locations of the horizons, 
\begin{align}
z(r) = \frac{r'_+ - r-}{r'_+ - r+} ~\frac{r - r_+}{r - r_-}\, .
\label{TMESol:Rad:Eq:IndepTrans}
\end{align}
These are transformed as $\{r'_-, r_-, r_+, r'_+, \infty\} \rightarrow \{z_r,\infty,0,1,z_{r\infty}\}$, where 
\begin{subequations}
\begin{align}
    z_{r\infty} &= z|_{r \rightarrow \infty} = \frac{r'_+ - r-}{r'_+ - r+} \, , \\
    z_r &= z|_{r \rightarrow r'_-} = \frac{r'_+ - r-}{r'_+ - r+} \frac{ r'_- - r_+}{ r'_- - r_-} \, .
\end{align}
\end{subequations}
The physical region of interest, the domain of outer communication, lies between the event horizon $r_+$ and the cosmological horizon $r'_+$, i.e., $z \in [0, 1]$. 
The regular singularity at $r = \infty$ can again be removed by an additional factor in the f-homotopic transformation of the dependent variable. 
Thus,
\begin{align}
    {}_sR^\text{(ij)}_{lm}(z) = &z^{B_1} (z - 1)^{B_2} (z - z_r)^{B_3} (z - z_{r\infty})^{B_5} \notag \\
&\times y^{(r)}_{ij}({}_s\lambda_{lm}; z).
\end{align}
The exponents take a particular form in order to complete the transformations to the canonical form of Heun's equation, in which their definition can be conveniently written by $K_m(r)$, giving
\begin{align}
    B_i = \pm i \left|B(r_i)\right|
\end{align}
with
\begin{align}
    B(r) = \frac{K_m(r)}{\Delta'_r(r)} \, ,
\end{align}
where $i \in \{1, 2, 3, 4\}$ and $r_i \in \{r_+, r'_+, r'_-, r_-\}$ and ${B_5 = 2 s + 1}$. 
As in the angular case, $B_4 = B^*_3$. 
Finally, the transformation results in 
\begin{align}
    &\frac{d^2 y}{d z^2} + \left(\frac{2 B_1 + s + 1}{z} + \frac{2 B_2 + s + 1}{z -1} + \frac{2 B_3 + s + 1}{z - z_r}\right)\frac{d y}{d z} \notag \\
    & + \frac{\sigma_+ \sigma_- z + v}{z (z -1) (z - z_r)}y = 0 \, , \label{TMESol:Rad:Eq:HeunForm}
\end{align}
where the remaining functions are
\begin{subequations}
\begin{align}
    \sigma_\pm = &\left(1 - B_4 + \frac{3}{2} s\right) \pm \left(B_4 + \frac{1}{2} s\right), \\
    v = &\frac{\lambda - 2s(1 - \alpha) - \frac{\Lambda}{3} (s + 1) (2s + 1) (r_+ r_- + r'_+ r'_-)}{\frac{\Lambda}{3}(r_- - r'_-) (r_+ r'_+)} \notag \\
    & - \frac{i(2s + 1) [2(1 + \alpha) \{\omega (r_+ r_- + a^2) - a m\}}{\frac{\Lambda}{3} (r_- - r'_-) (r_- - r_+) (r_+ - r'_+)} \, . 
\end{align}
\end{subequations}
A coefficient comparison between \cref{Heun:Eq:DEQ,TMESol:Rad:Eq:HeunForm} identifies the Heun parameters as
\begin{subequations}
\begin{align}
    \gamma &= 2 B_1 + 1 \, , \\
    \delta &= 2 B_2 + 1 \, , \\
    \epsilon &= 2 B_3 + 1 \, , \\
    \alpha &= \sigma_+ \, , \\
    \beta &= \sigma_- \, , \\
    a_H &= z_r \, , \\
    q &= -v
\end{align}
\end{subequations}
with the identity for the exponents 
\begin{align}
    B_1 + B_2 + B_3 + B_4 = 0 \, .
\end{align}
As in the angular case, the exponents have a sign ambiguity. 
While in the angular case the signs are resolved by requiring regularity at $x = \pm 1$, in the radial case they are resolved by a more complex boundary condition discussed in the next section. 

\subsubsection{Boundary condition and final solution}
\label{TMESol:Rad:BC}
\begin{figure}
\centering
\includegraphics[width=\linewidth]{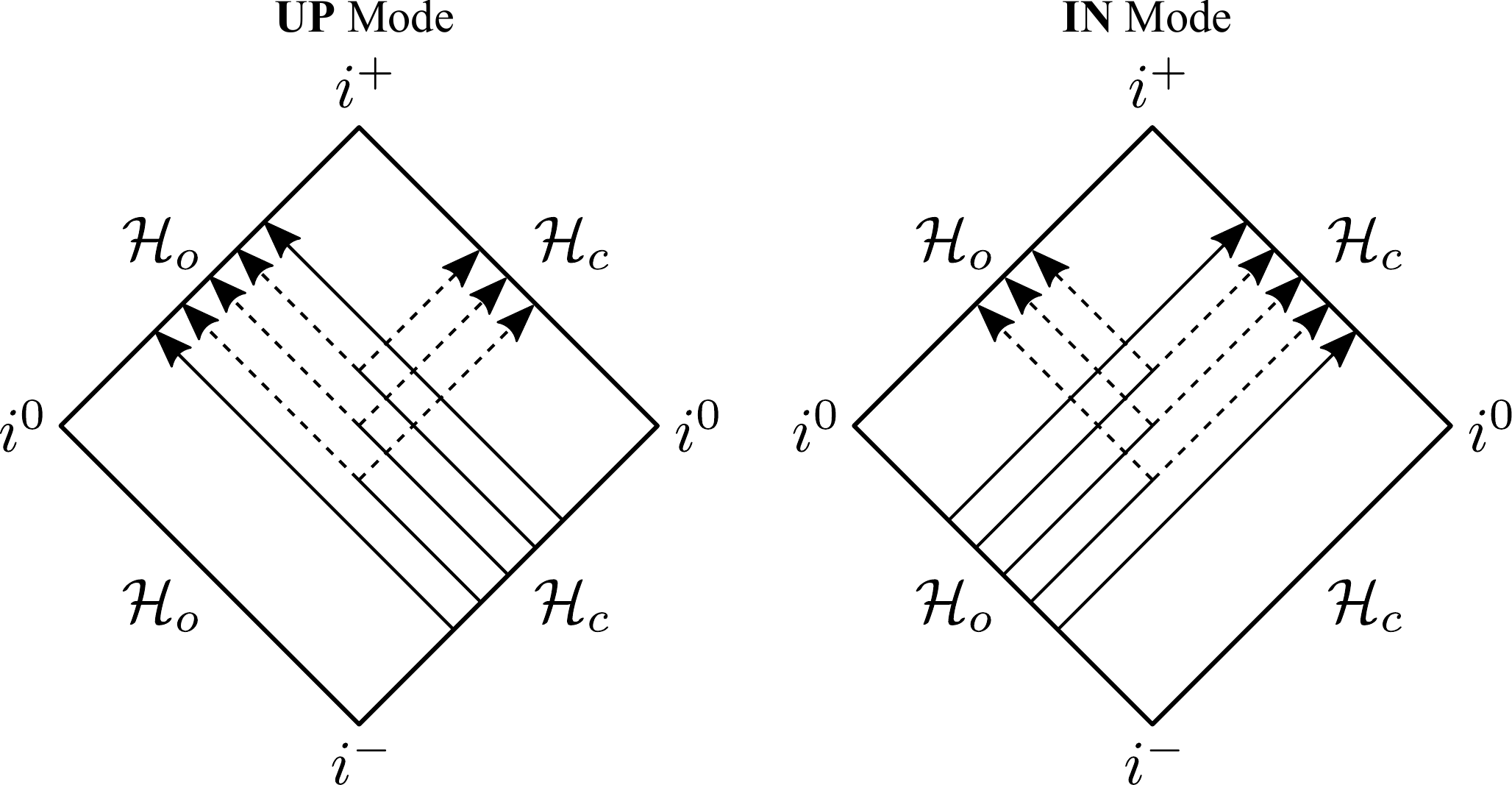}
\caption{Penrose diagram showing only an excerpt of the fully analytically extended diagram (domain of outer communication). 
Purely ingoing waves (\textbf{IN} mode) and purely outgoing waves (\textbf{UP} mode) are scattered to both horizons by the black hole. 
$\mathcal{H}_o$ is the event horizon, where $\mathcal{H}_c$ is the cosmological horizon. 
An extended analytical version can be found in \cite{Akcay_2011}.}
\label{TMESol:Rad:BC:Fig:InUpMode}
\end{figure}

The boundary condition imposed on the radial solution, which leads to the full solution of the radial Teukolsky equation, roots from the physical context of black holes. 
The main property of horizons is the semipermeability of the information flow, i.e., in the case of the event horizon $r_+$, that information falls into the black hole, but nothing escapes it in a classical sense. 
In the case of the cosmological horizon $r'_+$, information crossing it to the outside of the domain of outer communication will be unreachable for an observer due to the superluminal expansion behind it. 
Taking this into account, two linear independent solutions can be formulated: The In- and the Up-mode, as discussed in many publications (see e.g. \cite{Teukolsky_1973,Leaver_1986,Andersson2000,Glampedakis2001,Motohashi2021, Dolan2009,Nambu2016,Nambu2019}). 
A Penrose diagram depicts this in \cref{TMESol:Rad:BC:Fig:InUpMode}. 
Its mathematical construction is adequately done in terms of the asymptotic behavior of the radial solution at the horizons. 
To do this, the radial Teukolsky equation is transformed into a quasi-Schrödinger representation \cite{Motohashi2021}, which leads to
\begin{align}
    R(r) \sim \Delta_r^{-\frac{s}{2} \pm \left(B_h + \frac{s}{2}\right)}(r) \, , ~r \rightarrow r_h \, .
\label{TMESol:Rad:BC:Eq:SolAsymp}
\end{align}
This, in combination with the time-dependent term from the separation ansatz $e^{-i \omega t}$, determines the in- and outgoing properties of the respective waves by proper choice of signs. 
On the basis of this, the In- and Up-modes can be defined as
\begin{subequations}
\label{TMESol:Rad:BC:Eq:BCSetup}
\begin{align}
R_{\text{in}} \rightarrow &\begin{cases}
C^{(\text{trans})} \Delta^{-B_1}, &(r \rightarrow r_+) \\
C^{(\text{ref})} \Delta^{B_2} + C^{(\text{inc})} \Delta^{B_2}, &(r \rightarrow r'_+)
\end{cases} \, , \label{TMESol:Rad:BC:Eq:BCSetupRin} \\
R_{\text{up}} \rightarrow &\begin{cases}
D^{(\text{up})} \Delta^{B_1} + D^{(\text{ref})} \Delta^{-B_1}, &(r \rightarrow r_+) \\
D^{(\text{trans})} \Delta^{-B_2}, &(r \rightarrow r'_+) \\
\end{cases} \, . \label{TMESol:Rad:BC:Eq:BCSetupRup}
\end{align}
\end{subequations}
The six scattering coefficients $C^\text{(trans)}$, $C^\text{(ref)}$, $C^\text{(inc)}$, $D^\text{(up)}$, $D^\text{(ref)}$, $D^\text{(trans)}$ are to be determined. 
With ${}_sR^\text{(in)}_{lm}(r) = {}_sR^\text{(02)}_{lm}({}_s\lambda_{lm}; r)$ and ${}_sR^\text{(up)}_{lm}(r) = {}_sR^\text{(11)}_{lm}({}_s\lambda_{lm}; r)$ , the corresponding asymptotic behavior of \cref{TMESol:Rad:BC:Eq:SolAsymp} coincides with \cref{Heun:Hl:Eq:Ay01,Heun:Hl:Eq:Ay12}, respectively. 
Using \cref{Heun:CC:Eq:y02,Heun:CC:Eq:y11}, it can be concluded that
\begin{subequations}
\label{TMESol:Rad:BC:Eq:R}
\begin{align}
{}_sR^\text{(in)}_{lm}(r) &= \begin{cases}
R_{0 2}(r), &(r \rightarrow r_+) \\
C_{21} R_{11}(r) + C_{22}R_{12}(r), &(r \rightarrow r'_+)
\end{cases} \, , \label{TMESol:Rad:BC:Eq:Rin} \\
{}_sR^\text{(up)}_{lm}(r) &= \begin{cases}
D_{11} R_{01}(r) + D_{12}R_{02}(r), &(r \rightarrow r'_+) \\
R_{11}(r), &(r \rightarrow r'_+)
\end{cases} \, . \label{TMESol:Rad:BC:Eq:Rup}
\end{align}
\end{subequations}
A coefficient comparison of \cref{TMESol:Rad:BC:Eq:BCSetup,TMESol:Rad:BC:Eq:R} expresses the scattering coefficients in terms of the Heun connection coefficients, which can be looked up in \cite{Motohashi2021}.

It should be emphasized that this approach to wave scattering by a black hole allows for a fully analytical solution in terms of the local Heun functions and connection coefficients \cref{Heun:CC:Eq:C,Heun:CC:Eq:D}. 
Applications of the scattering coefficients include the derivation of reflection and transmission, which are important for S-matrices and evaluation of differential cross-sections. 
This becomes possible by the presence of a positive cosmological constant, i.e., in a de Sitter spacetimes. 
However, a negative cosmological constant, i.e., in anti-de Sitter spacetimes, leads to complex radial values of the negative and positive cosmological horizons $r'_-$, $r'_+$. 
Therefore, $r'_+$ no longer limits the domain of outer communication, but $r = \infty$. 
Because $r = \infty$ is not an irregular singularity for \cref{TME:Sep:Eq:Rad} in case of $\Lambda < 0$, a radial boundary condition can be formulated such that the solutions are analytical again. 
For a proper description in this case check Ref. \cite{Noda2022}.

\subsection{Scattering via Green's function}
\label{WavOpt:Green}
The scattering of monochromatic point sources by a KdS black hole is described by the spatial Green's function, for which the time-dependent part $e^{-i \omega t}$, acting as an $l$ and $m$ independent total complex phase shift, is omitted. 
It is calculated for all possible partial waves, which are written as a product of the solutions of the radial and angular Teukolsky equations, as well as the azimuthal part from \cref{TME:Sep:Eq:Sep}. 
In addition to $_sS_{l m}(\theta)$, the normalization constant $_s\zeta_{lm}$ from \cref{Heun:Hf:Normal:Eq:NormConst} ensures the existence of an orthonormal basis, similar to the case of spherical harmonics, allowing expansion of any square-integrable function in the manner of a Fourier series.
Finally, the Green's function is \cite{Motohashi2021}
\begin{align}
G(\vec{r}_O, \vec{r}_s, L) = \sum\limits^{L}_{l = s} \sum\limits^l_{m = -l} &{}_s\tilde{G}_{lm}(r_O, r_s) \frac{_sS_{l m}(\theta_O)}{{}_s\zeta_{lm}} \frac{_sS_{l m}^*(\theta_s)}{{}_s\zeta_{lm}} \notag \\
& \times e^{i m \phi_O} e^{-i m \phi_s} \, ,
\label{WavOpt:Green:Eq:GreenFun}
\end{align}
where a star marks the complex conjugate. 
The $s$-subscripts of the Boyer-Lindquist coordinates indicate the coordinates of the point source, while the $O$-subscripts indicate the observer's coordinates, and
$i,j$ are selected depending on \cref{TMESol:Ang:Final:Eq:IndexCond}.
The complete analytical solution is defined by an infinite sum $L = \infty$\footnote{For computing the results, a small discussion of convergence and truncation of the sum can be found in \cref{Appendix:Convergence}.}. 
In the case of the radial Teukolsky equation, the solution for a radiating point source, arbitrarily placed around the black hole, roots from the modified differential equation
\begin{multline}
    \left(\Delta^{-s} \frac{d}{dr}\left(\Delta^{s+1} \frac{d}{dr} \right) + {}_sV_{lm}^\text{(rad)}(r)\right) {}_s\tilde{G}_{lm}(r,r_s) \\ 
    = -\delta(r - r_s) \, .
\end{multline}
For In- and Up-solutions satisfying the boundary conditions introduced in \cref{TMESol:Rad:BC:Eq:BCSetup}, it is required that the linear independent solutions coincide at the location of the source. 
In the context of Green's functions, this is done in the standard way via
\begin{align}
    {}_s\tilde{G}_{lm}(r_O, r_s) = 
    &\frac{-\Delta^s(r_s)}{\Delta^{s+1} W_r\left[R_\text{in},R_\text{out}\right]} \notag \\
    & \times \bigg\{R_\text{in}(r_s) R_\text{up}(r_O) \Theta(r_O - r_s) \label{WavOpt:Green:Eq:radGreen} \\
    &~~~~+ R_\text{in}(r_O) R_\text{up}(r_s) \Theta(r_s - r_O)\bigg\} \, , \notag
\end{align}
where $\Theta$ is the Heaviside step function. 
Therefore, depending on the order of $r_O$ and $r_s$, one or the other term becomes relevant. 
It is important to note that $\Delta^{s+1} W_r\left[R_\text{in},R_\text{out}\right]$ is constant and is evaluated at some point in the overlapping convergence domain of the radial solutions.

\section{Wave-optical imaging of black holes}
\label{Results}
In the following, the wave-optical imaging of black holes is discussed based on the results above. 
The Green's function allows to describe the scattering of a point source and interference effects at arbitrary points around a black hole. 
However, the interference itself is not sufficient to achieve wave-optical imaging. 
For this, a short revision of this missing step is given below. 
After that, we investigate black hole scattering of scalar wave point sources in Schwarzschild-de Sitter and Kerr-de Sitter.
The main goal is to validate the wave-optical images with previous results from the ray-optical approaches.
The formation of an Einstein ring, frame-dragging, the wave-optical shadow, and additional image splitting in the presence of a rotating black hole, which is remotely predicted in the weak-gravitational case, are considered.

\subsection{Wave-optical imaging}
The choice of scalar ($s = 0$) point sources in first-order perturbations simplifies the study of scattering as a useful model for other spin fields \cite{Andersson2000}, where polarization degrees of freedom become negligible.
Furthermore, scalar waves allow access to wave-optical imaging via the Kirchhoff-Fresnel diffraction theorem \cite{Sharma2006}. 
A typical setup is shown in \cref{WavOpt:FourierOpt:Fig:Scheme}, also used in previous works \cite{Kanai2013,Nambu2016,Schneider1999,Turyshev2020,Turyshev2021}. 

Waves coming from a source pass through the lens plane and are diffracted to the image plane, resulting in the wave-optical image. 
In the lens plane, different shapes and types can be considered, e.g., a simple aperture or a convex lens, giving significantly different results.
In our case, we focus on the latter, as done in \cite{Kanai2013}, with the aperture size $d = 2M$ assumed in all further calculations. 
The far-field approximation of the Kirchhoff-Fresnel diffraction, also called Fourier optics, states that the diffraction integral in the image plane is approximately the two-dimensional Fourier transform of the lens plane, written as
\begin{align}
    \psi\left(\vec{r''}\right) \propto \int\int_\Sigma \Phi\left(\vec{r}\right) e^{-i \frac{\omega x''}{R_0''} x} e^{-i \frac{\omega y''}{R_0''} y} dx\, dy\, ,
\label{WavOpt:FourierOpt:Eq:FourierDiffInt}
\end{align}
where $\Sigma = x^2 + y^2 \leqslant d^2$ is the lens shape and $\Phi(\vec{r})$ is the field of the source in the lens plane. 
Note that $Z''$ is identical to the focal length $f$ of the lens in our considerations. 
The wave-optical image is evaluated by the absolute square of the Fourier transform \cref{WavOpt:Green:Eq:GreenFun} $\left|\mathcal{F}\left(G(\vec{r}, \vec{r}_s, l_\text{max})\right)\right|^2$, where $\mathcal{F}(x)$ is a short-hand notation for the Fourier transform. 

\begin{figure}
\centering
\includegraphics[width=\linewidth]{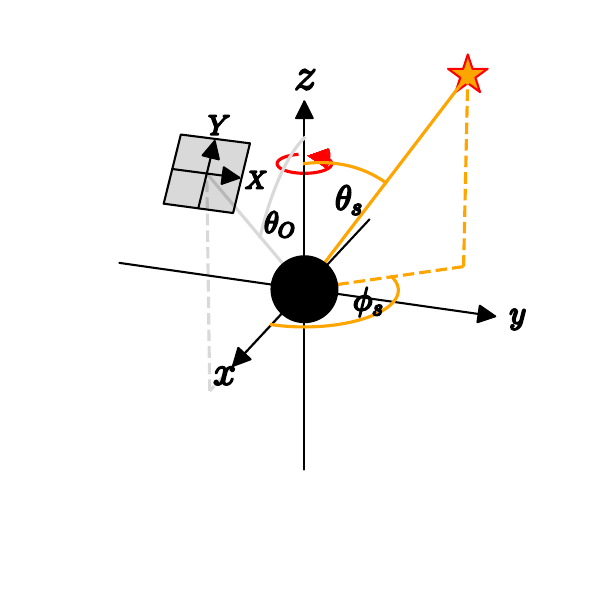}
\caption{Schematic illustration emphasizing the relationships between the polar coordinate $\theta_O$ of the observer plane and the point source coordinates $\left(\theta_s, \phi_s\right)$ in the black hole coordinate system. 
The gray plane is the observer plane on which the Green's function is evaluated. 
It coincides with the lens plane in \cref{WavOpt:FourierOpt:Fig:Scheme} with its respective plane coordinate system defined by capital $X,Y$.}
\label{Results:Fig:ObsSrcLoc}
\end{figure}

\begin{figure}
\centering
\includegraphics[width=\linewidth]{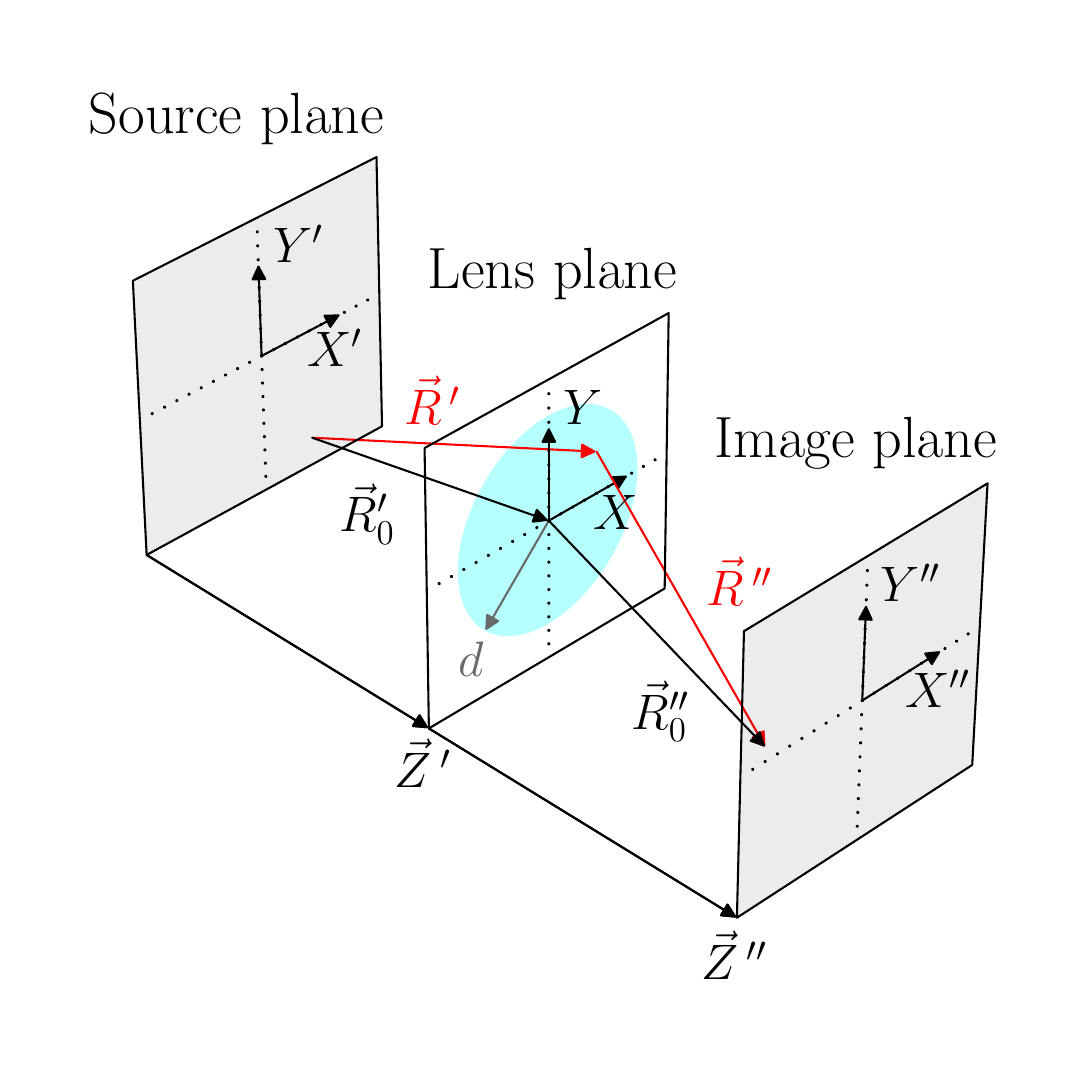}
\caption{Schematic plane arrangement. 
From a point in the source plane a wave is emitted to the lens plane, described by $\vec{r}'$, where $\vec{R}'_0$ is the distance to the origin of the lens plane. 
$\vec{r}''$ and $\vec{R}''_0$ have the same properties for the lens plane-image plane relation. 
The faint blue circle indicates the thin convex lens with radius $d$.}
\label{WavOpt:FourierOpt:Fig:Scheme}
\end{figure}

\subsection{Notes on the construction}
A two-dimensional observer plane placed around the black hole is considered, always facing towards the black hole regardless of its location.
Details can be found in \cref{Appendix:WavOpt:Coord} and the plane coincides with the lens plane mentioned in the last subsection.

Assuming that the distance between the observer and the black hole is much larger than the radial position in the observer plane $r_O \ll d$, the radius of each point can be considered as constant, and thus one radial solution suffices.
Note that the azimuthal coordinate of the observer plane $\phi_O$ is not listed in the set of free parameters because it is sufficient to modify only $\phi_s$.
Therefore, $\phi_O = 0$ is assumed.
Varying $\phi_S$ instead of $\phi_O$ also avoids recalculating the angular solutions of the observer plane, which drastically reduces computational cost.

The number of discrete points in the observer plane is reduced to a grid of $x_N = y_N = 71$ to balance resolution and computational cost. 
To increase the spectral resolution of the Fourier transformed observer plane, zero-padding of the result is applied as a post-processing step, which essentially appends zeros in both spatial dimensions of the discrete calculations.
Here, we will zero-pad the input to an output of $1071 \times 1071$ discrete points.
As another post-processing step, a Tukey filter\footnote{see \cref{Appendix:Tukey} for details} is applied to reduce aliasing effects in the Fourier transformation process. 

For the cosmological constant, a small value of $\Lambda M^2 = 10^{-3}$ is considered. 
Thus, the set of free parameters to examine is $\{M\omega, a, r_O, r_s, \theta_O, \theta_s, \phi_s\}$. 

Primarily, the computation time depends on the frequency, due to the cut-off of the Green's function (cf. \cref{Appendix:Convergence}), and the resolution of the observer plane. 
For reference, the most time-consuming calculation for $M\omega = 15$ and $a = 0.99M$ required 10 days of computation parallelized over 40 kernels. 

Note that the plots in the image plane are normalized to their respective maximum absolute value for each result shown. 
Consequently, it appears that each wave-optical image shown here has the same luminosity, which is not actually the case and limits the comparability of the results shown in terms of magnitude. 
\cref{Results:ShadowCompare:Fig:SourceSuperpositionSingleStepMaxStat} shows the maximum $\max(|G(\vec{r}, \vec{r}_s, l_\text{max}|)$ for different locations of the point source to give an impression of different normalizations.

\subsection{Schwarzschild-de Sitter}
\label{Results:SdS}
In the case of the SdS metric ($a = 0$), the normalized solution of the angular Teukolsky equation reduces to spin-weighted spherical harmonics \cref{TMESol:Ang:Eq:spinSpherHarmonic}, which reduces computational complexity and is independent of the choice of $\Lambda$. 
The radial Teukolsky equation depends on the $m$-multipole index in \cref{TME:Sep:Eq:RadK} through the product with the Kerr parameter $a$. 
Therefore, setting $a = 0$, the radial solution becomes independent of $m$, which considerably reduces the number of terms that must be computed.

\begin{figure*}
\centering
\subfloat[$M \omega = 4$]{\includegraphics[width=0.33\linewidth]{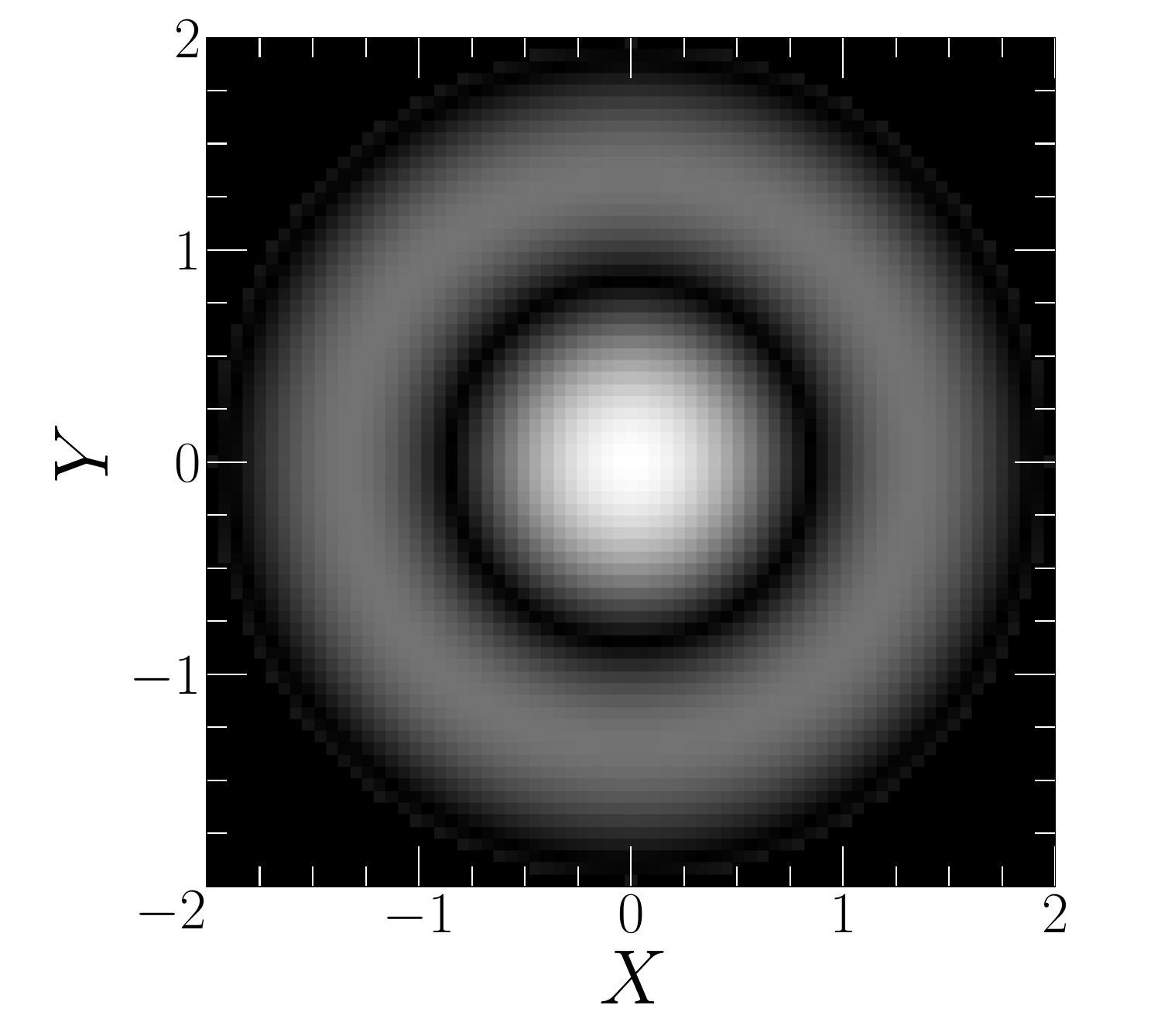}\label{Results:SdS:Fig:Mwvar:IMMw4}} \hfill
\subfloat[$M \omega = 12$]{\includegraphics[width=0.33\linewidth]{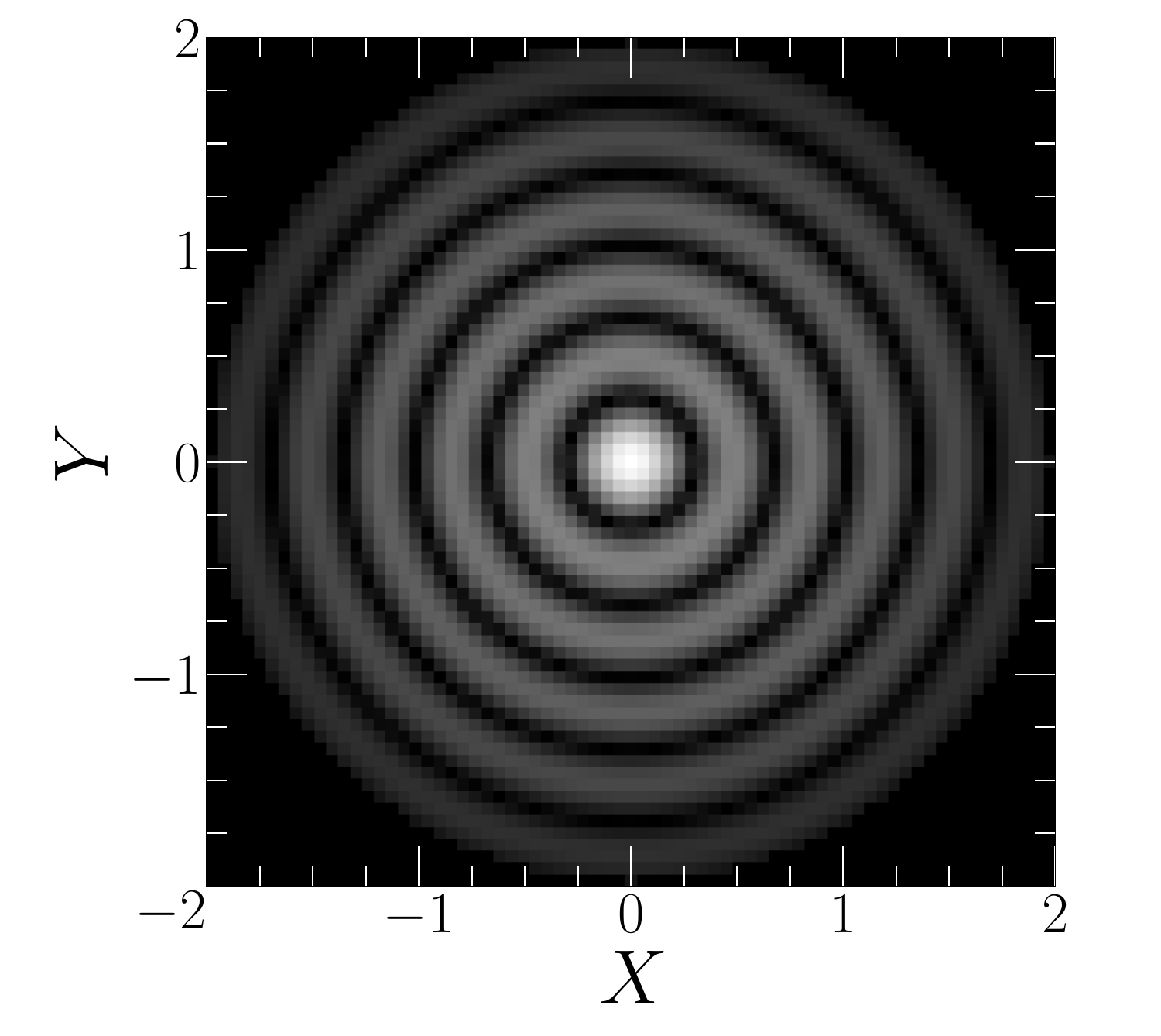}\label{Results:SdS:Fig:Mwvar:IMMw12}} \hfill
\subfloat[$M \omega = 18$]{\includegraphics[width=0.33\linewidth]{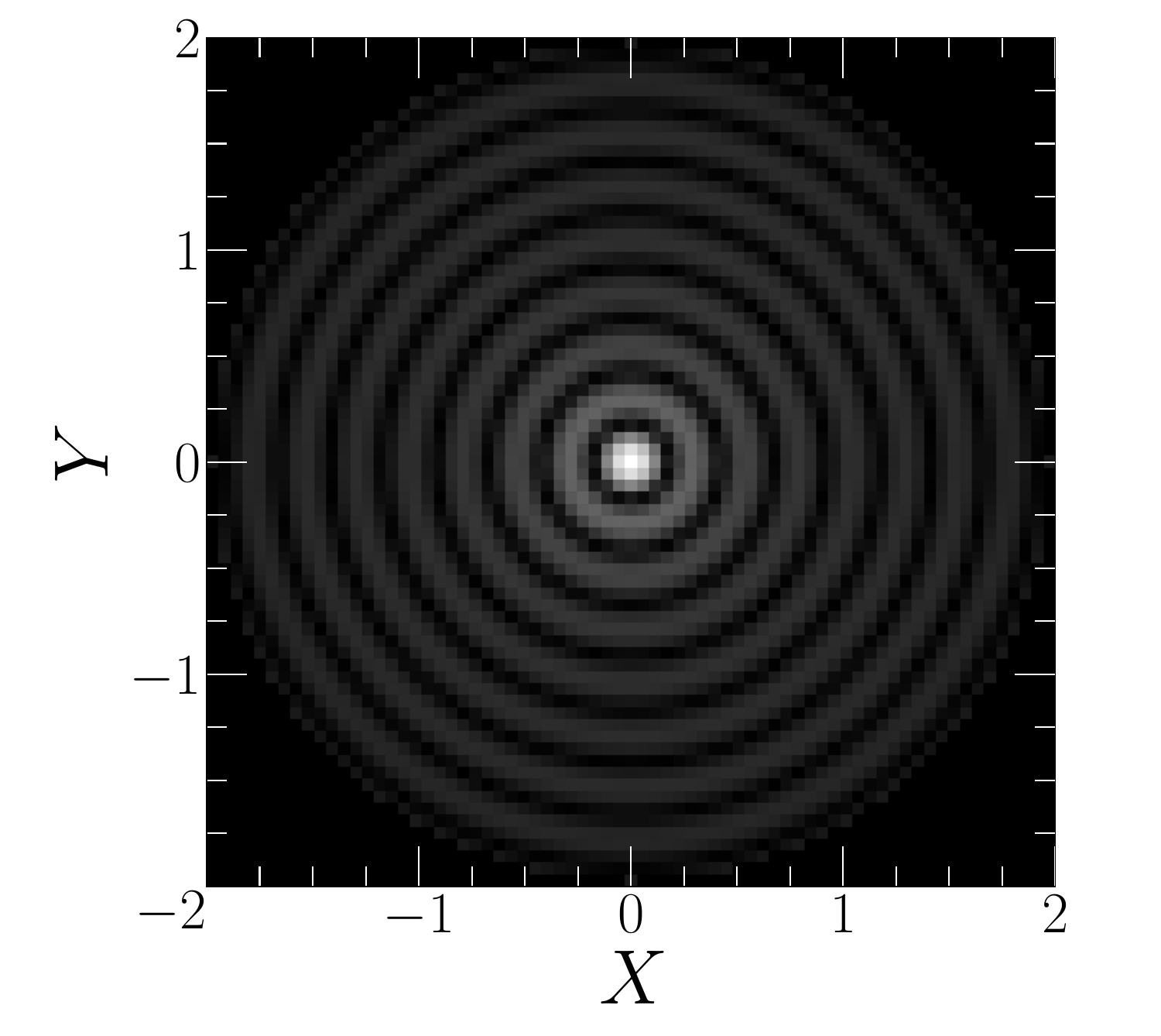}\label{Results:SdS:Fig:Mwvar:IMMw18}} \\
\subfloat[$M \omega = 4$]{\includegraphics[width=0.33\linewidth]{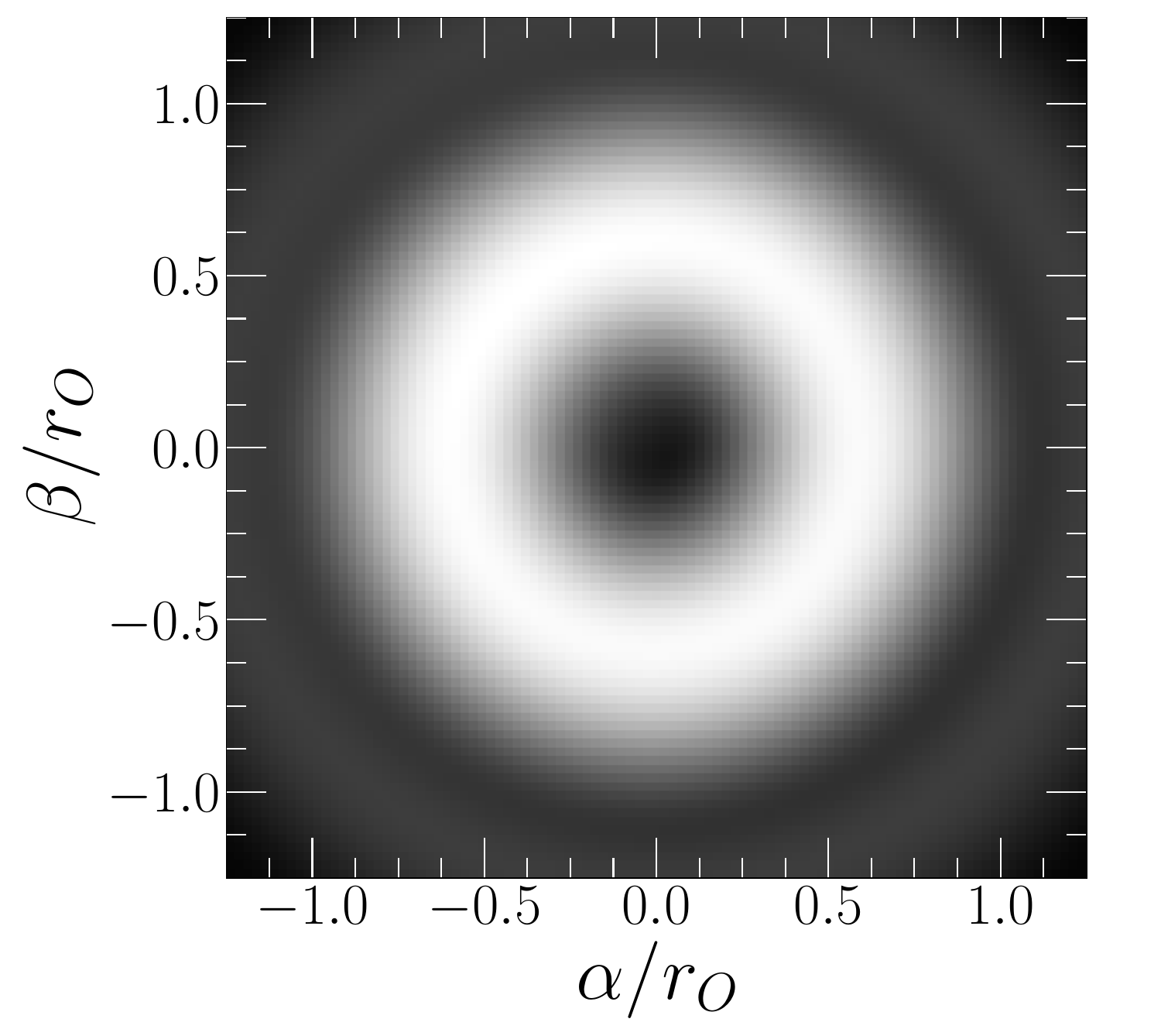}\label{Results:SdS:Fig:Mwvar:FFTMw4}} \hfill
\subfloat[$M \omega = 12$]{\includegraphics[width=0.33\linewidth]{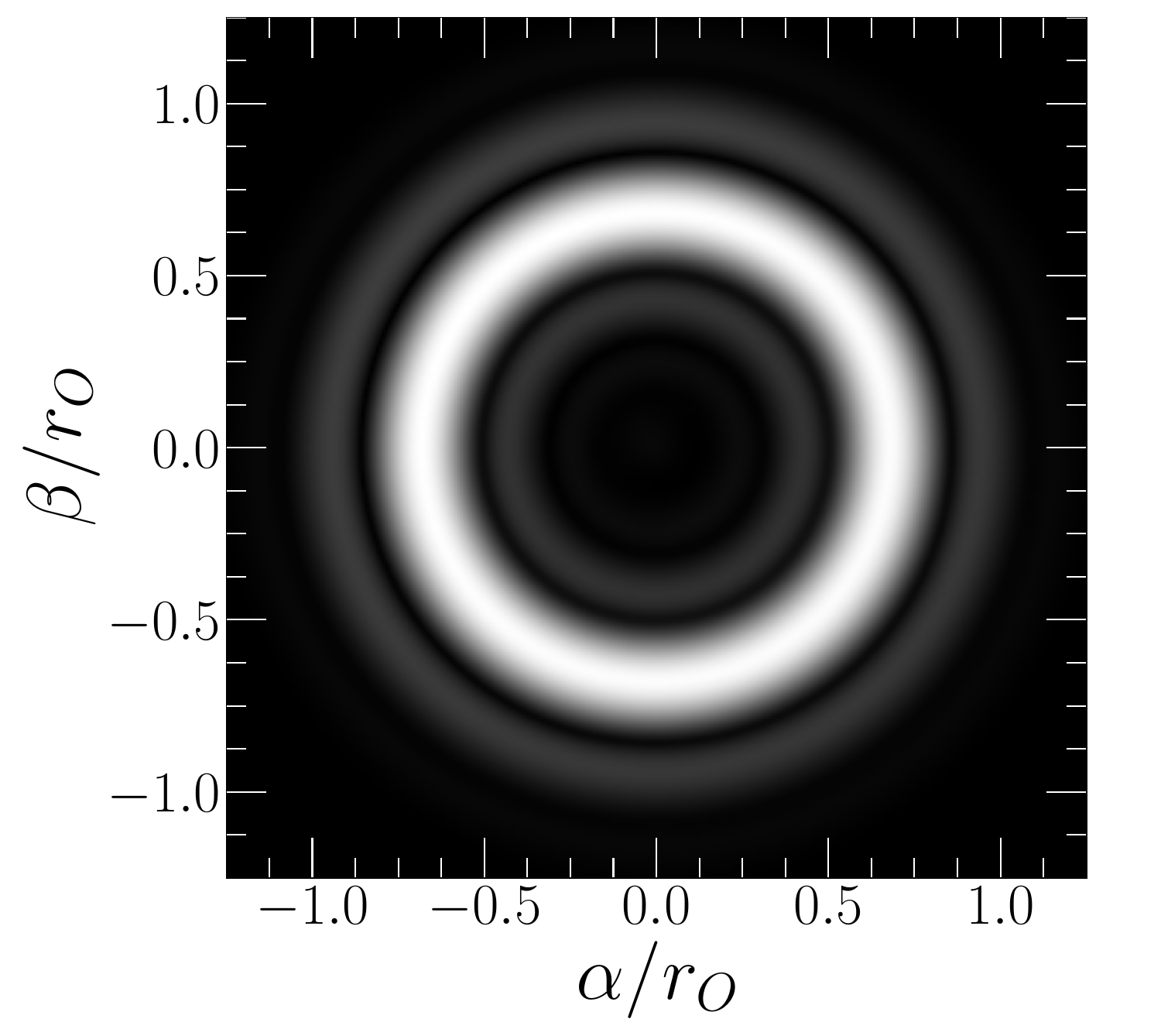}\label{Results:SdS:Fig:Mwvar:FFTMw12}} \hfill
\subfloat[$M \omega = 18$]{\includegraphics[width=0.33\linewidth]{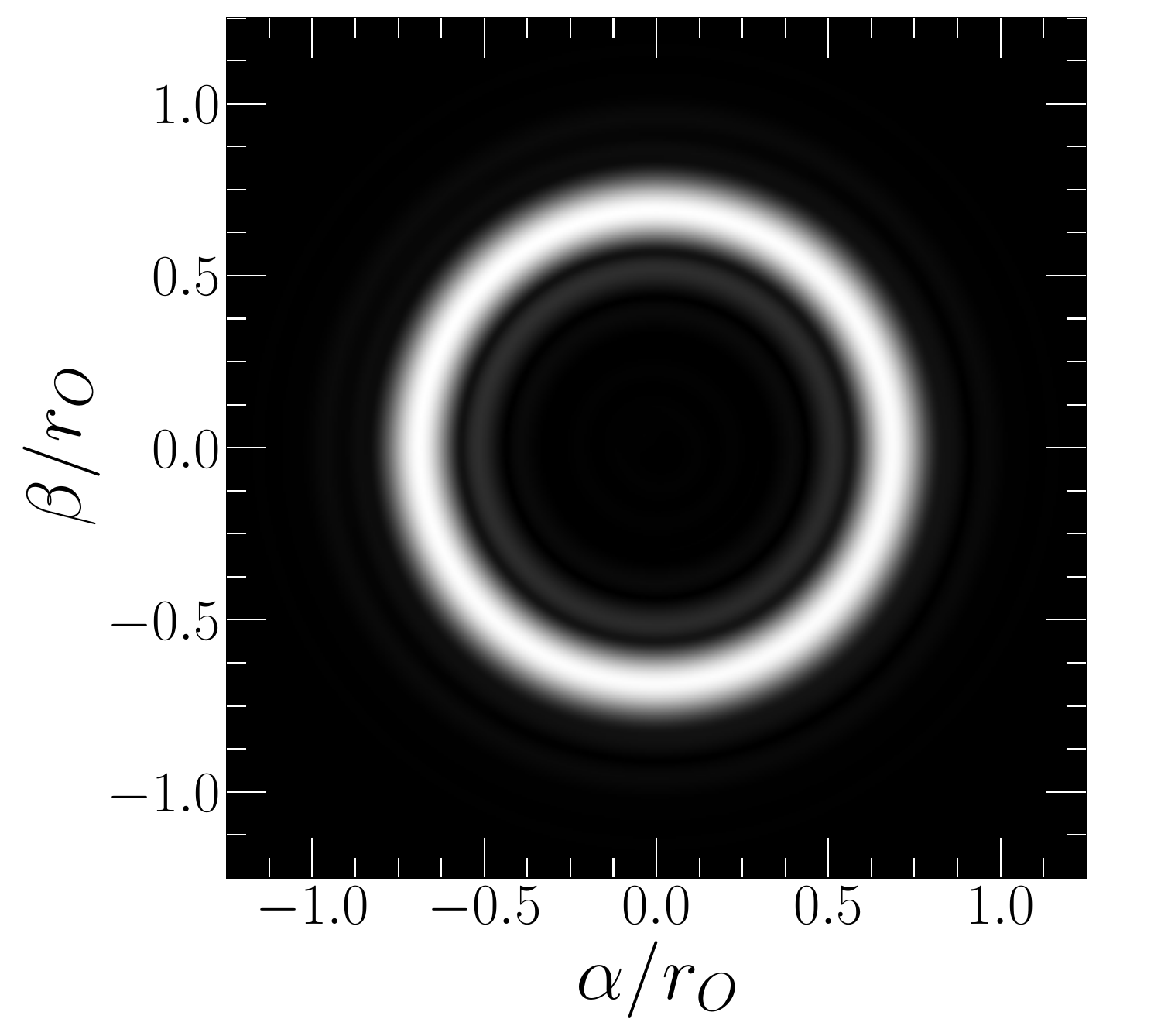}\label{Results:SdS:Fig:Mwvar:FFTMw18}}
\caption{Variation of the source frequency $M\omega \in \{4, 12, 18\}$ for SdS, using parameters $a = 0.00 M$, $\Lambda M^2 = 10^{-3}$, $r_O = 10 M$, $r_s = 20 M$, $\theta_O = \frac{\pi}{2}$, $\theta_s = \frac{\pi}{2}$, $\phi_s = \pi$. 
In this constellation, the Einstein ring forms. 
The upper row shows the imaginary part $\Im\left(G(\vec{r}, \vec{r}_s, l_\text{max}\right)$, whereas in the lower row the resulting image from $\left|\mathcal{F}\left(G(\vec{r}, \vec{r}_s, l_\text{max})\right)\right|^2$ is shown, respectively. 
An apparent effect of increasing frequency is a sharper Einstein ring in the image plane and finer structures of interference in the observer plane.}
\label{Results:SdS:Fig:Mwvar}
\end{figure*}

\cref{Results:SdS:Fig:Mwvar} shows the frequency variation for an SdS black hole. 
The observer plane and the point source are located in the equatorial plane and aligned antipodally ($\theta_O = \frac{\pi}{2}$, $\theta_s = \frac{\pi}{2}$ and $\phi_s = \pi$). 
For three different results of $M\omega \in \{4, 12, 18\}$ and radial locations $r_O = 10 M$, $r_s = 20 M$, the wave-optical images are computed. 
In this alignment, previous results predict the formation of a so-called Einstein ring. 
The upper row of the figure shows the imaginary part of the Green's function. 
The concentric circles become finer as $M\omega$ increases. 
Actual wave-optical images resulting from the scattering are shown in the lower row, respectively. 
The resulting images are consistent with the prediction of Einstein rings, which become sharper as the frequency increases.
However, unlike the ray-optical approach, these images show both an expansion and a distribution of magnitude.

\begin{figure*}
\centering
\subfloat[$\theta_s = \frac{4}{10}\pi$, $\phi_s = \frac{11}{10} \pi$]{\includegraphics[width=0.33\linewidth]{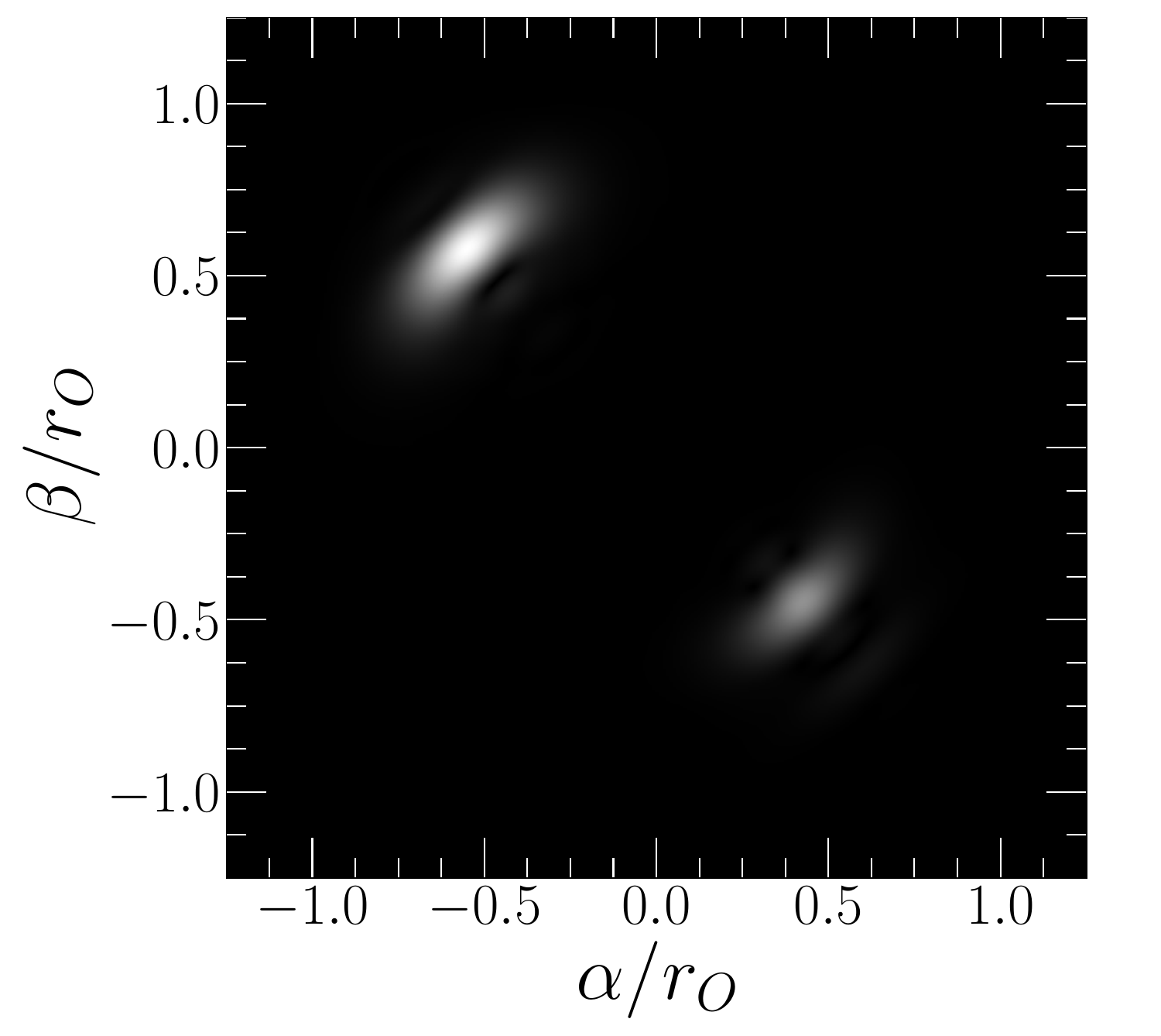}\label{Results:SdS:Fig:DiffSource:9}} \hfill
\subfloat[$\theta_s = \frac{4}{10}\pi$, $\phi_s = \frac{10}{10} \pi$]{\includegraphics[width=0.33\linewidth]{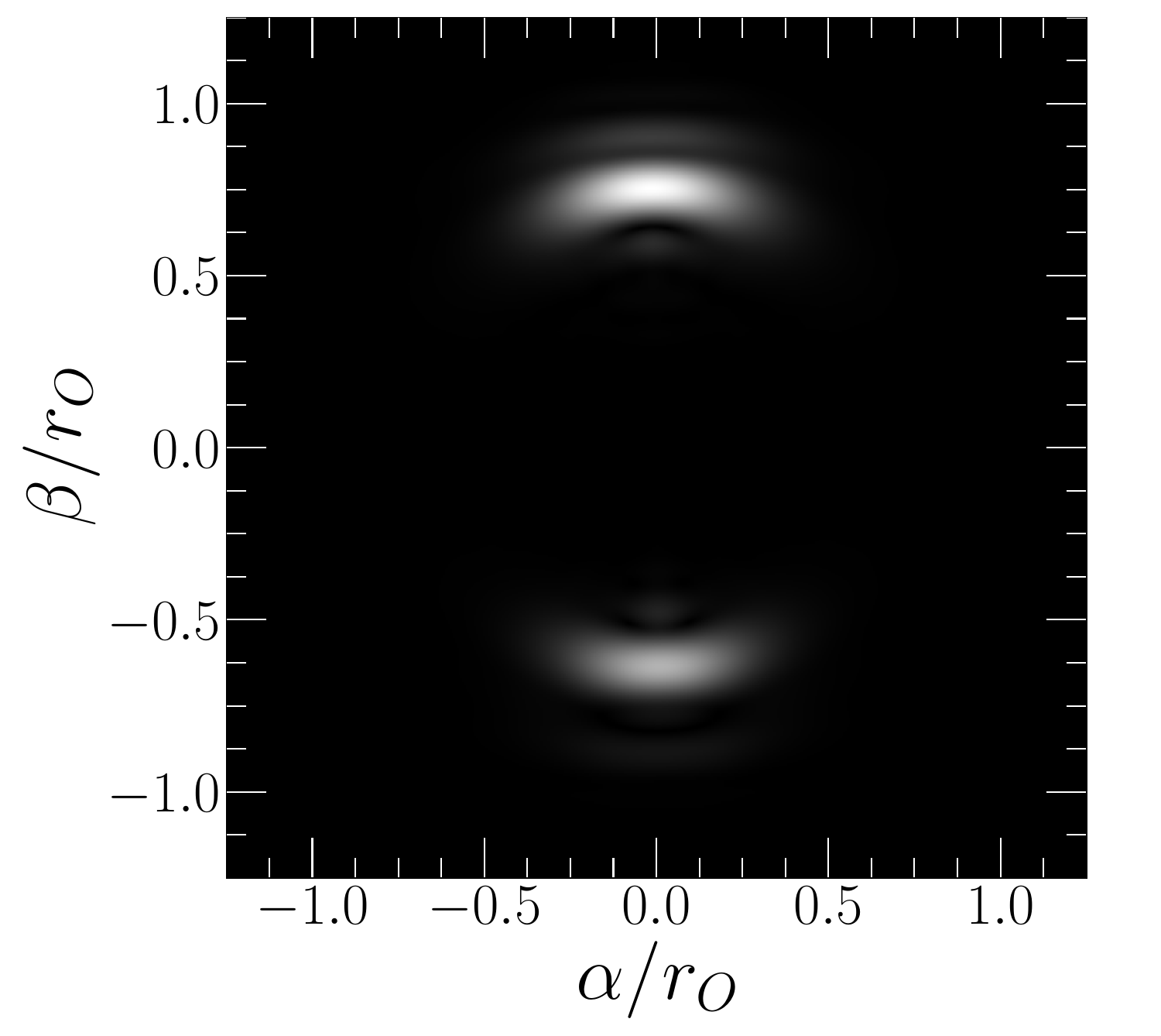}\label{Results:SdS:Fig:DiffSource:8}} \hfill
\subfloat[$\theta_s = \frac{4}{10}\pi$, $\phi_s = \frac{9}{10} \pi$]{\includegraphics[width=0.33\linewidth]{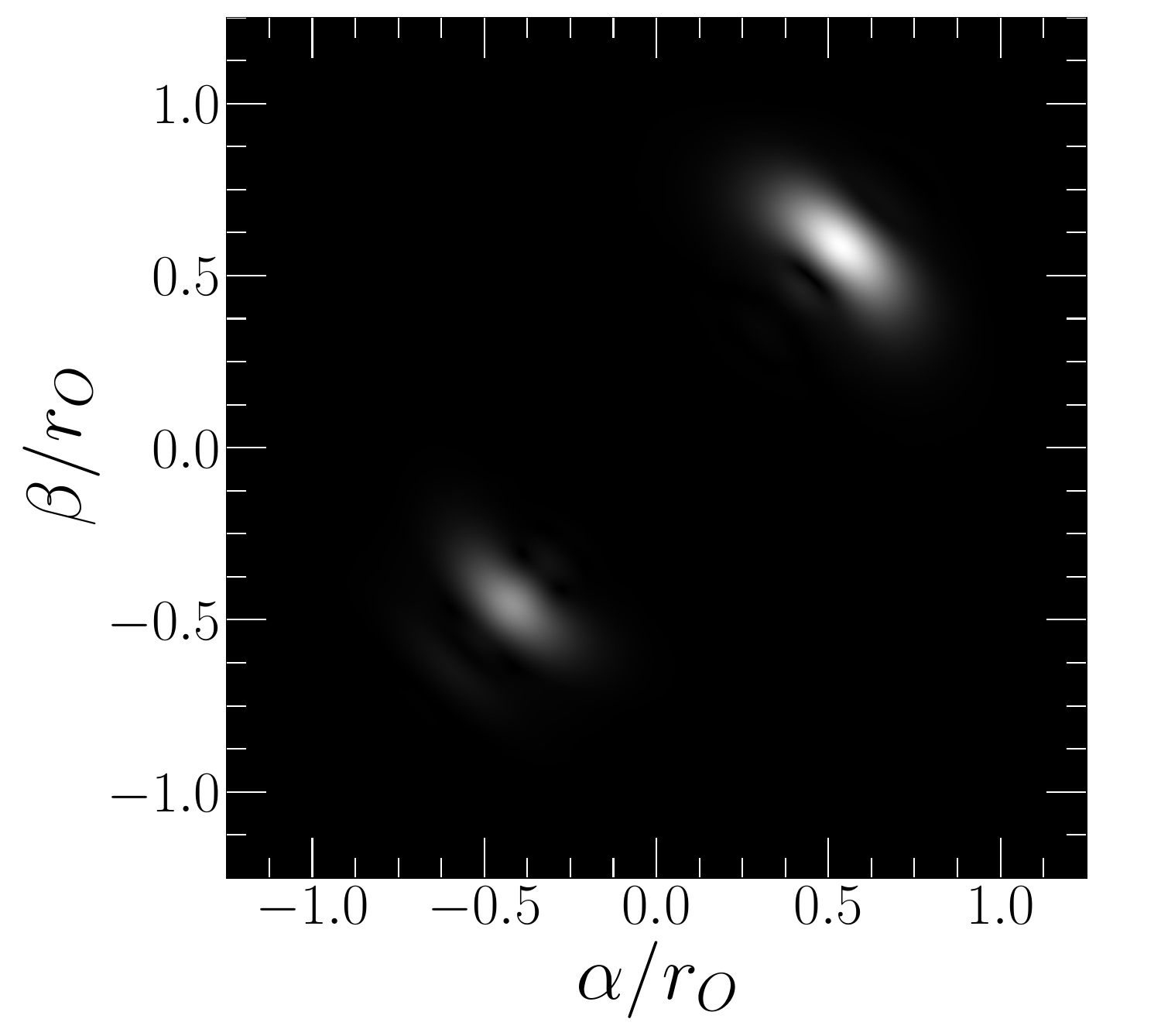}\label{Results:SdS:Fig:DiffSource:7}} \\
\subfloat[$\theta_s = \frac{5}{10}\pi$, $\phi_s = \frac{11}{10} \pi$]{\includegraphics[width=0.33\linewidth]{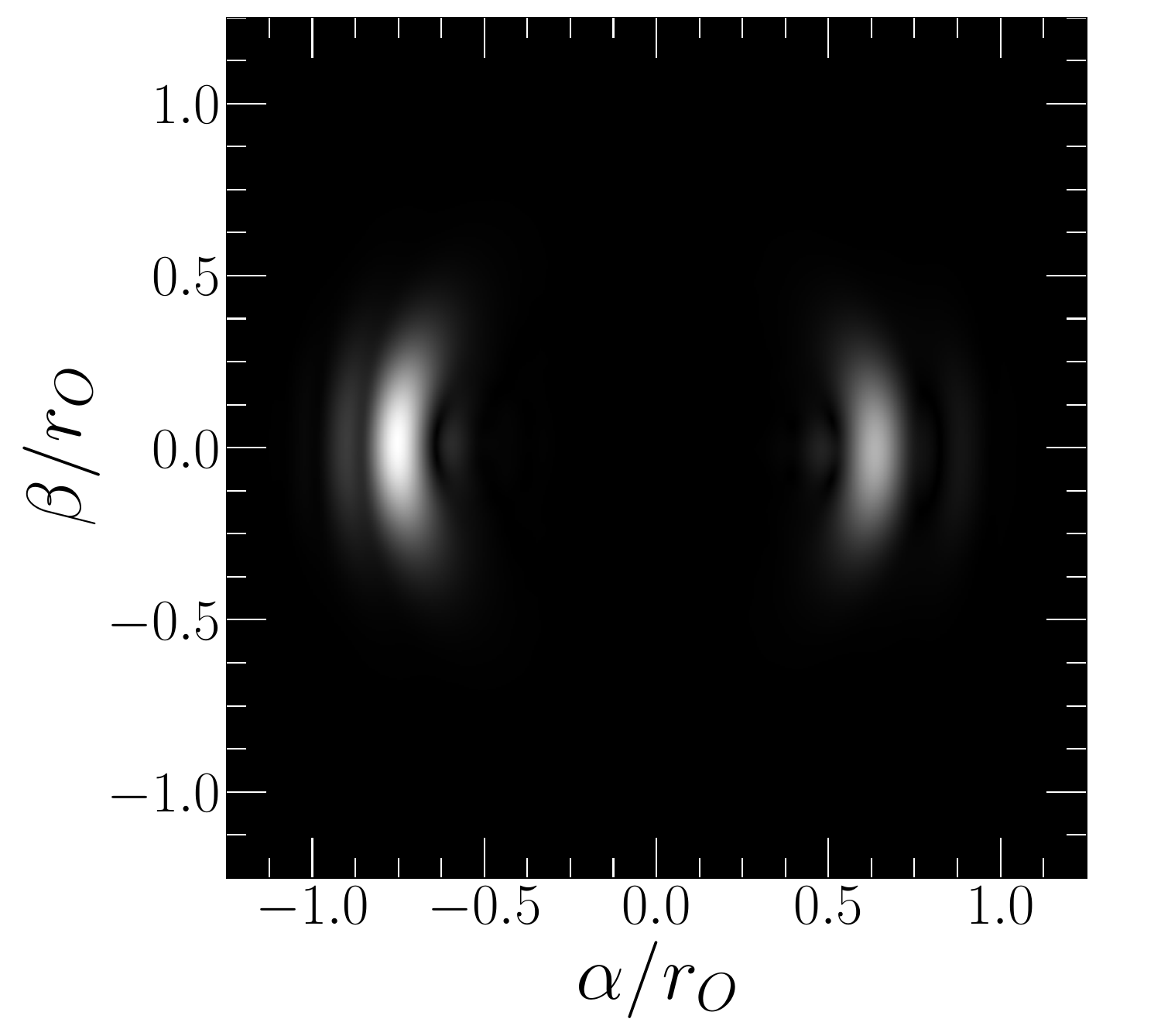}\label{Results:SdS:Fig:DiffSource:6}} \hfill
\subfloat[$\theta_s = \frac{5}{10}\pi$, $\phi_s = \frac{10}{10} \pi$]{\includegraphics[width=0.33\linewidth]{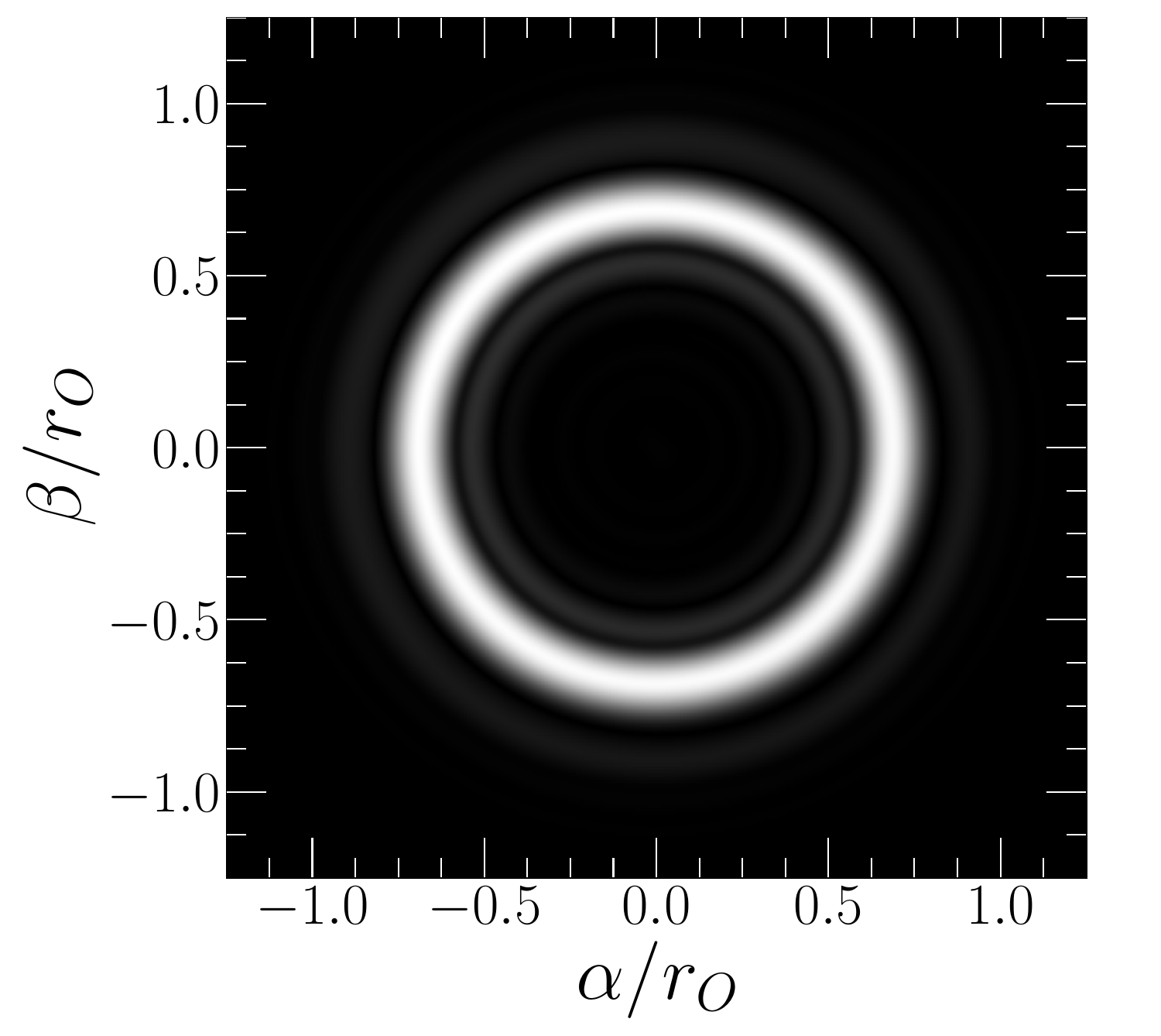}\label{Results:SdS:Fig:DiffSource:5}} \hfill
\subfloat[$\theta_s = \frac{5}{10}\pi$, $\phi_s = \frac{9}{10} \pi$]{\includegraphics[width=0.33\linewidth]{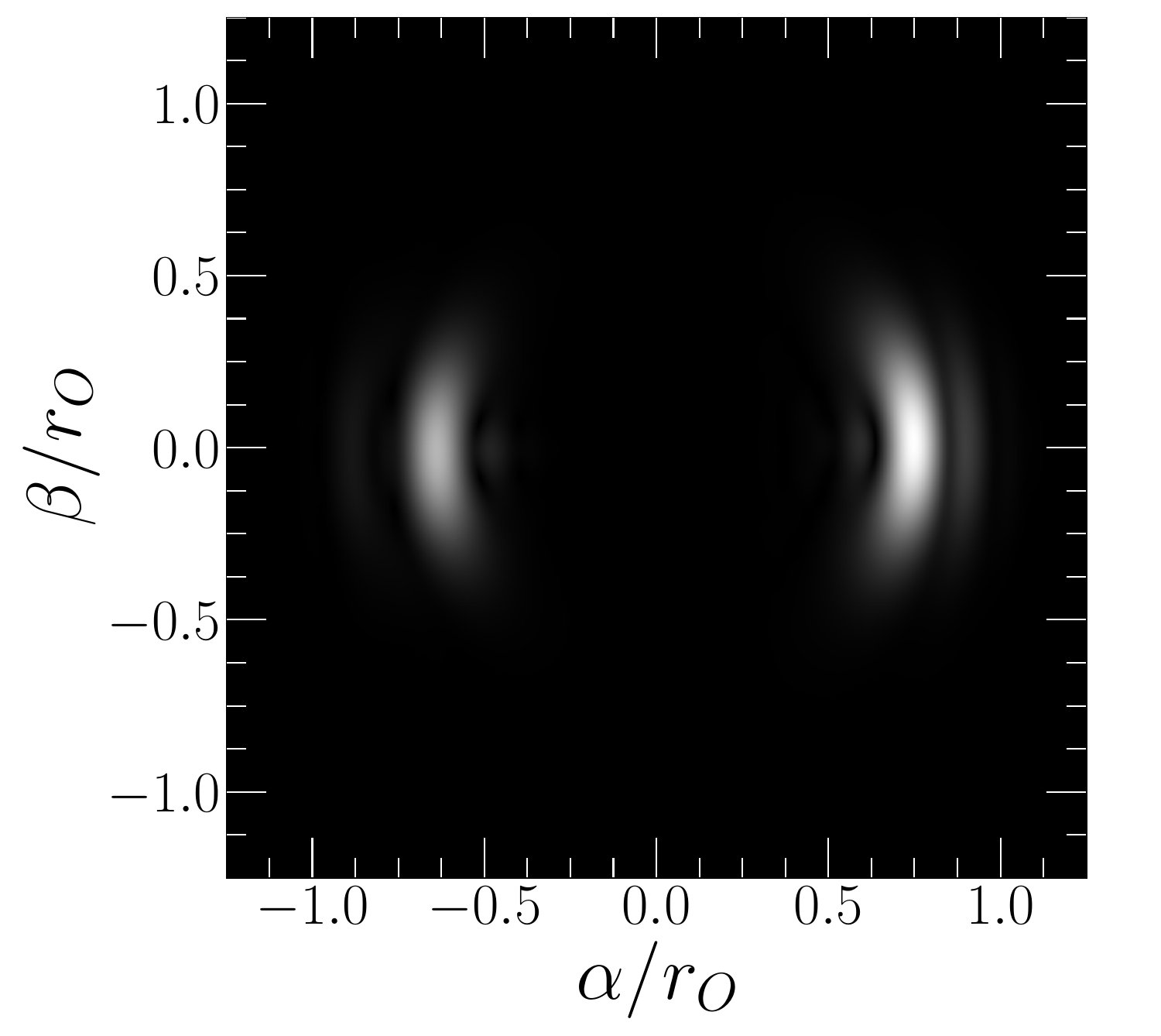}\label{Results:SdS:Fig:DiffSource:4}} \hfill
\subfloat[$\theta_s = \frac{6}{10}\pi$, $\phi_s = \frac{11}{10} \pi$]{\includegraphics[width=0.33\linewidth]{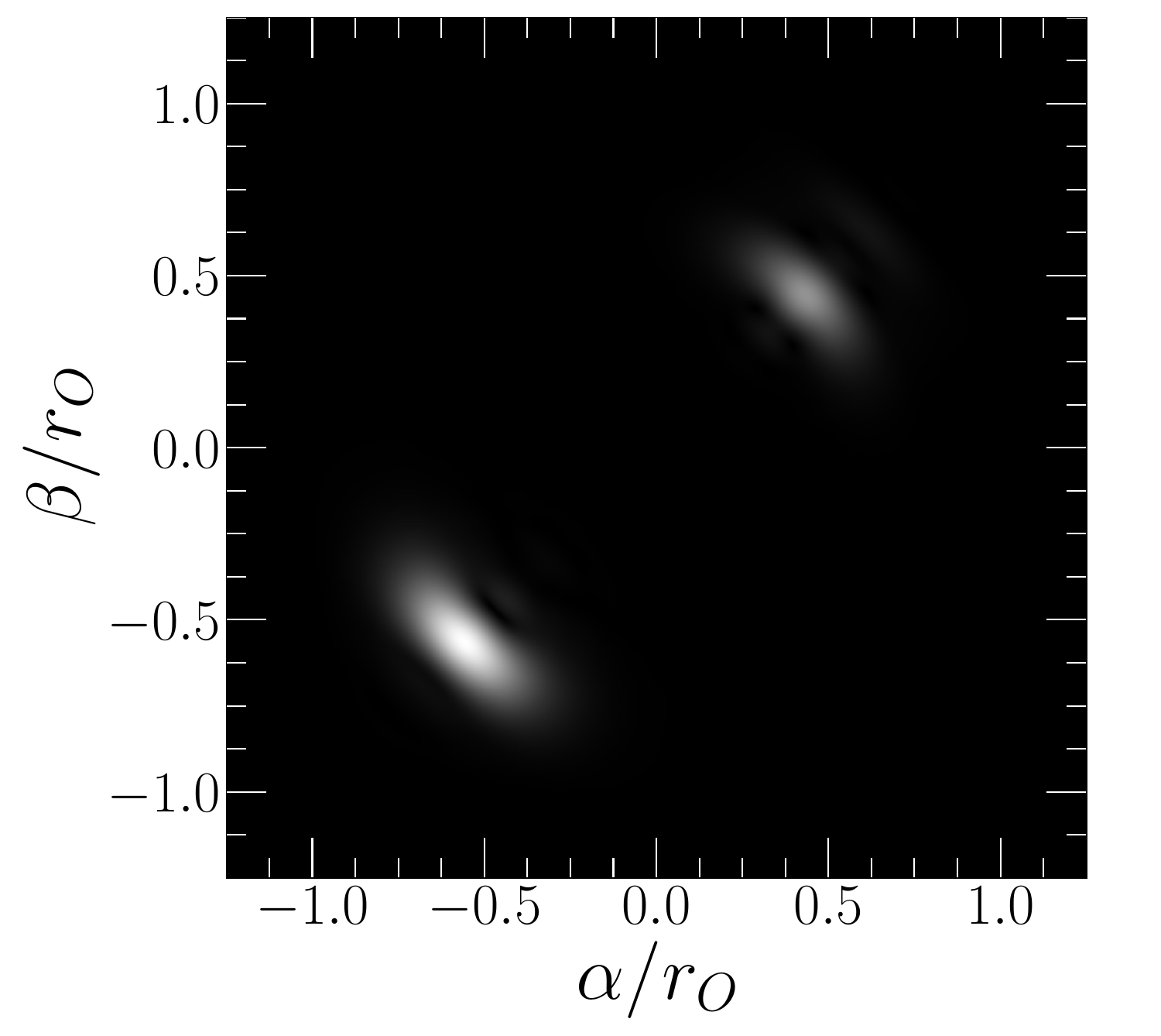}\label{Results:SdS:Fig:DiffSource:3}} \hfill
\subfloat[$\theta_s = \frac{6}{10}\pi$, $\phi_s = \frac{10}{10} \pi$]{\includegraphics[width=0.33\linewidth]{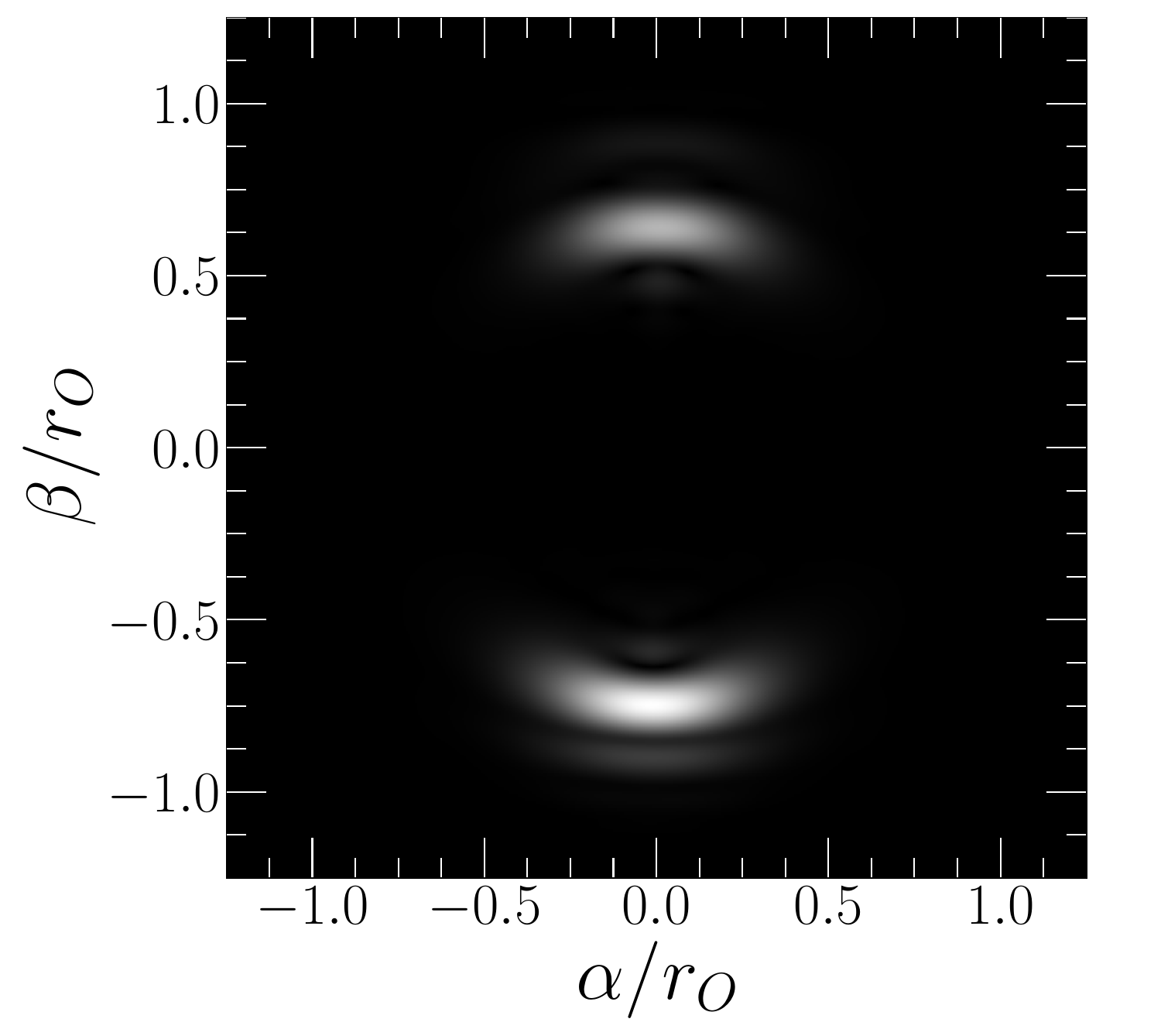}\label{Results:SdS:Fig:DiffSource:2}} \hfill
\subfloat[$\theta_s = \frac{6}{10}\pi$, $\phi_s = \frac{9}{10} \pi$]{\includegraphics[width=0.33\linewidth]{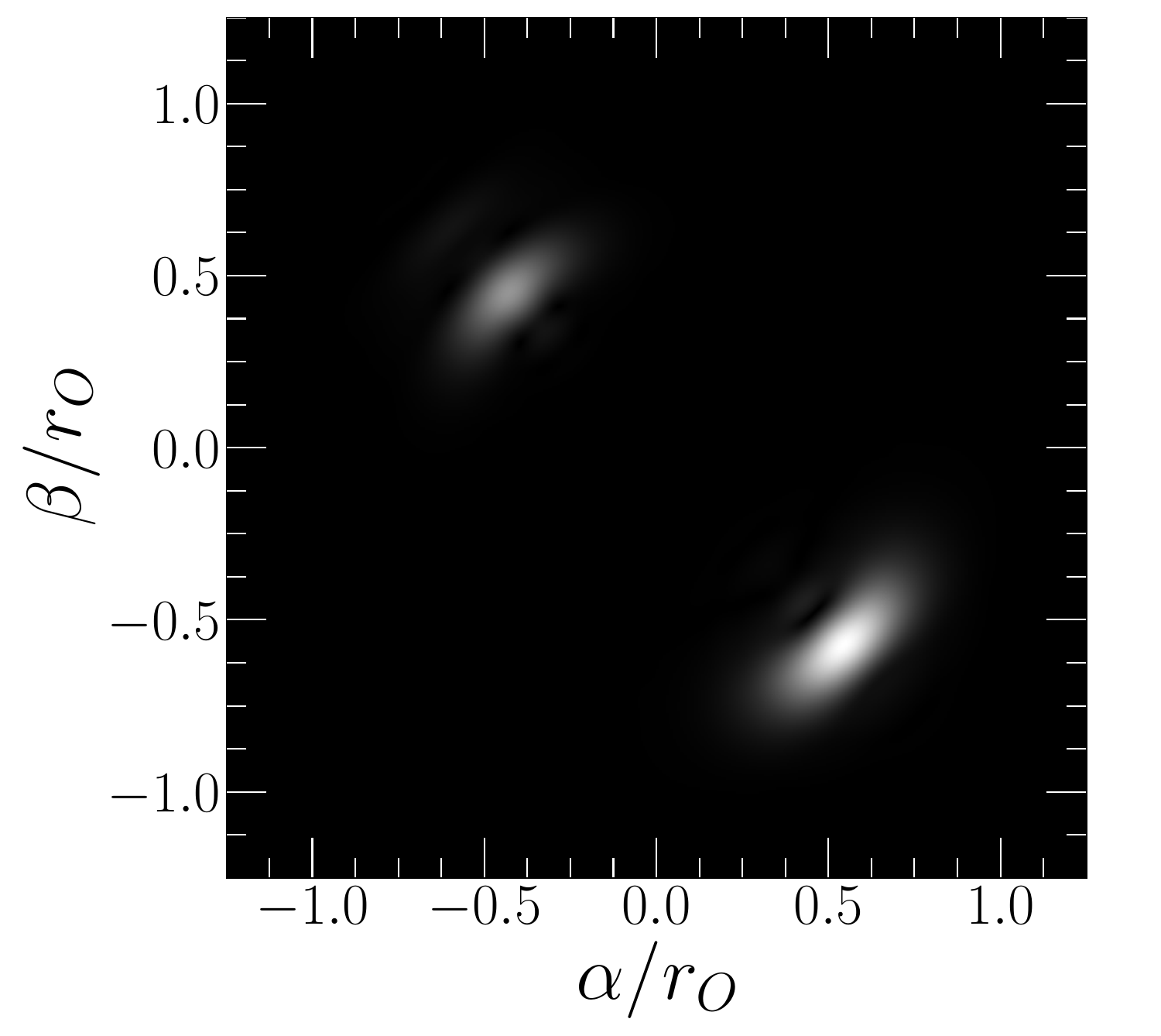}\label{Results:SdS:Fig:DiffSource:1}} 
\caption{Variation of the point source's location for $a = 0.00 M$, $M\omega = 20$, $\Lambda M^2 = 10^{-3}$, $r_O = 10 M$, $r_s = 20 M$, $\theta_O = \frac{1}{2} \pi$. 
Here, only the image $\left|\mathcal{F}\left(G(\vec{r}, \vec{r}_s, l_\text{max})\right)\right|^2$ is shown. 
Primary and secondary images of the source are clearly visible.}
\label{Results:SdS:Fig:DiffSource}
\end{figure*}

Varying the angular position $\phi_S$ of the source for fixed $r_s$ changes the resulting image. 
For nine alignments that deviate from the antipodal alignment in steps of $\frac{\pi}{10}$ for $\theta_s$ and $\phi_s$ respectively, the resulting images are shown for $M\omega = 20$ in \cref{Results:SdS:Fig:DiffSource}.
Moving the source breaks the Einstein ring and the formation of primary and secondary images around the center. 
The primary images face outwards, and the secondary images face inward. 
The bending of these images is a natural consequence of the imaging by the wave-optical ansatz. 
The secondary image is, trivially, fainter than the primary image.

\subsection{Kerr-de Sitter}
\label{Results:KdS}
\begin{figure*}
\centering
\subfloat[$a = 0.90M$]{\includegraphics[width=0.33\linewidth]{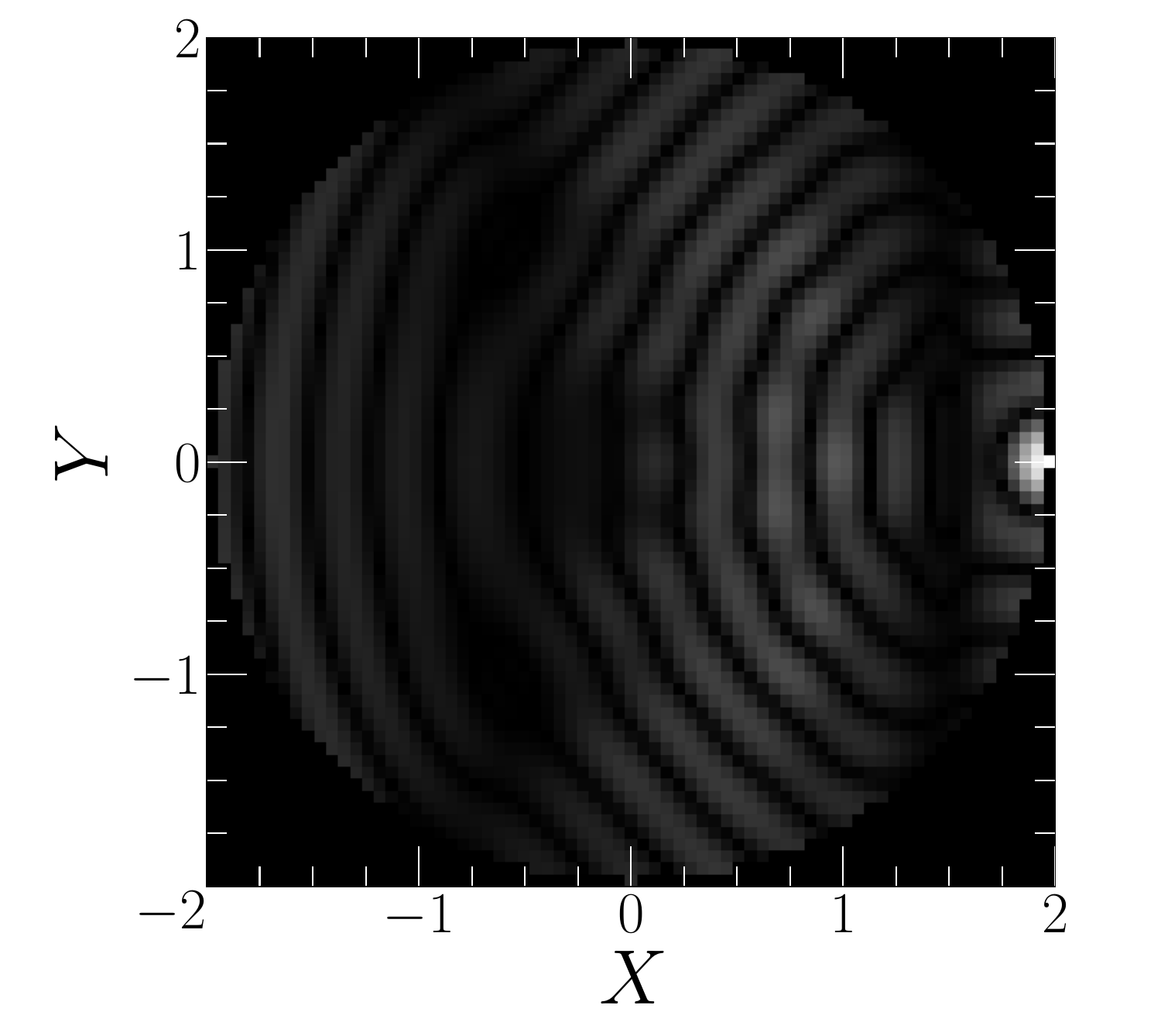}\label{Results:KdS:Fig:aVar:IMa0.90}} \hfill
\subfloat[$a = 0.60M$]{\includegraphics[width=0.33\linewidth]{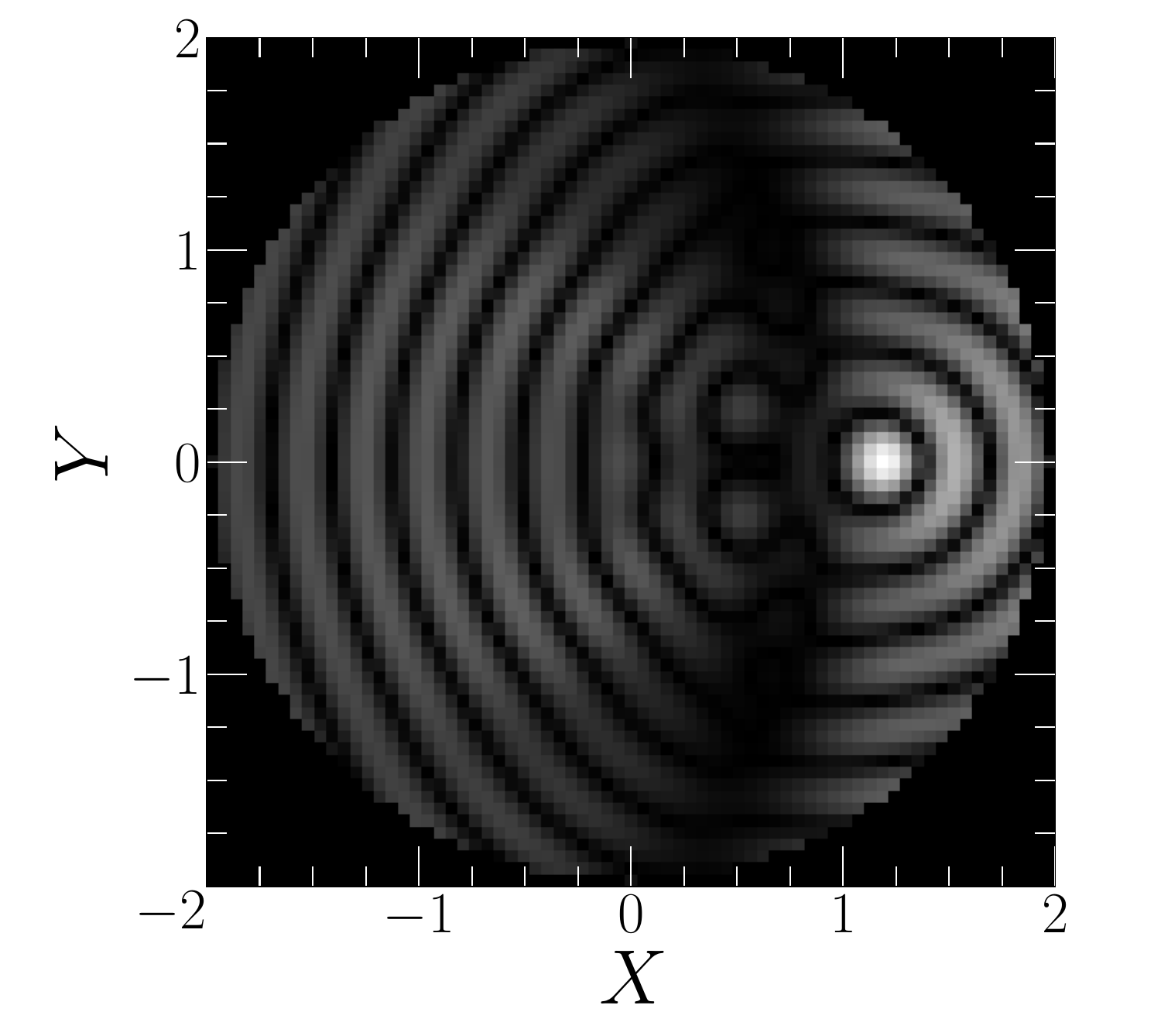}\label{Results:KdS:Fig:aVar:IMa0.60}} \hfill
\subfloat[$a = 0.00M$ (SdS)]{\includegraphics[width=0.33\linewidth]{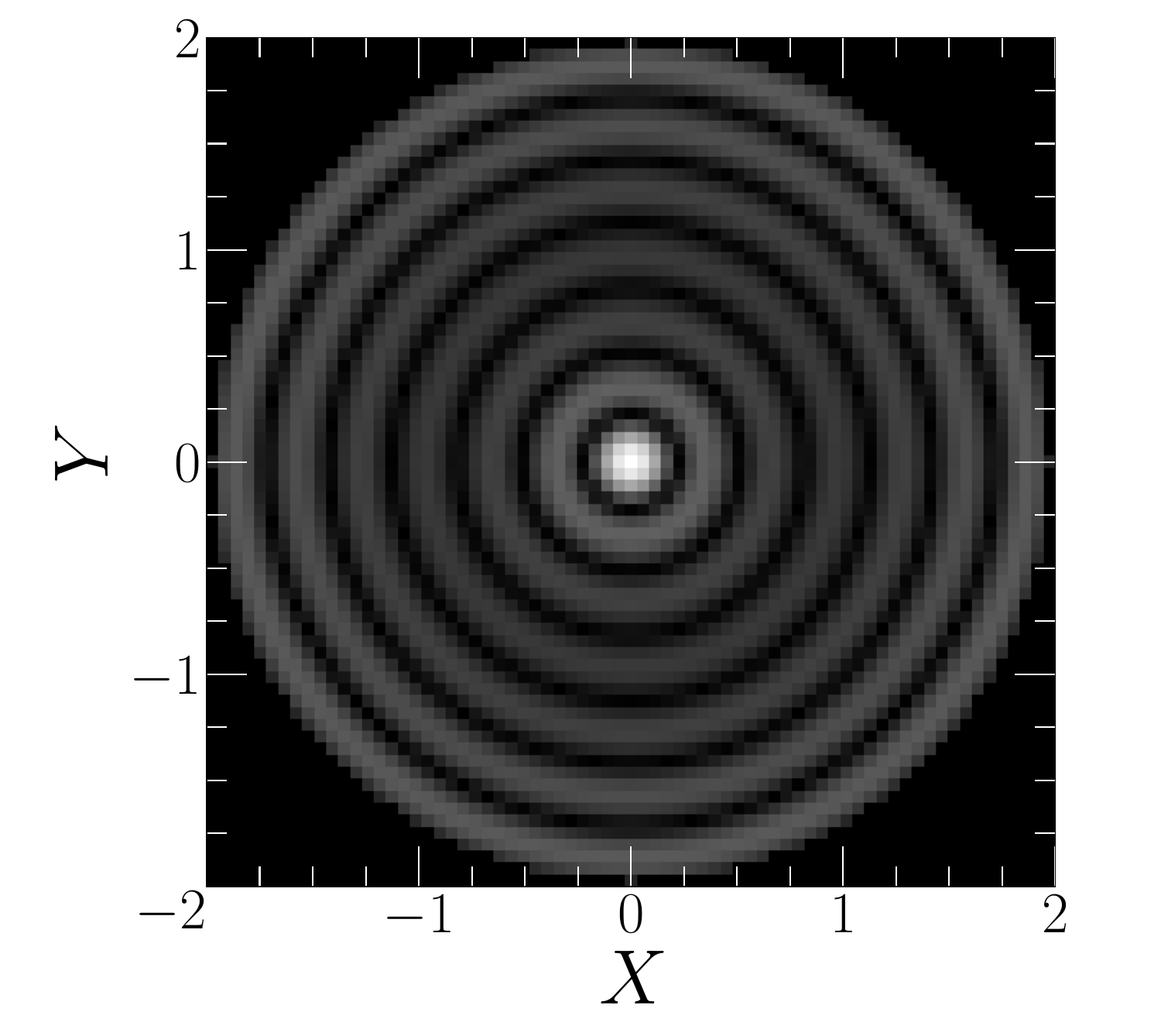}\label{Results:KdS:Fig:aVar:IMa0.00}} \\
\subfloat[$a = 0.90M$]{\includegraphics[width=0.33\linewidth]{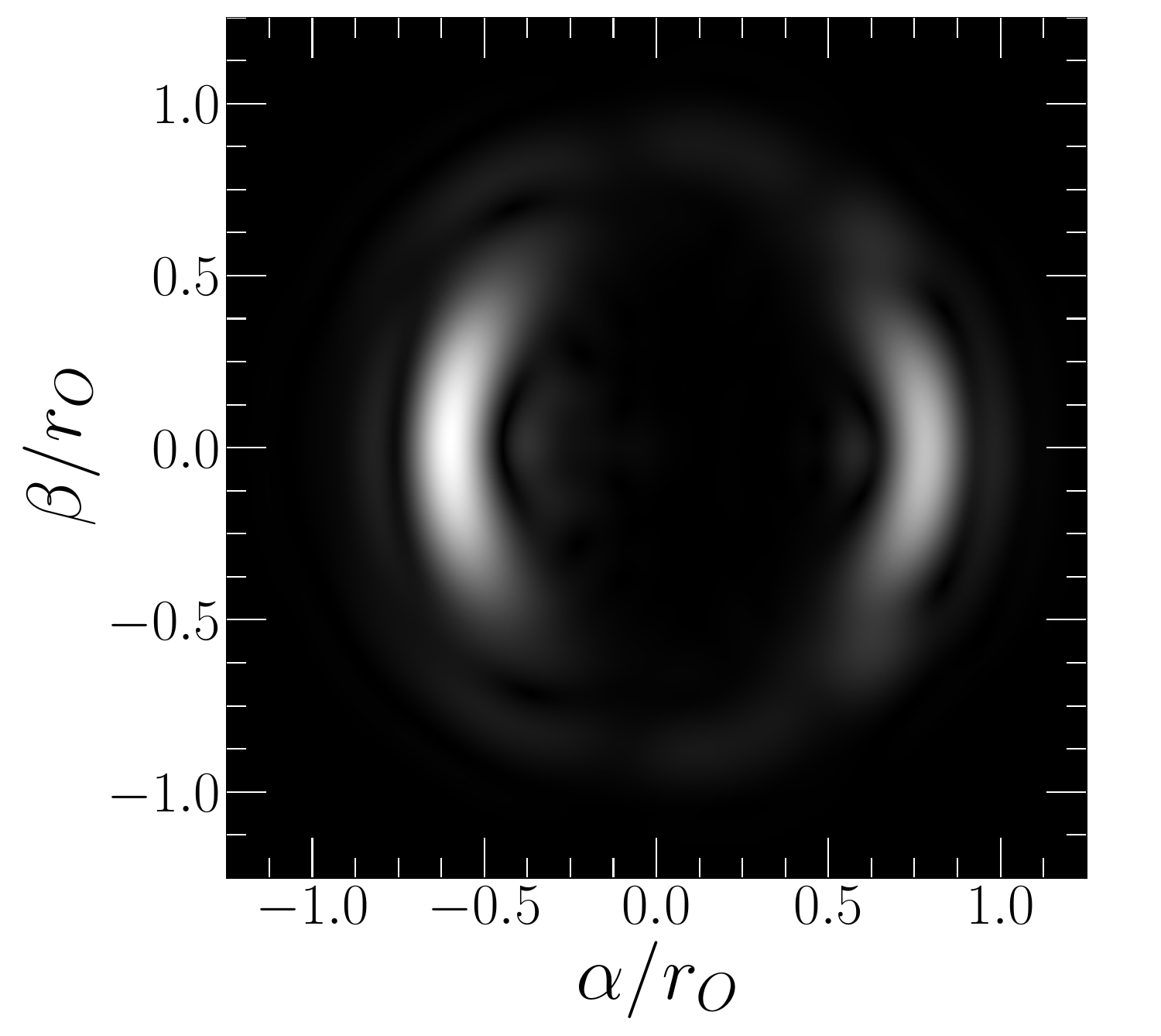}\label{Results:KdS:Fig:aVar:FFTa0.90}} \hfill
\subfloat[$a = 0.60M$]{\includegraphics[width=0.33\linewidth]{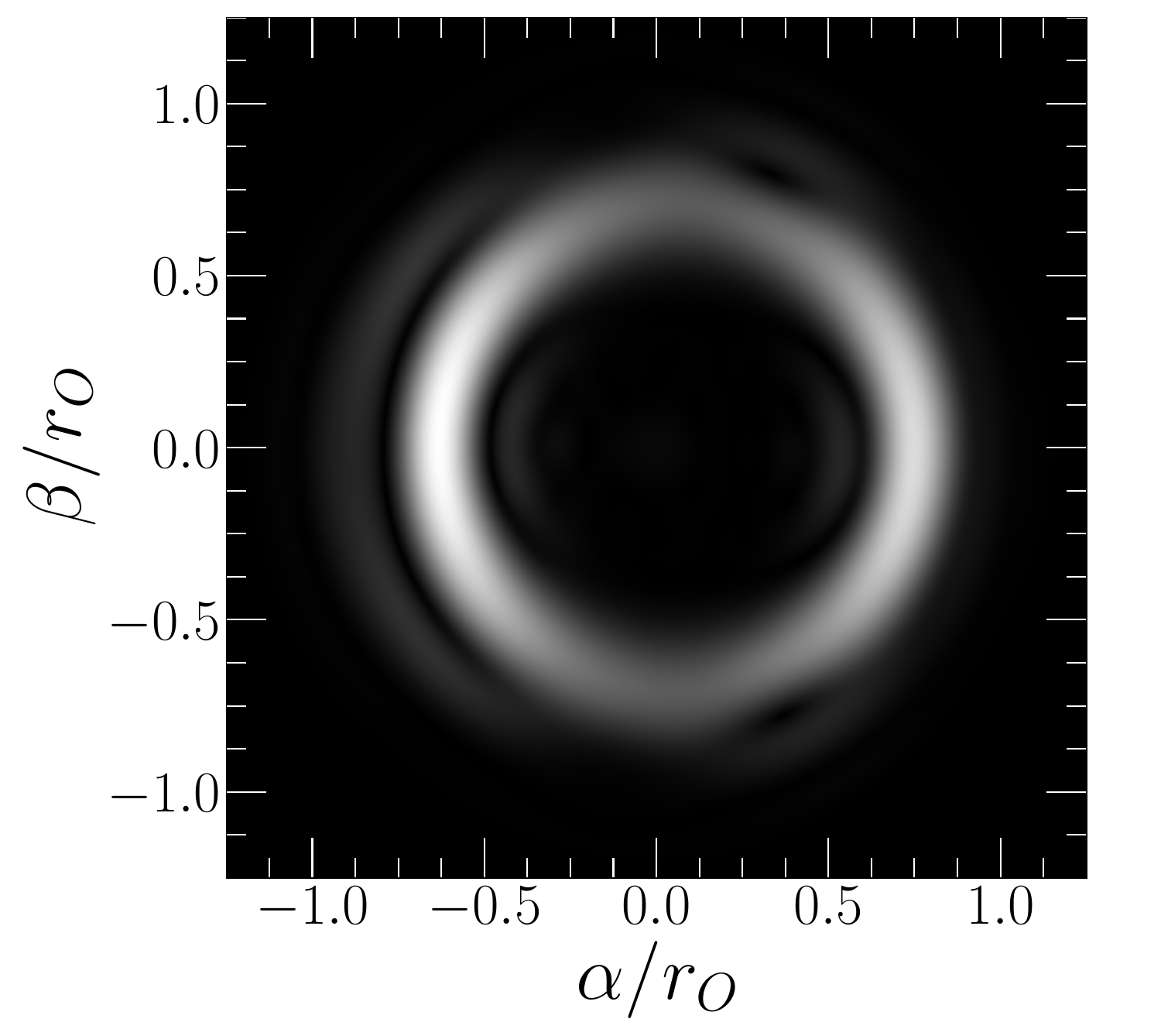}\label{Results:KdS:Fig:aVar:FFTa0.60}} \hfill
\subfloat[$a = 0.00M$ (SdS)]{\includegraphics[width=0.33\linewidth]{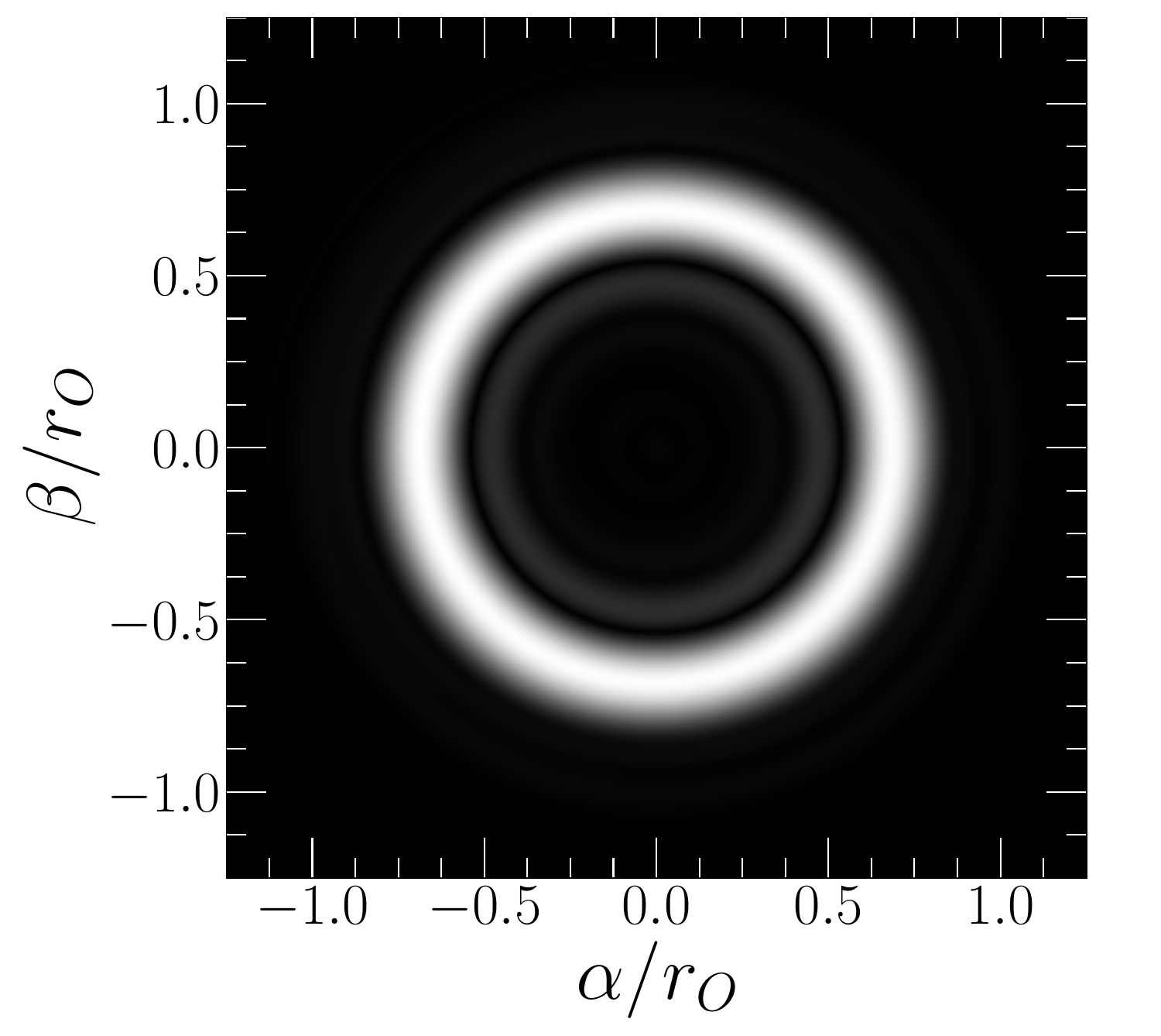}\label{Results:KdS:Fig:aVar:FFTa0.00}}
\caption{Variation of the Kerr parameter $a / M \in \{0.00, 0.60, 0.90\}$ for $M\omega = 15$, $\Lambda M^2 = 10^{-3}$, $r_O = 10 M$, $r_s = 20 M$, $\theta_O = \frac{\pi}{2}$, $\theta_s = \frac{\pi}{2}$, $\phi_s = \pi$. 
Increasing $a$ results in a shift of the apparent source position, as well as additional structures breaking the symmetry in the observer plane. 
Even if the source is moved so that it appears directly behind the black hole (the center of the concentric circles is in the middle of the observer plane, as in \cref{Results:KdS:Fig:aVar:IMa0.00}), no clean Einstein ring emerges, but a distorted one with additional features appears. 
The upper row shows again $\Im\left(G(\vec{r}, \vec{r}_s, l_\text{max})\right)$ in the observer plane and the lower row shows the images in the image plane, $\left|\mathcal{F}\left(G(\vec{r}, \vec{r}_s, l_\text{max})\right)\right|^2$.}
\label{Results:KdS:Fig:aVar}
\end{figure*}

The KdS case involves a parameter choice of $0 < a < a_\text{max}$, where the full solutions of the radial and angular Teukolsky equations (\cref{WavOpt:Green:Eq:radGreen,TMESol:Ang:Final:Eq:Final} respectively) come into play. 
Its effect on the wave-optical image is studied in \cref{Results:KdS:Fig:aVar} for three different values of $a / M \in \{0, 0.60, 0.99\}$ in an antipodal alignment of source and observer in the equatorial plane. 
Increasing $a$ shifts the apparent point source location, which can be intuitively explained by the resulting frame-dragging caused in the vicinity of rotating black holes. 
Therefore, an antipodally aligned observer-source constellation no longer appears antipodal in the presence of a nonzero Kerr parameter. 
In addition, the results from the KdS observer planes gain a symmetry breaking feature when compared to the SdS results; see \cref{Results:KdS:Fig:aVar:IMa0.00,Results:KdS:Fig:aVar:IMa0.60,Results:KdS:Fig:aVar:IMa0.90}. 
The comparison between the results with nonzero Kerr parameters and $a = 0$ is shown in \cref{Results:KdS:Fig:aVar:IMa0.00,Results:KdS:Fig:aVar:FFTa0.00}. 
The wave-optical images limit towards the SdS case.
The two cases differ mainly in the solution of the angular Teukolsky equation.
Omitting the normalization constant, a crucial step in solving the angular Teukolsky equation, results in unusable lens and image planes that neither match at all the results shown, nor the limit towards the SdS case.

\begin{figure*}
\centering
\subfloat[$\theta_s = \frac{4}{10}\pi$, $\phi_s = \frac{11}{10} \pi$]{\includegraphics[width=0.33\linewidth]{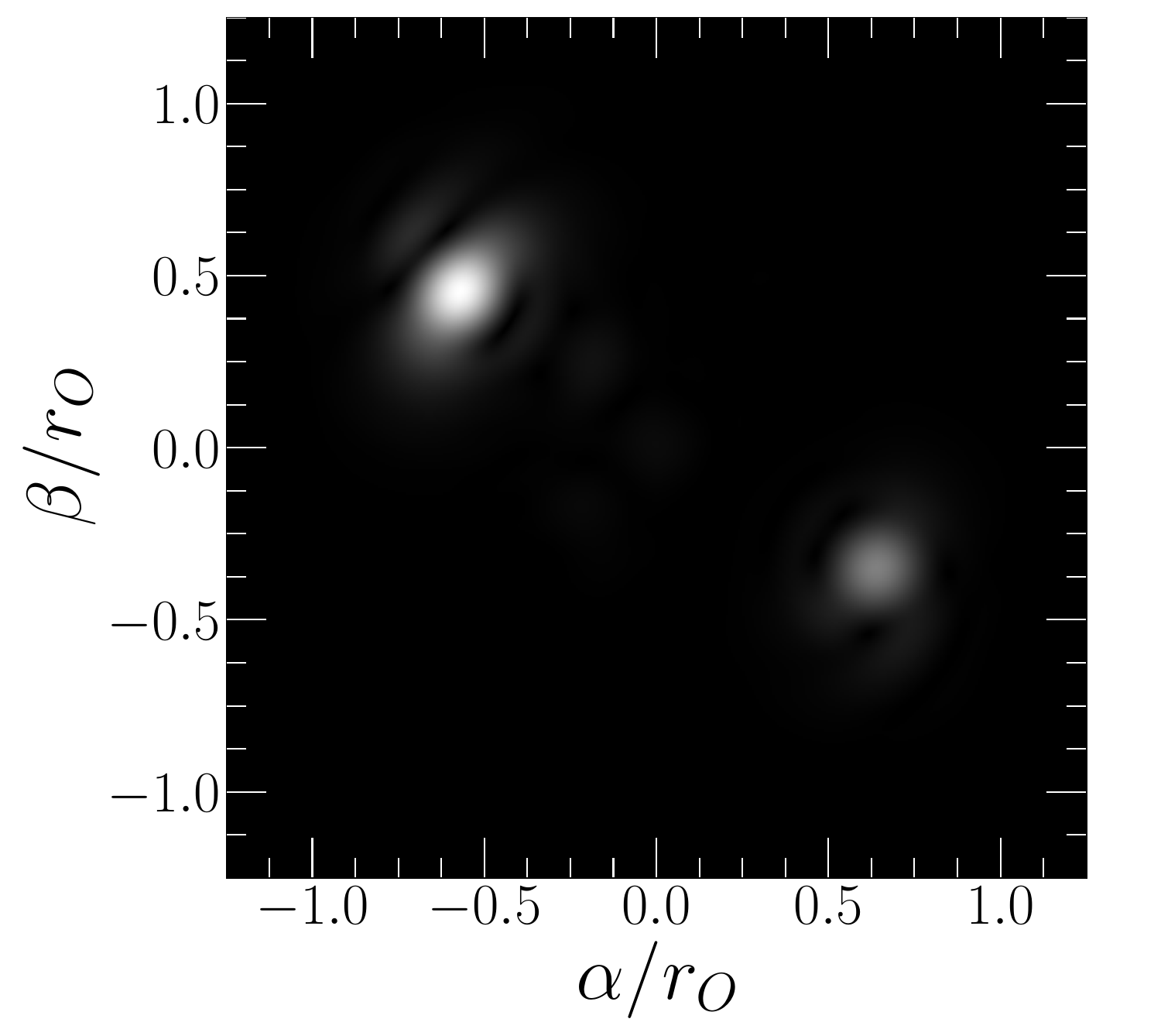}\label{Results:KdS:Fig:DiffSource:9}} \hfill
\subfloat[$\theta_s = \frac{4}{10}\pi$, $\phi_s = \frac{10}{10} \pi$]{\includegraphics[width=0.33\linewidth]{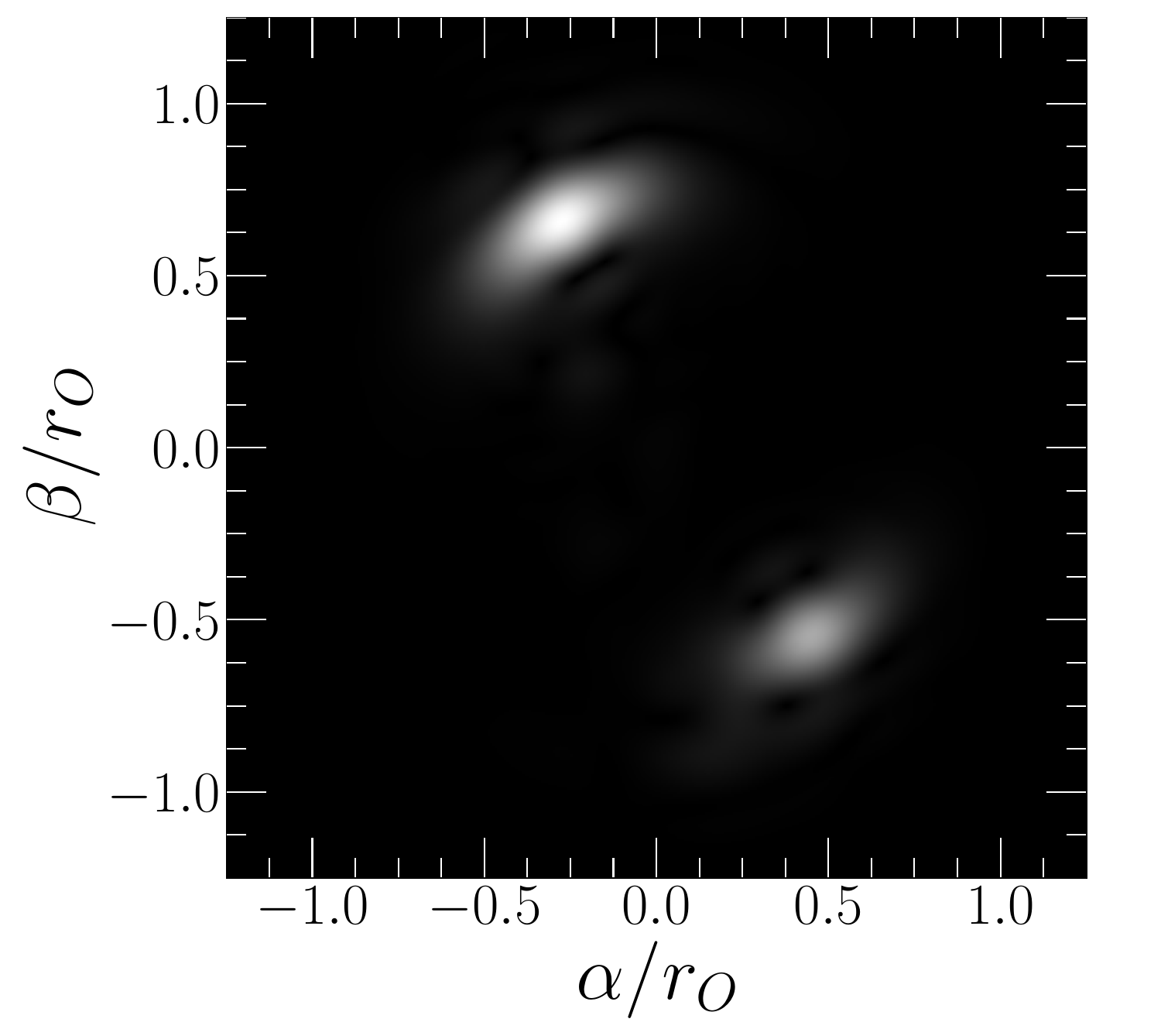}\label{Results:KdS:Fig:DiffSource:8}} \hfill
\subfloat[$\theta_s = \frac{4}{10}\pi$, $\phi_s = \frac{9}{10} \pi$]{\includegraphics[width=0.33\linewidth]{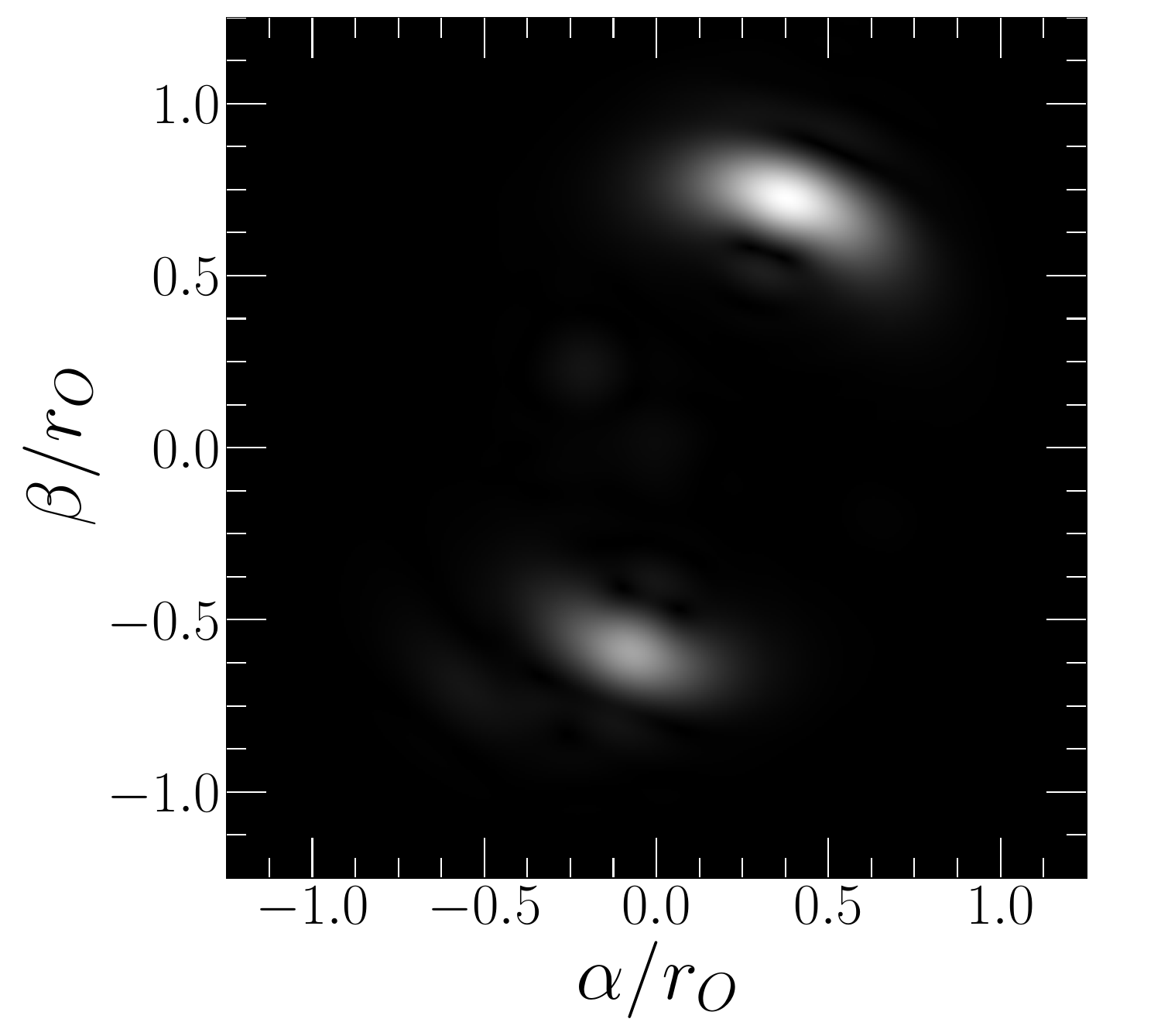}\label{Results:KdS:Fig:DiffSource:7}} \\
\subfloat[$\theta_s = \frac{5}{10}\pi$, $\phi_s = \frac{11}{10} \pi$]{\includegraphics[width=0.33\linewidth]{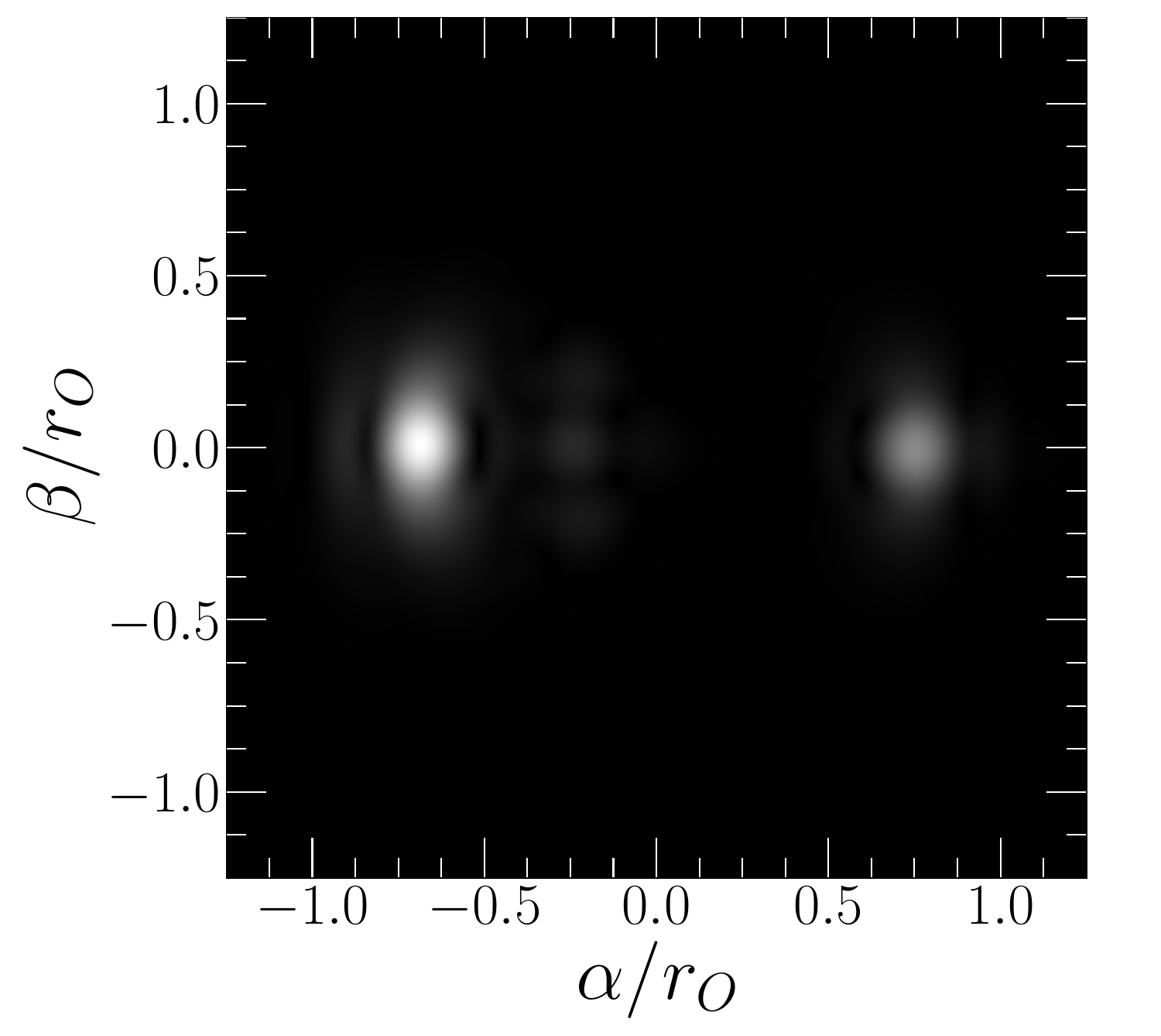}\label{Results:KdS:Fig:DiffSource:6}} \hfill
\subfloat[$\theta_s = \frac{5}{10}\pi$, $\phi_s = \frac{10}{10} \pi$]{\includegraphics[width=0.33\linewidth]{sKdS-FFT-a0.99M_Mw15.0_ro10M_rs20M_to0.50pi_ts0.50pi_po0.00pi_ps1.00pi.pdf}\label{Results:KdS:Fig:DiffSource:5}} \hfill
\subfloat[$\theta_s = \frac{5}{10}\pi$, $\phi_s = \frac{9}{10} \pi$]{\includegraphics[width=0.33\linewidth]{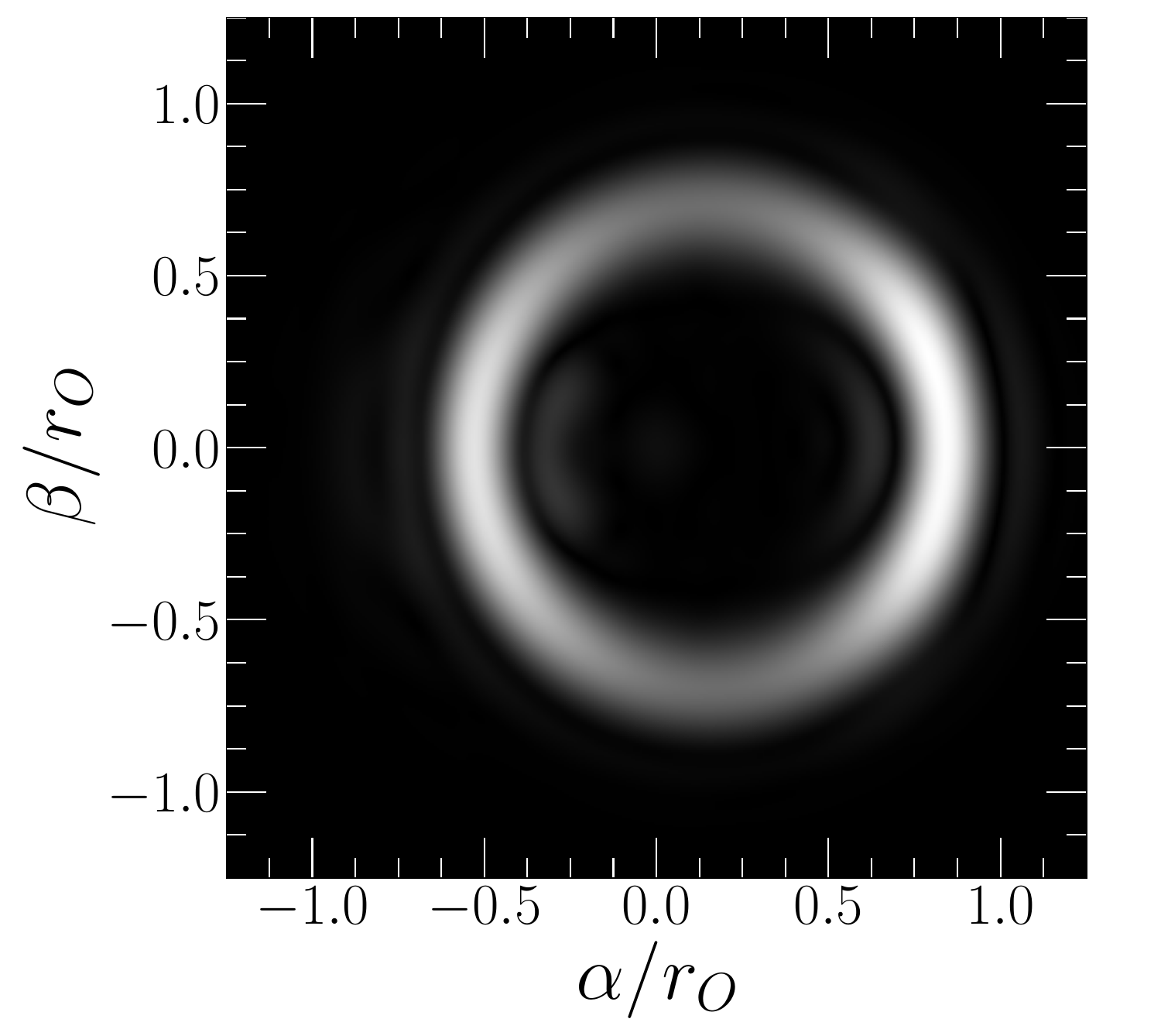}\label{Results:KdS:Fig:DiffSource:4}} \\
\subfloat[$\theta_s = \frac{6}{10}\pi$, $\phi_s = \frac{11}{10} \pi$]{\includegraphics[width=0.33\linewidth]{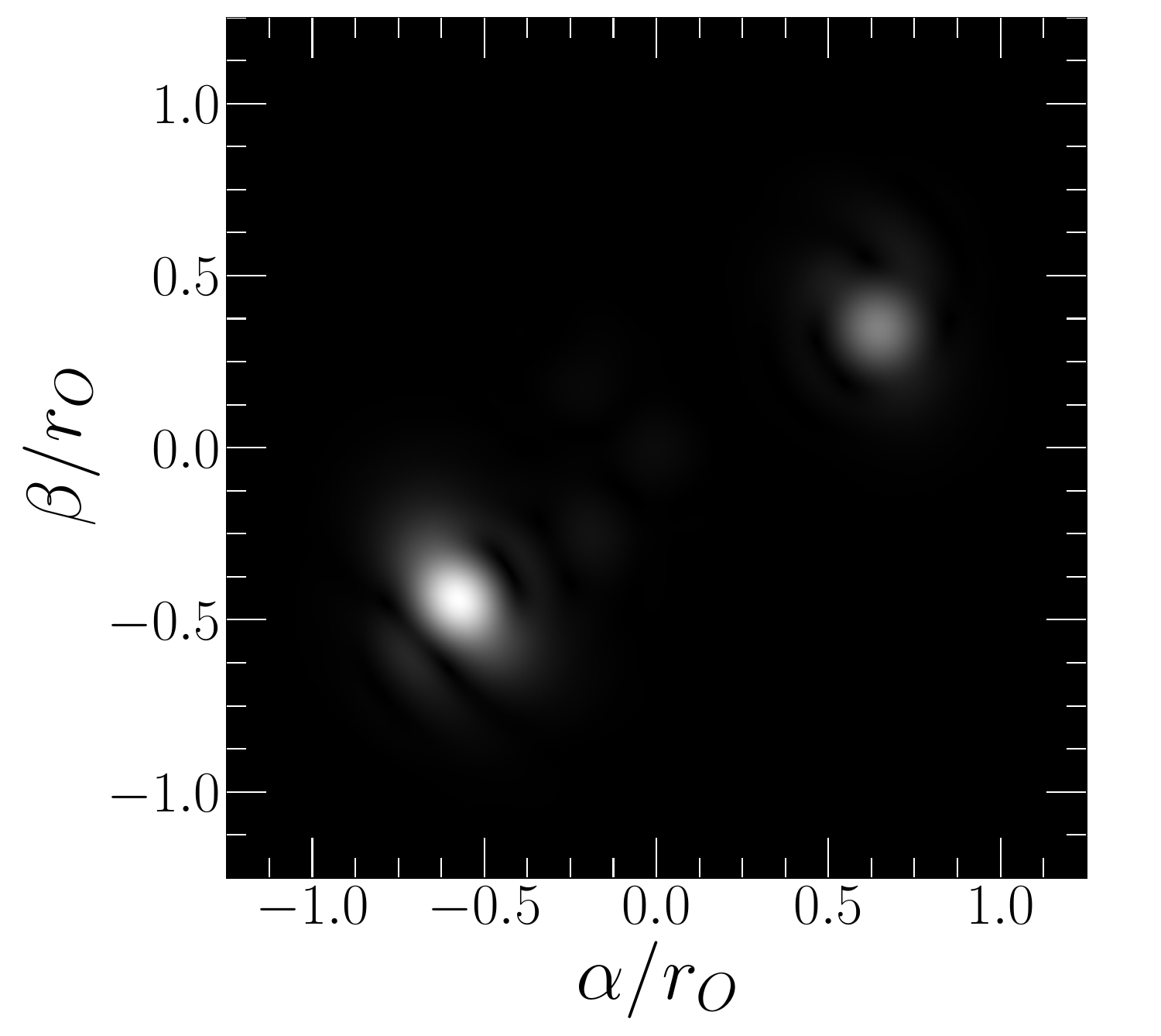}\label{Results:KdS:Fig:DiffSource:3}} \hfill
\subfloat[$\theta_s = \frac{6}{10}\pi$, $\phi_s = \frac{10}{10} \pi$]{\includegraphics[width=0.33\linewidth]{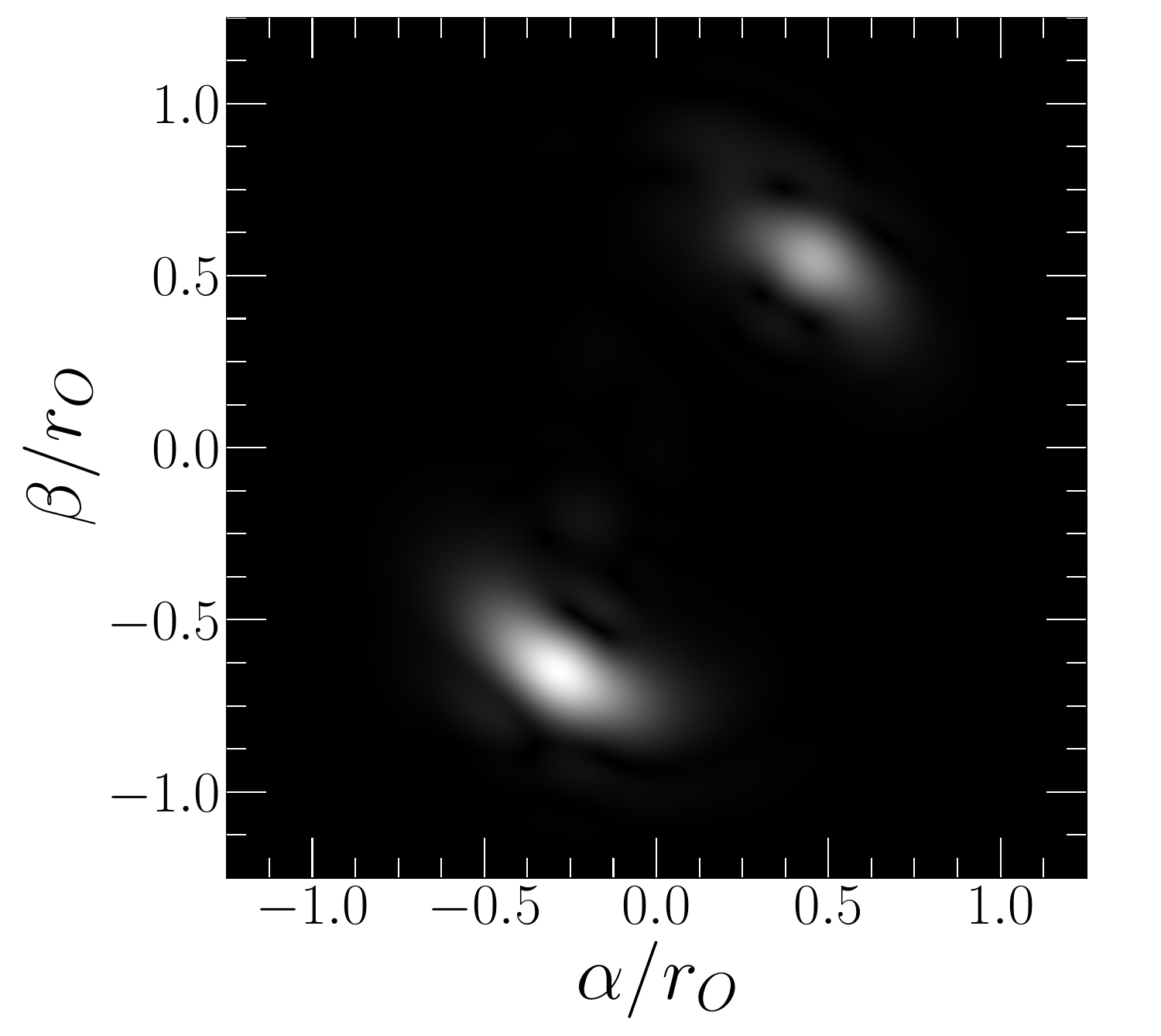}\label{Results:KdS:Fig:DiffSource:2}} \hfill
\subfloat[$\theta_s = \frac{6}{10}\pi$, $\phi_s = \frac{9}{10} \pi$]{\includegraphics[width=0.33\linewidth]{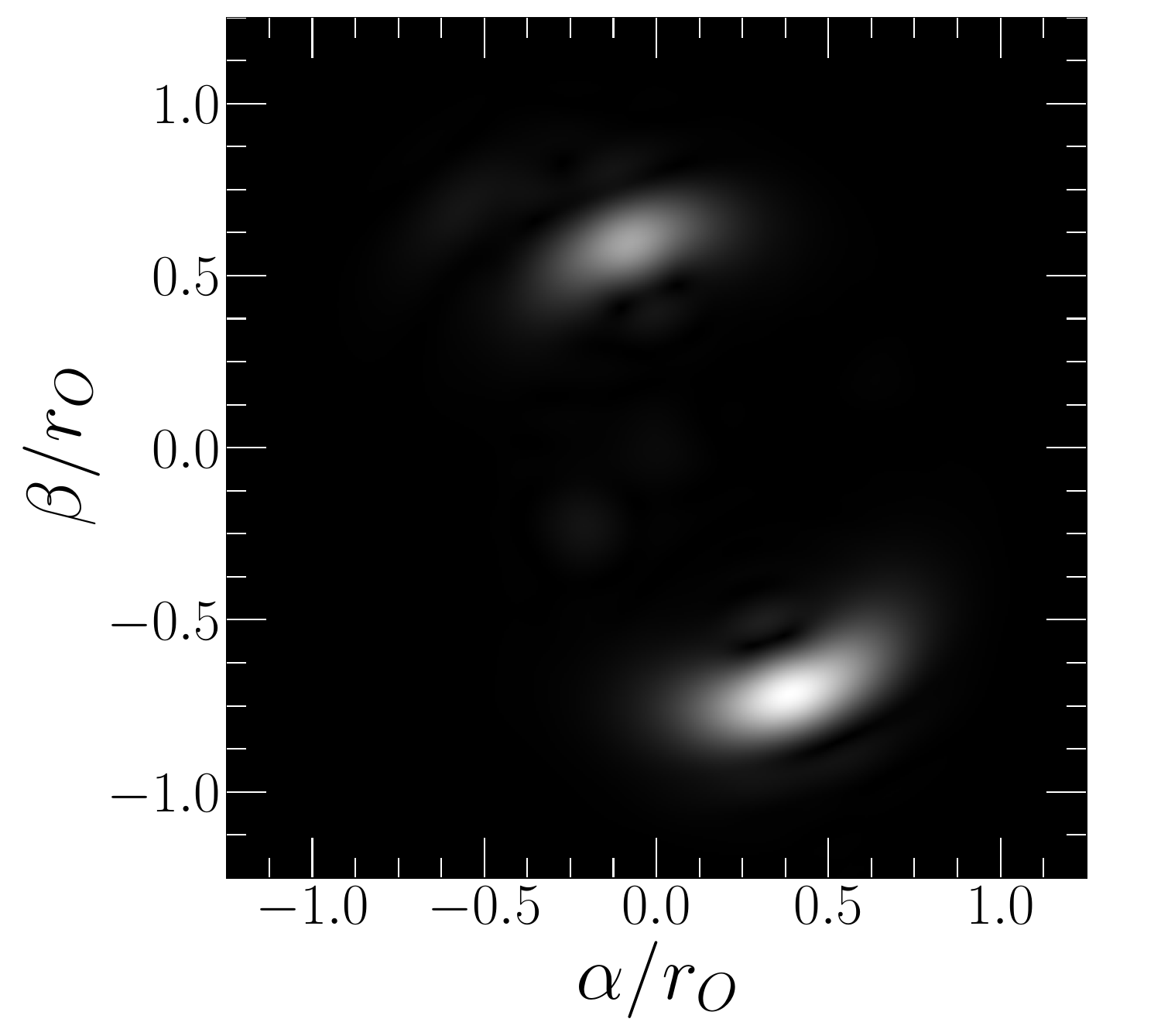}\label{Results:KdS:Fig:DiffSource:1}}
\caption{Variation of the source position according to the captions of the subfigures for $a = 0.99 M$, $M\omega = 15$, $\Lambda M^2 = 10^{-3}$, $r_O = 10 M$, $r_s = 20 M$, $\theta_O = \frac{1}{2} \pi$, similar to \cref{Results:SdS:Fig:DiffSource}, showing the image $\left|\mathcal{F}\left(G(\vec{r}, \vec{r}_s, l_\text{max})\right)\right|^2$. 
The shift in apparent source location due to the frame-dragging is observable.}
\label{Results:KdS:Fig:DiffSource}
\end{figure*}

To once again visualize the frame-dragging effects, \cref{Results:SdS:Fig:DiffSource} is recreated for the KdS case with $a = 0.99M$ in \cref{Results:KdS:Fig:DiffSource}.
Choosing the azimuthal position $\phi_S$ of the point source on the equatorial plane so that it appears antipodal in the KdS case, the Einstein ring forms. 
However, it has a broken symmetry and additional features compared to the SdS case, e.g., a structure resembling a Kerr black hole inside the Einstein ring. 
\cref{Results:ShadowCompare} will show that the current setup is not sufficient to fully reveal the black hole shadow in the wave-optical regime. 

\begin{figure*}
\centering
\subfloat[$a = 0.00M$ (SdS)]{\includegraphics[width=0.33\linewidth]{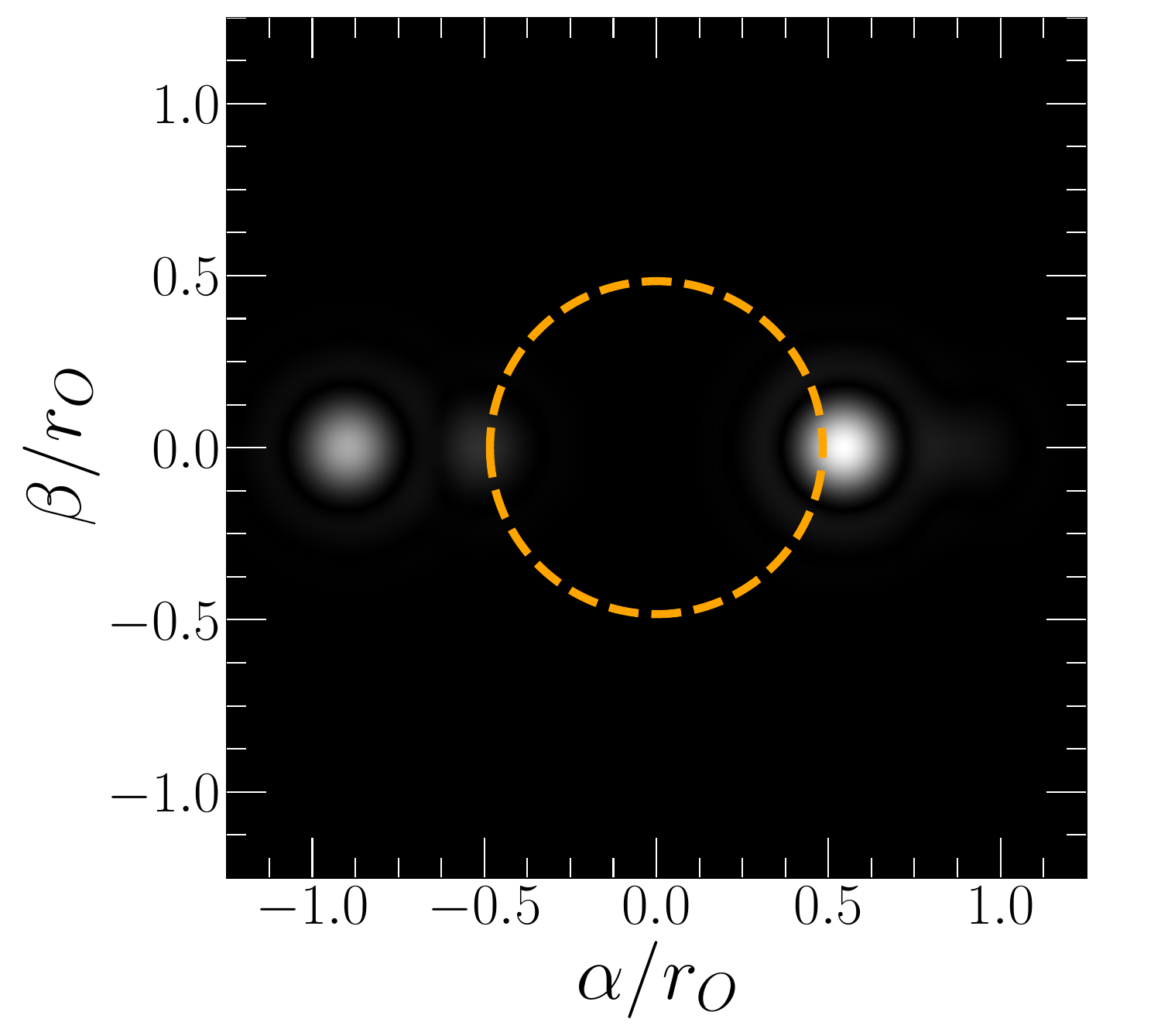}\label{Results:KdS:Fig:aVar2:FFTa0.00}} \hfill
\subfloat[$a = 0.20M$]{\includegraphics[width=0.33\linewidth]{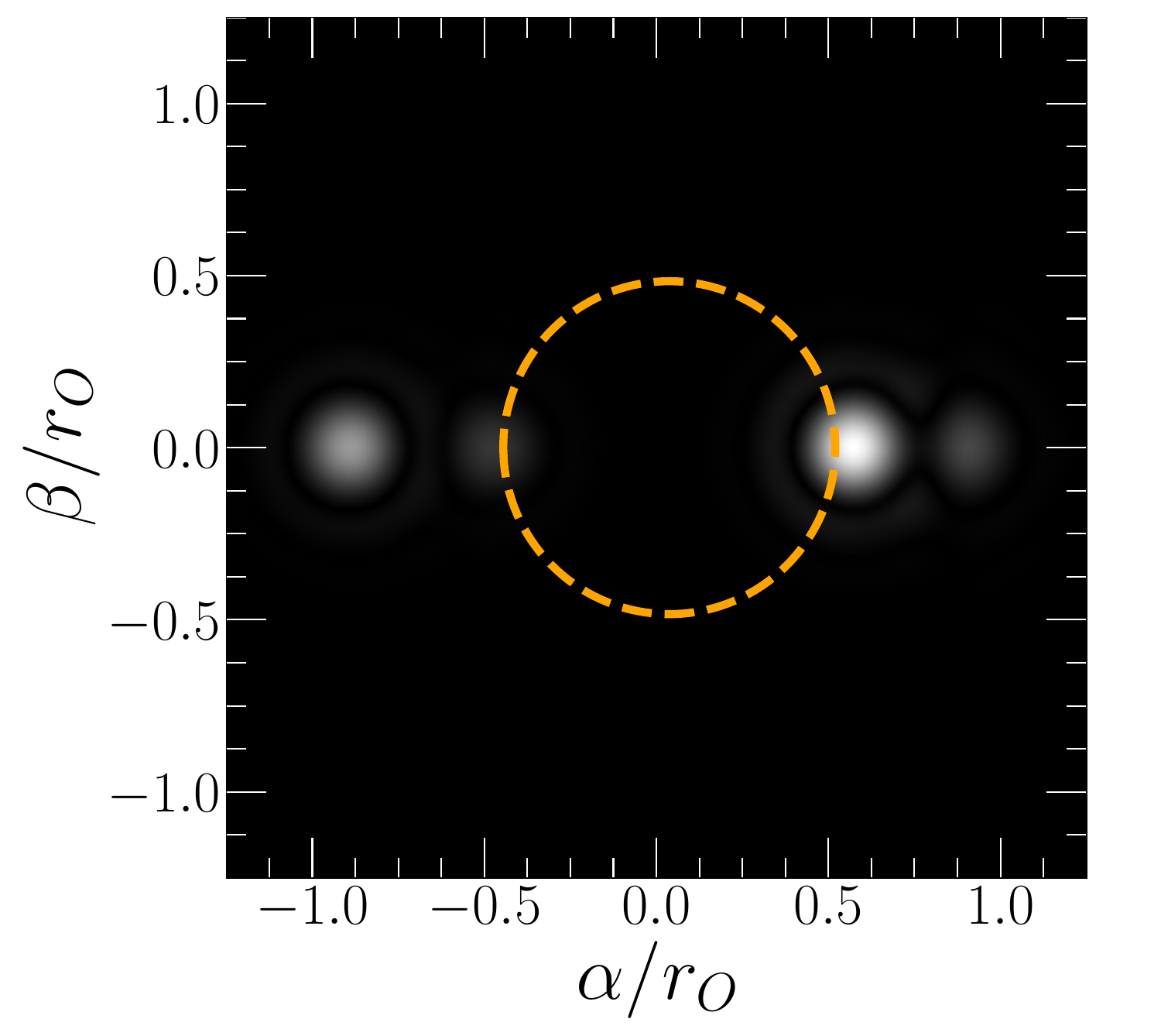}\label{Results:KdS:Fig:aVar2:FFTa0.20}} \hfill
\subfloat[$a = 0.60M$]{\includegraphics[width=0.33\linewidth]{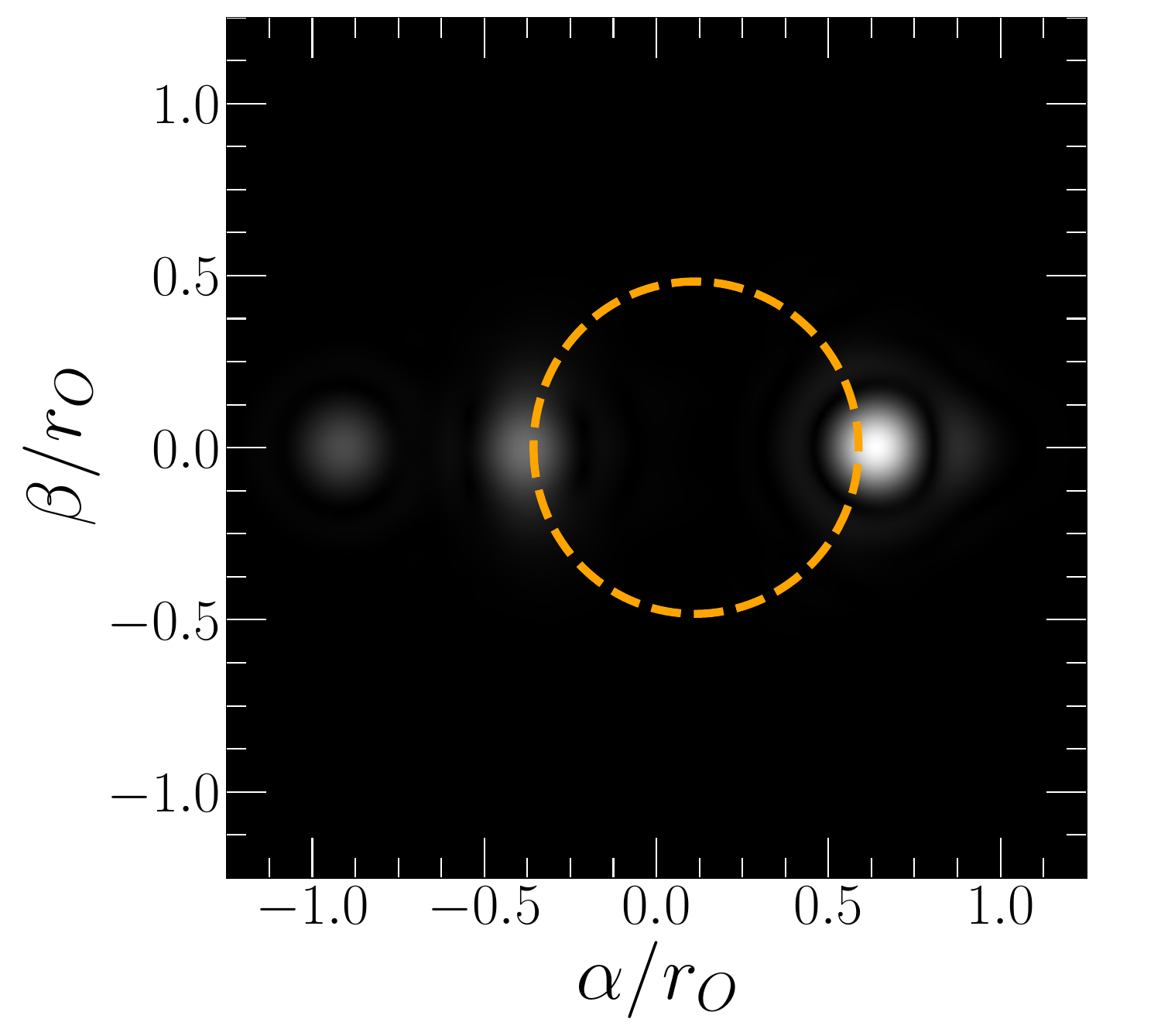}\label{Results:KdS:Fig:aVar2:FFTa0.60}} \\
\subfloat[$a = 0.80M$]{\includegraphics[width=0.33\linewidth]{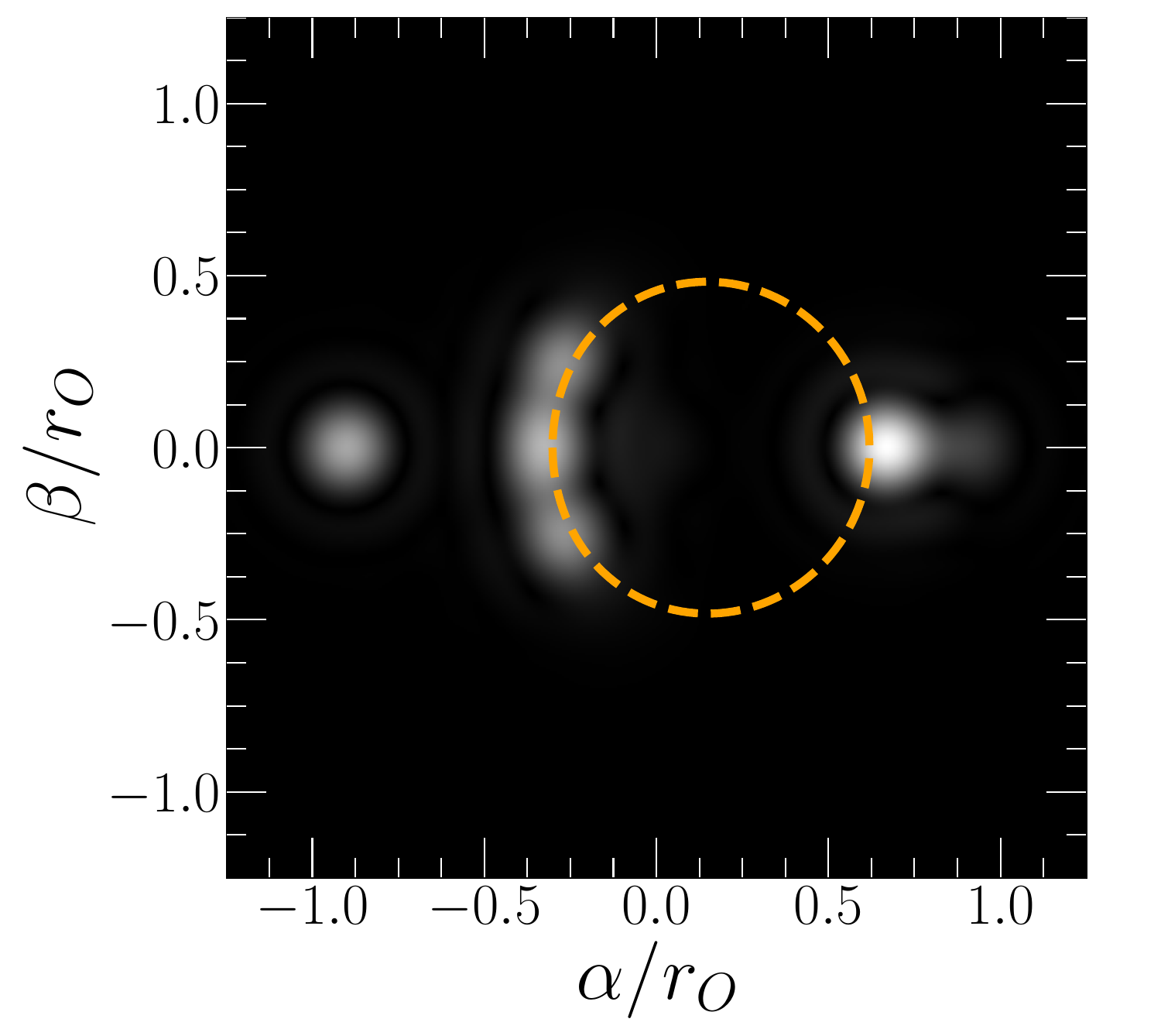}\label{Results:KdS:Fig:aVar2:FFTa0.80}} \hfill
\subfloat[$a = 0.90M$]{\includegraphics[width=0.33\linewidth]{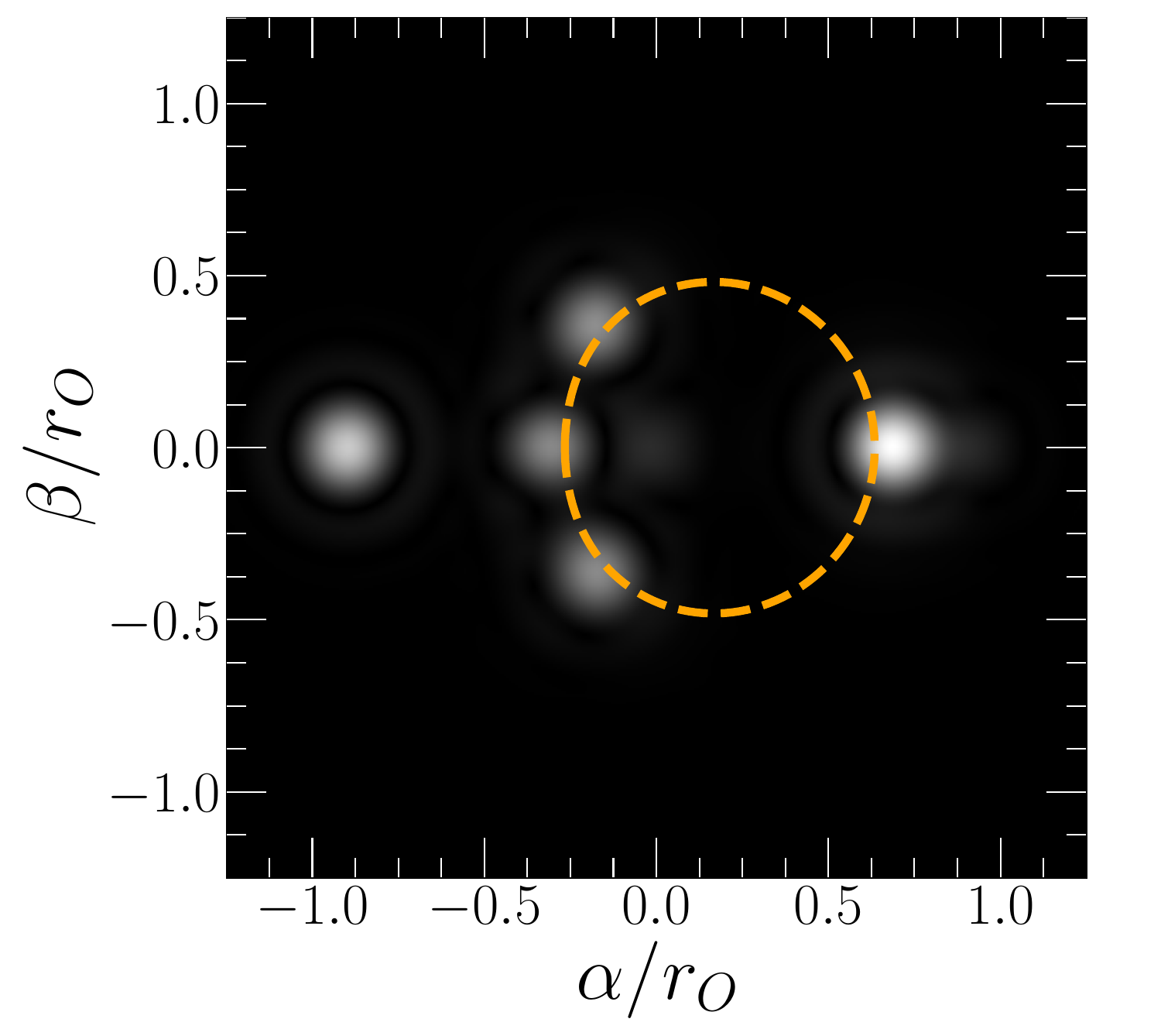}\label{Results:KdS:Fig:aVar2:FFTa0.90}} \hfill
\subfloat[$a = 0.99M$]{\includegraphics[width=0.33\linewidth]{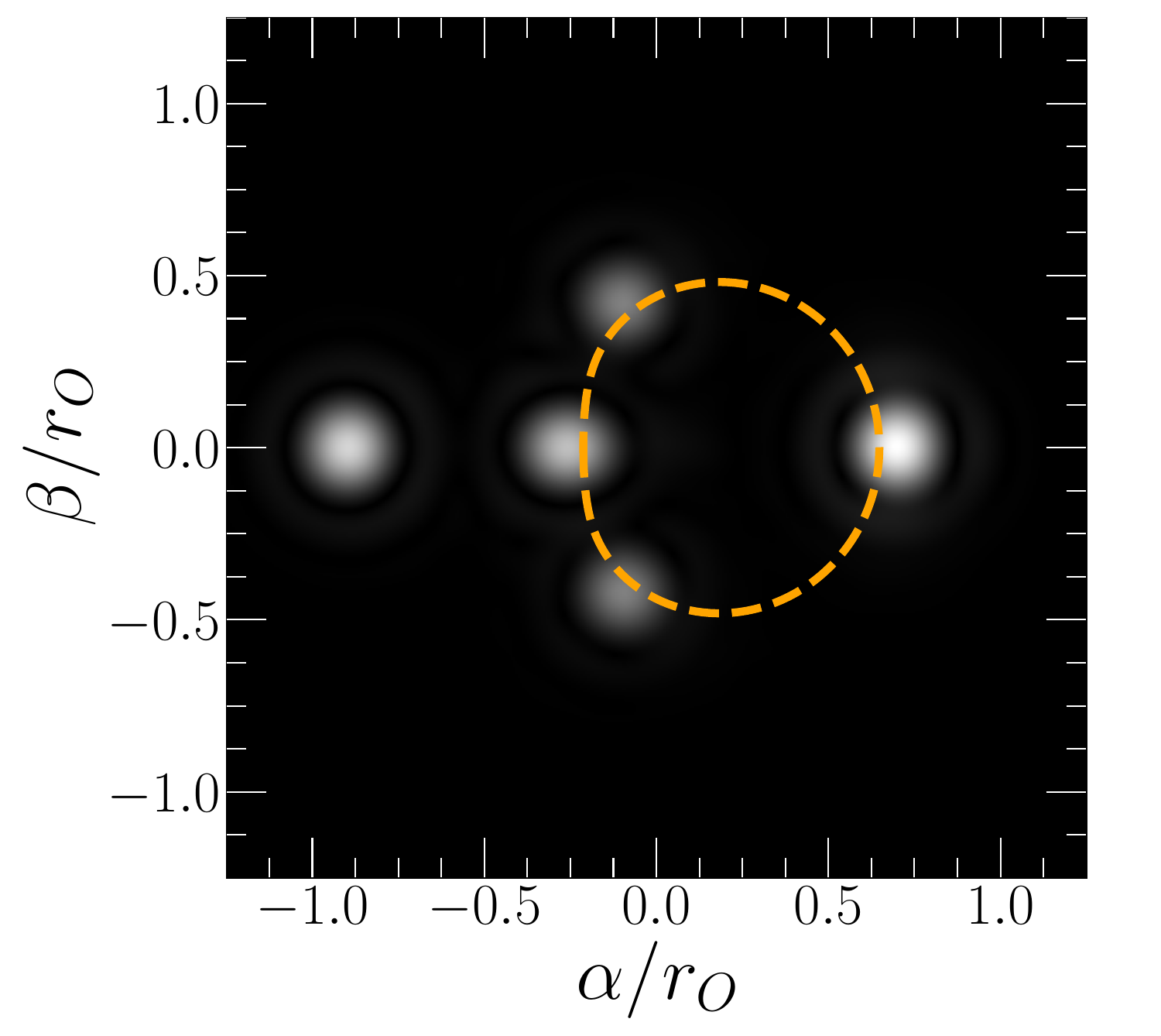}\label{Results:KdS:Fig:aVar2:FFTa0.99}}
\caption{Variation of the Kerr parameter $a/M = \{0.00, 0.20, 0.60, 0.80, 0.90, 0.99\}$ for $M\omega = 15$, $\Lambda M^2 = 10^{-3}$, $r_O = 10 M$, $r_s = 20 M$, $\theta_O = \frac{\pi}{2}$, $\theta_s = \frac{\pi}{2}$, $\phi_s = \frac{\pi}{2}$. 
The image $\left|\mathcal{F}\left(G(\vec{r}, \vec{r}_s, l_\text{max})\right)\right|^2$ is shown. 
The point source is in the equatorial plane, but at $\phi_s = \frac{\pi}{2}$ instead of $\phi_s = \pi$ as before. 
The ray-optical shadow of the black hole is shown (orange-dotted line), as described in \cref{Results:ShadowCompare}.
In this setup, in addition to the primary and secondary images, two additional images separate from the infinitely stacked images at the shadow's boundary with increasing $a$.}
\label{Results:KdS:Fig:aVar2}
\end{figure*}

If the source and observer plane are placed in the equatorial plane and $\phi_s = \frac{\pi}{2}$ is chosen, another effect is apparent, as shown in \cref{Results:KdS:Fig:aVar2}. 
In the case of $a = 0.00 M$, three images aligned at $\beta / r_O = 0$ can be observed.
The lensing approach to imaging and black hole shadows describes the appearance of infinitely stacked images close to the shadow\footnote{Shown as an orange-dotted curve, see \cref{Results:ShadowCompare} for the description of the ray-optical shadow} in \cref{Results:KdS:Fig:aVar2}.
The wave-optical image has these infinitely stacked images near the shadow of the black hole on the left half of \cref{Results:KdS:Fig:aVar2}. 
Increasing the Kerr parameter $a$ causes two more distinct images to separate from the infinitely stacked images, resulting in a total of five visible images of the point source.
However, they appear to be one point. 
Varying the frequency reveals the reason for this. 
For example, when $M\omega = 3$, the four projected points of the source merge into one very large point, brighter than the primary image on the right. 
As the frequency decreases, the projections merge. 
The same argument applies to the infinitely stacked images near the shadow's boundary on the left side, where scattered wave merges into a single point. 
As $l$ increases in \cref{WavOpt:Green:Eq:GreenFun}, the amplitudes of \cref{WavOpt:Green:Eq:radGreen} decrease to zero as $l \rightarrow \infty$. 
Physically, this can be explained by the fact that the modes are more and more absorbed by the black hole.

Our results can be checked against previous work. 
Lens maps computed for the Kerr spacetime \cite{Bohn2015}\footnote{See p. 5, Figure 4 bottom right excerpt. Zooming in very close to the location of our observed additional images, one can barely see the edges of the green and blue planes, which is the location of our point source (a few pixels) for the color coding of the authors work.} confirm the results. 
The appearance of more images in the presence of angular momentum of a massive body is also discussed in previous literature \cite{Bliokh1989,Ibanez1983}.
The described observation cannot be made if $\phi_s = \frac{3}{2} \pi$. 
Also, if the observer plane is placed on one of the poles ($\theta_O \in \{0, \pi\}$) when $\phi_s = \frac{1}{2} \pi$. 
In the latter case, however, there is another effect worth mentioning: due to frame-dragging, the apparent location of the source rotates around the center of the image along the shadow boundary\footnote{This is also shown in \cite{Bohn2015}, p.5, Figure 4, bottom left excerpt} (cf. \cref{Results:ShadowCompare}).

\subsection{Black hole shadow and wave optical imaging}
\label{Results:ShadowCompare}
We have shown that the method outlined above produces images with an intensity distribution that cannot be reproduced using the ray optics approach that solves for null geodesics. 
The latter was useful to verify the wave-optical results, e.g., by observing the frame-dragging or, in case of \cref{Results:KdS:Fig:aVar2}, the appearance of additional images of the source. 

However, a genuine property of black holes has not yet been reproduced with wave optics: the shadow of a black hole. 
In particular the Kerr case is appealing, in which large Kerr parameters $a$ produce a very characteristic shadow in the equatorial plane $\theta_O = \pi/2$, which deviates significantly from a circle and is shifted from its proper origin \cite{Perlick2022, Grenzebach2015}. 
The simple reason for this lies in the construction: In the case of the shadow description originating from the ray-optical approach, the boundary of the shadow can be computed by null geodesics, emitted into the past from the observer's position, which get infinitely close to the photon region of the black hole.
The set of these geodesics covers the whole photon region.
Here, however, the approach is reversed: a point source emits waves that are scattered by the black hole to interfere at the observer's plane. 
All geodesics that would theoretically hit an observer coming from the same point source do not cover the entire photon region. 
Thus, the characteristic shadow will not be revealed in its full nature in the wave-optical imaging of a single source.

Examples of equations describing the apparent shadow for an observer were first given by Bardeen \cite{Bardeen1973} for Kerr black holes. 
Grenzebach et al. \cite{Grenzebach2015} give a description of the shadows for the entire PB spacetime class for arbitrary observers. 
See \cite{Perlick2022} for a comprehensive comparison between the two approaches.
The main difference between the two approaches is the shift of the origin of the shadow depending on $a$, which is due to different definitions of the observer. 
A principal null ray for Bardeen's observer has an angular momentum of $L = p_\phi = 0$. 
Therefore, the observer is corotating and is referred to as a Zero-Angular Momentum Observer (ZAMO) or Locally Nonrotating Frame (LNRF). 
In Grenzebach's case, the principal null ray has an angular momentum of $L = -a \sin\theta_O$ and the observer is called a standard observer, who is a non-corotating, static observer, as viewed from infinity. 
Furthermore, Bardeen's formula differs in its applicability, as it is only valid for observers at large distances. 
In the following, the shadows are calculated using the equations of Grenzebach et al. 
At an observer's location, the shadow is described by the celestial angles 
\begin{subequations}
\label{Results:ShadowCompare:Eq:GrenzebachCelestAngle}
\begin{align}
    \sin \vartheta(r) &= \left.\frac{\sqrt{\Delta_r K_E}}{r^2 + a^2 - a L_E}\right|_{r = r_O} \, , \label{Results:ShadowCompare:Eq:GrenzebachCelestAngleTheta} \\
    \sin \psi(r) &= \left.\frac{L_E - a \sin^2 \theta}{\sqrt{\Delta_\theta K_E} \sin\theta}\right|_{\theta = \theta_O} \, , \label{Results:ShadowCompare:Eq:GrenzebachCelestAnglePsi}
\end{align}
\end{subequations}
where
\begin{subequations}
\begin{align}
    K_E(r) &= \frac{16 r^2 \Delta_r}{\left(\Delta'_r\right)^2} \, ,\label{Results:ShadowCompare:Eq:GrenzebachCelestSupp1} \\
    a L_E(r) &= (\Sigma + a \chi) - \frac{4 r \Delta_r}{\Delta'_r} \, .
\label{Results:ShadowCompare:Eq:GrenzebachCelestSupp2}
\end{align}
\end{subequations}
The stereographic projection is employed to construct the shadow on a plane, resulting in
\begin{subequations}
\label{Results:ShadowCompare:Eq:GrenzebachCelestAng}
\begin{align}
    x(r_p) &= -2 \tan\left(\frac{1}{2}\vartheta(r_p)\right) \sin\psi(r_p) \label{Results:ShadowCompare:Eq:GrenzebachCelestAngx} \, , \\
    y(r_p) &= -2 \tan\left(\frac{1}{2}\vartheta(r_p)\right) \cos\psi(r_p) \label{Results:ShadowCompare:Eq:GrenzebachCelestAngy} \, , 
\end{align}
\end{subequations}
where $r_p \in \left[r_p^\text{min},r_p^\text{max}\right]$ is the radius of the photon region seen from $\theta_O$. 
At the poles $\theta_O \in \{0, \pi\}$ the photon region degenerates to a single point ($r_p^\text{min} = r_p^\text{max}$), while in the equatorial plane it has the maximal extension.
For $a = 0$, the photon region becomes a photon sphere (a topological sphere of radius $r_p = 3 M$). 
However, in our case we still have a cosmological constant that introduces corrections to the radius of the photon sphere.
These corrections are discussed, for example, in \cite{Adler2022}.
The projections $x(r_p)$, $y(r_p)$ can be translated into the origin reference of Bardeen's projection. 
In the previous paragraph, it was mentioned that the difference in origins depends on $a$ and $\theta_O$. 
Shifting $x$ by
\begin{subequations}
\label{Results:ShadowCompare:Eq:Bardeen}
\begin{align}
    \frac{\alpha(r_p)}{r_O} &= x(r_p) - \frac{a \sin\theta_O}{r_O}, \label{Results:ShadowCompare:Eq:BardeenAlpha}\\
    \frac{\beta(r_p)}{r_O} &= y(r_p) \label{Results:ShadowCompare:Eq:BardeenBeta}
\end{align}
\end{subequations}
gives the correct translation between the two observer definitions \cite{Perlick2022}. 
$\alpha(r_p),\beta(r_p)$ are the expressions computed by the Bardeen formula, which divided by $r_O$ gives the celestial angles. 

In the next step, the shadow computed from the ray-optical approach has to be included in the wave-optical image for comparison.
Hence, the relationship between the aforementioned projections and the image plane must be further examined. 
In \cref{Appendix:WavOpt:Coord} we discuss how the coordinates of the image plane are calculated from the viewing angles in the observer plane.
Assuming that observers are at large distances, the viewing angle $\vartheta$ becomes small. 
The projection on the celestial sphere by \cref{Results:ShadowCompare:Eq:GrenzebachCelestAngle}, the stereographic projection \cref{Results:ShadowCompare:Eq:GrenzebachCelestAng} and the angles derived from the coordinates of the image plane \cref{Appendix:WavOpt:Coord:Eq:DetPlaneCoord,Appendix:WavOpt:Coord:Eq:OpeningAngle,Appendix:WavOpt:Coord:Eq:ImPlane} agree and approximate the projections up to the first order in $\vartheta$.
Under this assumption, the center of the observer plane of \cref{WavOpt:FourierOpt:Fig:Scheme} coincides with the location of the observer in \cite{Grenzebach2016}. 

\begin{figure}
\centering
\includegraphics[width=\linewidth]{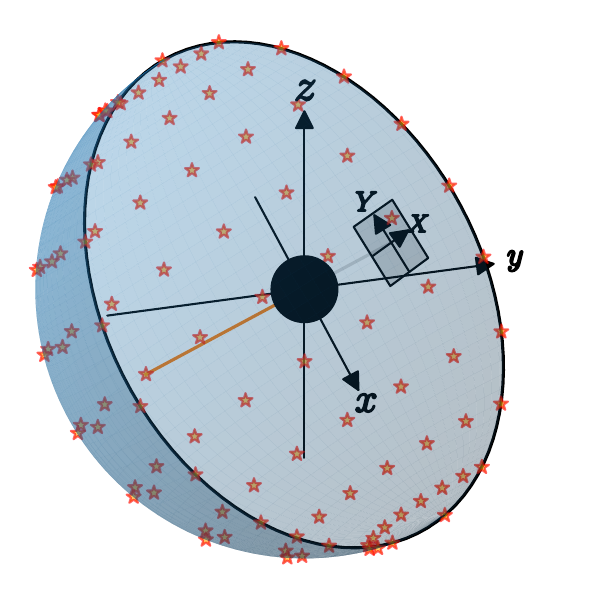}
\caption{Schematic representation of the hemisphere of sources aligned antipodally to the observer plane. 
The orange line points to the antipodal point of the center of the observer plane. 
Sources are aligned such that they are distributed from pole to pole in steps of $\frac{\pi}{10}$ for their coordinates $\theta$ and $\phi$, resulting in a total of 101 sources. 
The relationship between the expansion of the source sphere and the distance to the black hole of the observer plane is true to scale for the evaluated parameter sets used in \cref{Results:ShadowCompare:Fig:MainFig1,Results:ShadowCompare:Fig:MainFig2}. 
The rotation angles are defined in the same way as in \cref{Results:Fig:ObsSrcLoc}, but are omitted here for clarity.}
\label{Results:ShadowCompare:Fig:ObsMultiSrcLoc}
\end{figure}

\begin{figure}
\centering
\includegraphics[width=\linewidth]{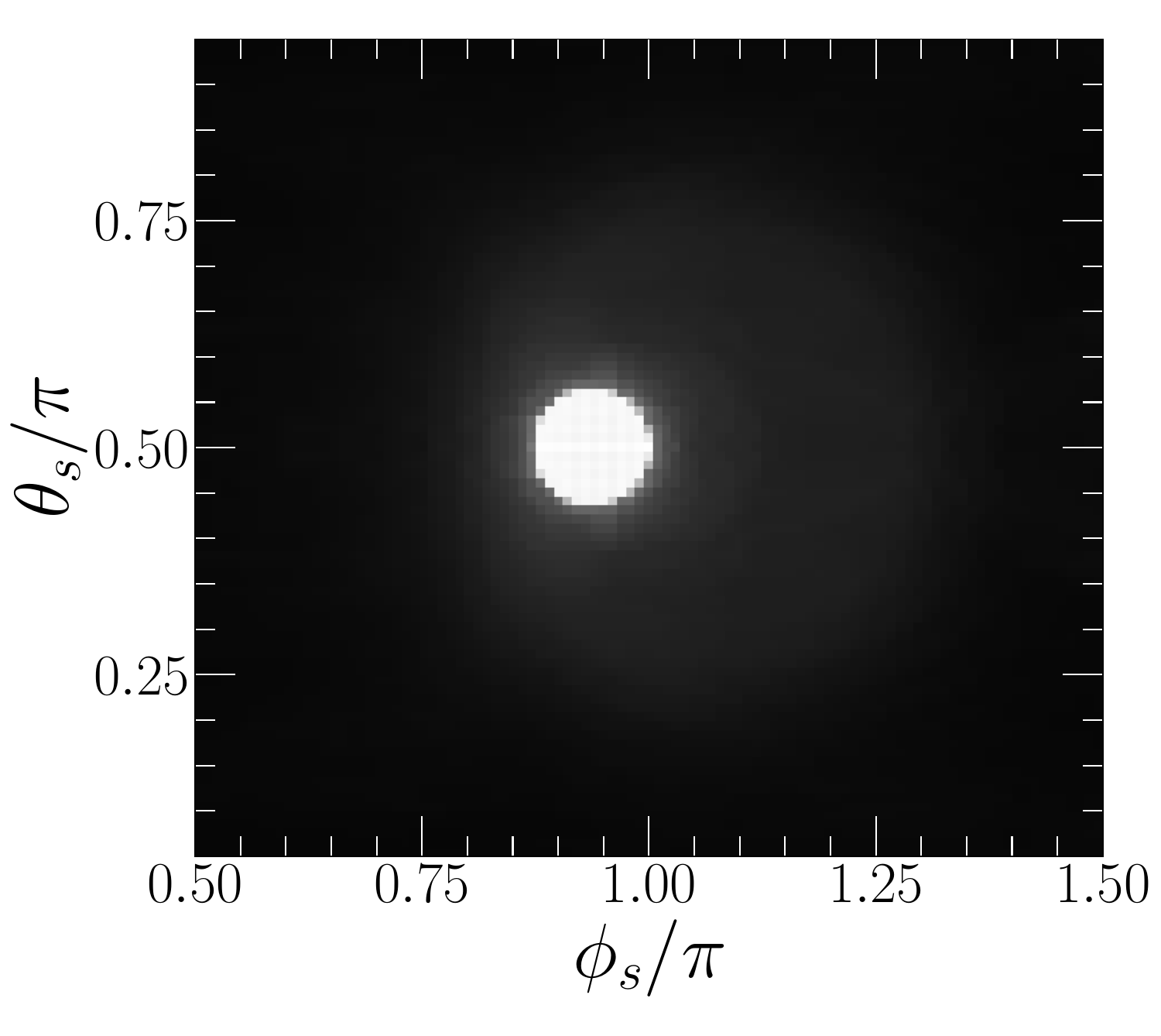}
\caption{Evaluation of $\max\left(\left|G(\vec{r}, \vec{r}_s, L)\right|\right)$ for one single source located at $\left(\phi_s, \theta_s\right)$ . 
Using $a = 0.99M$, $\Lambda M^2 = 10^{-3}$, $M\omega = 15$, $r_O = 10M$, $r_s = 20M$, $\theta_O = \frac{\pi}{2}$, a total of 10.001 point sources are calculated. 
The magnitudes are minimum-maximum normalized for highest contrast. 
The choice of parameters correlates with the results of \cref{Results:ShadowCompare:Fig:MainFig1:0.99,Results:KdS:Fig:aVar2:FFTa0.99,Results:KdS:Fig:DiffSource} and gives an impression of the magnitude and how the normalization of each figure is roughly correlates with different positions of the source and the resulting observable brightness.}
\label{Results:ShadowCompare:Fig:SourceSuperpositionSingleStepMaxStat}
\end{figure}

To model the wave-optical shadow of a black hole, a slight modification to the previous approach must be considered.
The nature of the TME and Green's function as linear differential equations allow for the superposition principle to be applied.
Instead of a single point source as used in \cref{Results:Fig:ObsSrcLoc}, a superposition of many point sources with the same frequency and amplitude aligned on a hemisphere opposite the observer is considered; see \cref{Results:ShadowCompare:Fig:ObsMultiSrcLoc}. 
On this hemisphere, scalar point sources have angular separations of $\frac{\pi}{10}$ in both angular directions. 
In total, 101 sources are taken into account for the observation of the wave-optical shadow.
In this way, the photon region is sufficiently well covered.

\begin{figure*}
\centering
\subfloat[$a = 0.99M$]{\includegraphics[width=0.33\linewidth]{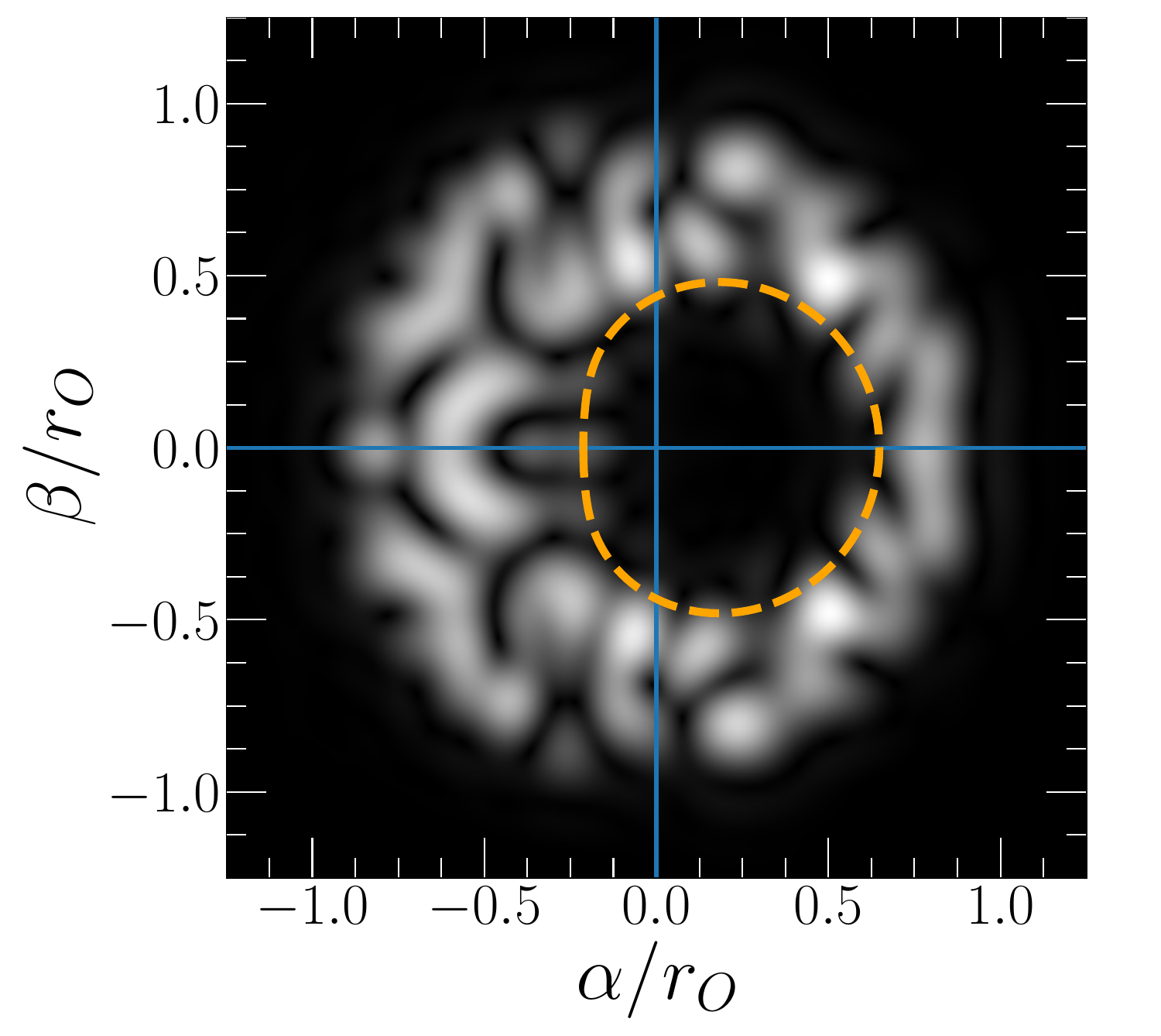}\label{Results:ShadowCompare:Fig:MainFig1:0.99}}
\subfloat[$a = 0.60M$]{\includegraphics[width=0.33\linewidth]{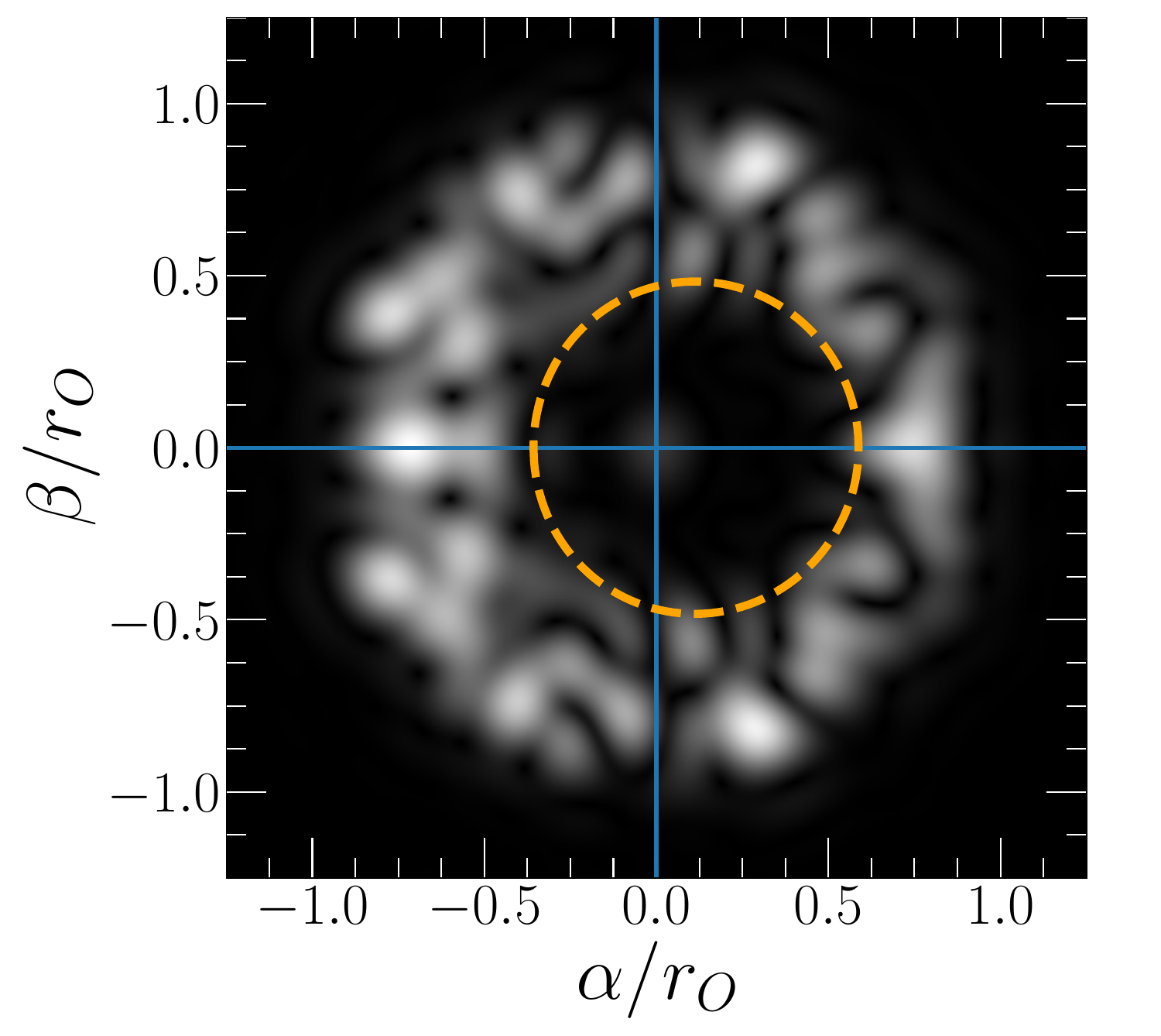}\label{Results:ShadowCompare:Fig:MainFig1:0.60}}
\subfloat[$a = 0.00M$ (SdS)]{\includegraphics[width=0.33\linewidth]{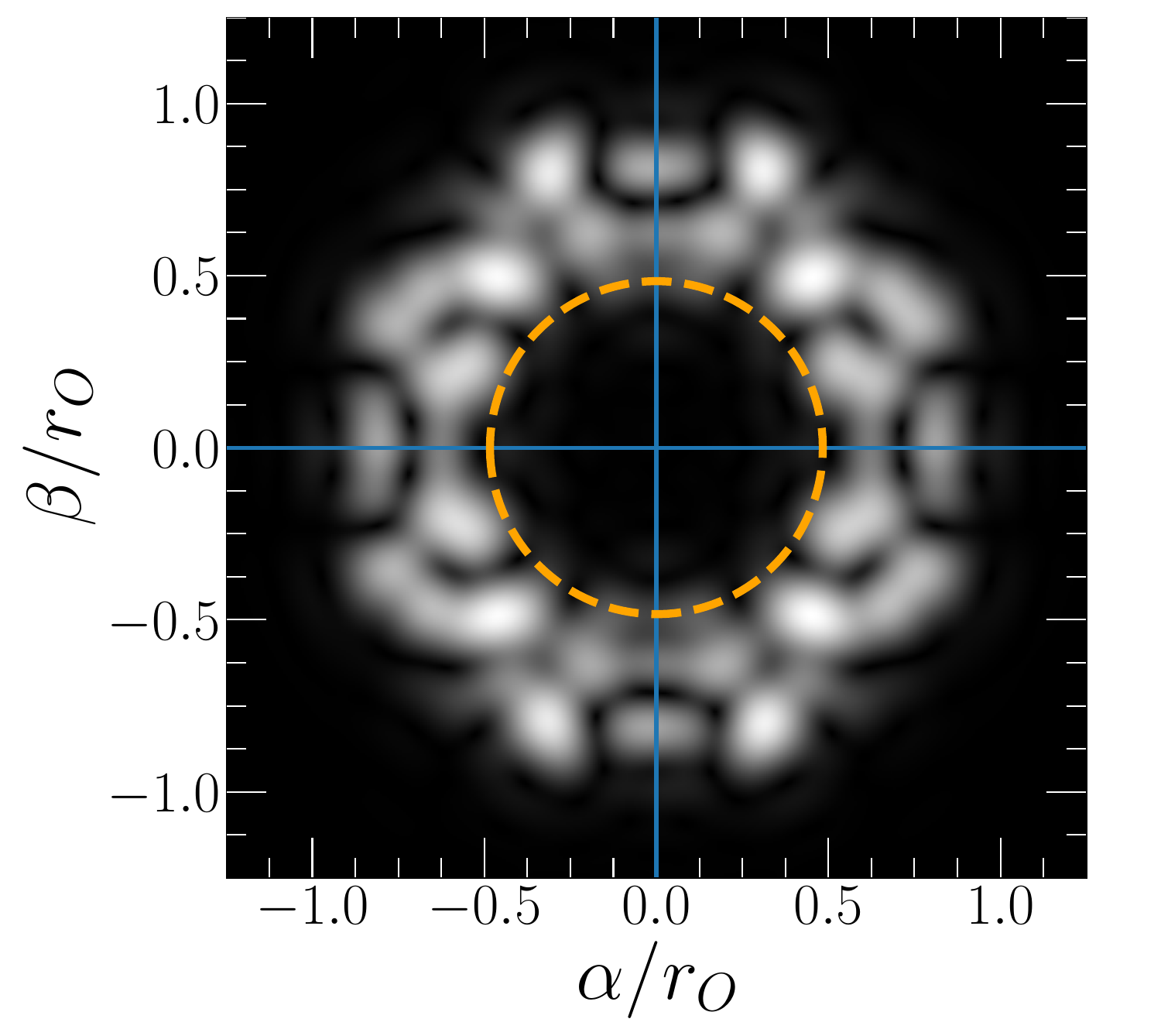}\label{Results:ShadowCompare:Fig:MainFig1:0.00}}
\caption{Validation of the wave-optical results by equation \cref{Results:ShadowCompare:Eq:GrenzebachCelestAng} of \cite{Grenzebach2015}, which is used to produce the dashed orange lines.
$\left|\mathcal{F}\left(G(\vec{r}, \vec{r}_s, l_\text{max})\right)\right|^2$ is shown and the blue lines at $\alpha / r_O = 0$ and $\beta / r_O = 0$ highlight the center.
For different $a$ the shadows are plotted according to the result in \cref{Appendix:WavOpt:Coord:Eq:ImPlane} for the parameters $M\omega = 15$, $\Lambda M^2 = 10^{-3}$, $r_O = 10 M$, $r_s = 20 M$, $\theta_O = \frac{\pi}{2}$, see \cref{Results:ShadowCompare} for the distribution of the sources.}
\label{Results:ShadowCompare:Fig:MainFig1}
\end{figure*}

The first test to construct the characteristic shadow is performed with $M\omega = 15$ and the center of the observer plane placed in the equatorial plane $\theta_O = \frac{\pi}{2}$ with $r_O = 10M$. 
The hemisphere of the sources has a radius of $r_S = 20M$. 
As $a$ increases, the deformation and the shift from the origin should also increase.
\cref{Results:ShadowCompare:Fig:MainFig1} shows the wave imaging results supplemented by the boundary curve of the shadow given by ray optics. 
It can be seen that the shadows have a darker inner region surrounded by an interfering structure. 
Although the inner region appears dark, it is important to note that the intensity is not zero. 
As mentioned above, the images are normalized to the maximum of their respective magnitudes. 
In the inner region, diffraction still leads to illumination, which is observable on the observer plane, as can be seen in, e.g. \cref{Results:ShadowCompare:Fig:MainFig1:0.60}. 
Comparing the results for different $a$ leads to agreement with the results for ray-optical shadows.
The same arguments as for the KdS examination apply to the SdS case as well.
The degeneration of the photon region to a photon sphere results in a circular shadow.

\begin{figure*}
\centering
\subfloat[$\theta_O = \frac{4}{6} \pi$]{\includegraphics[width=0.33\linewidth]{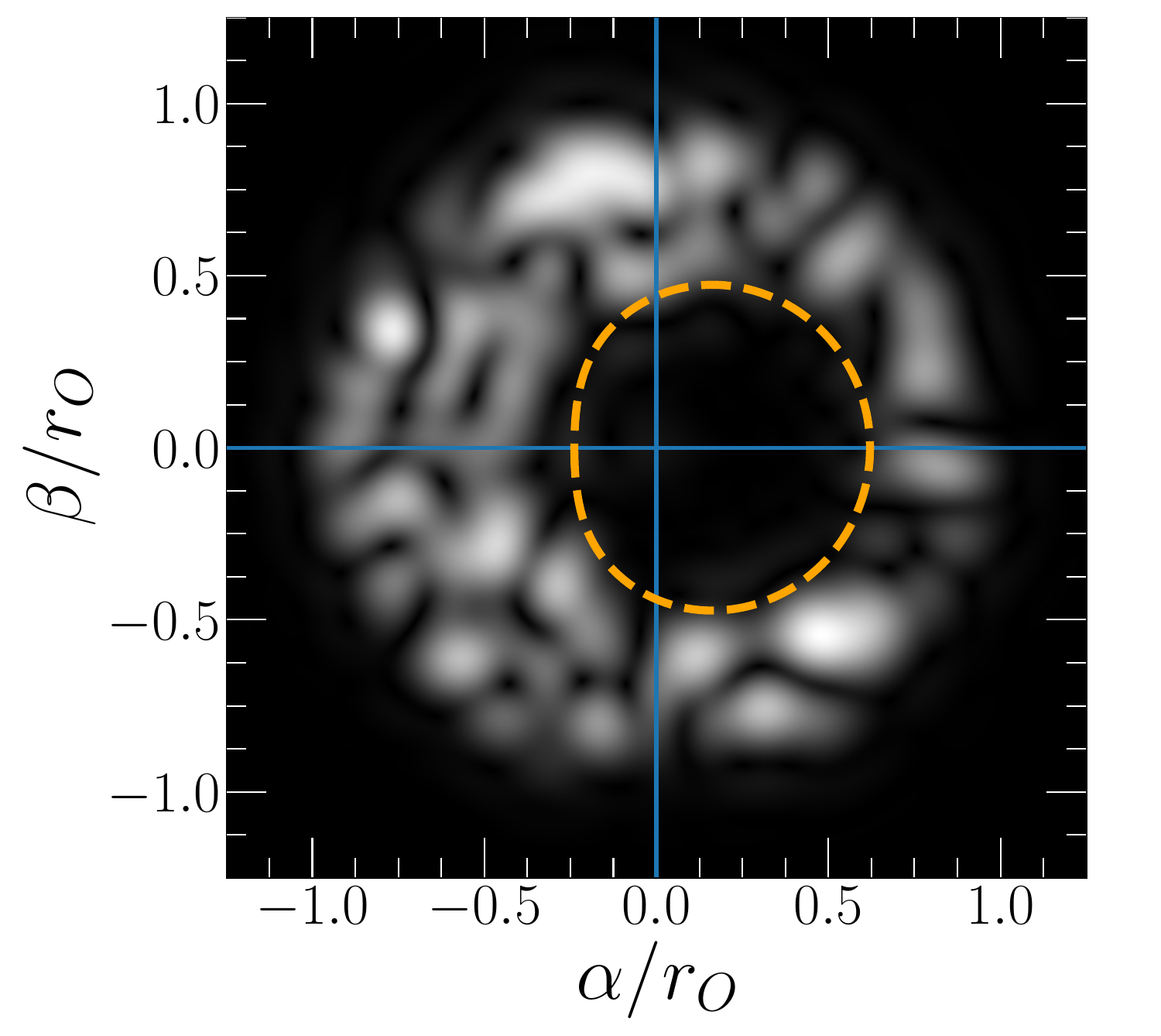}\label{Results:ShadowCompare:Fig:MainFig2:5-8}}
\subfloat[$\theta_O = \frac{5}{6} \pi$]{\includegraphics[width=0.33\linewidth]{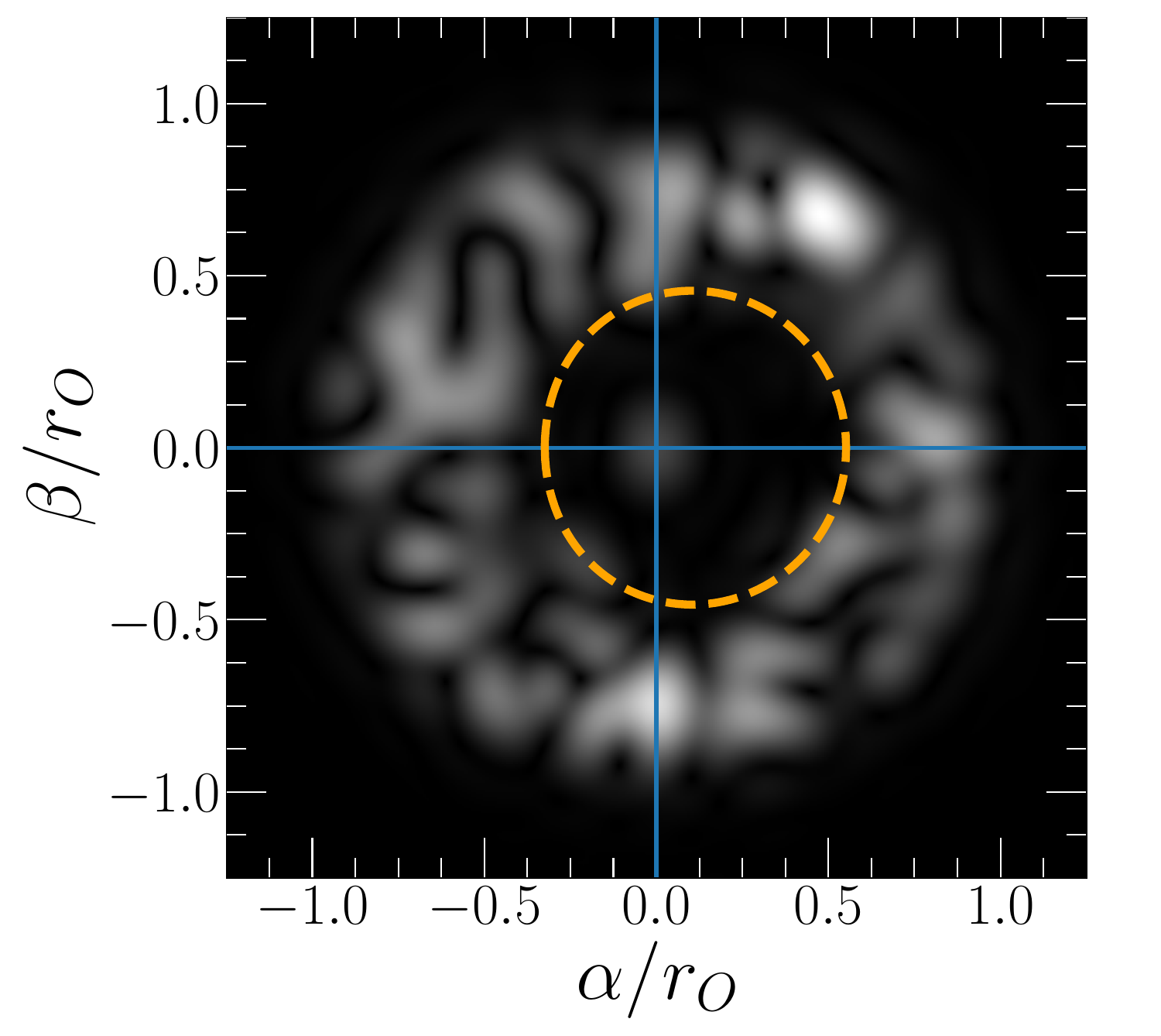}\label{Results:ShadowCompare:Fig:MainFig2:6-8}}
\subfloat[$\theta_O = \frac{6}{6} \pi$]{\includegraphics[width=0.33\linewidth]{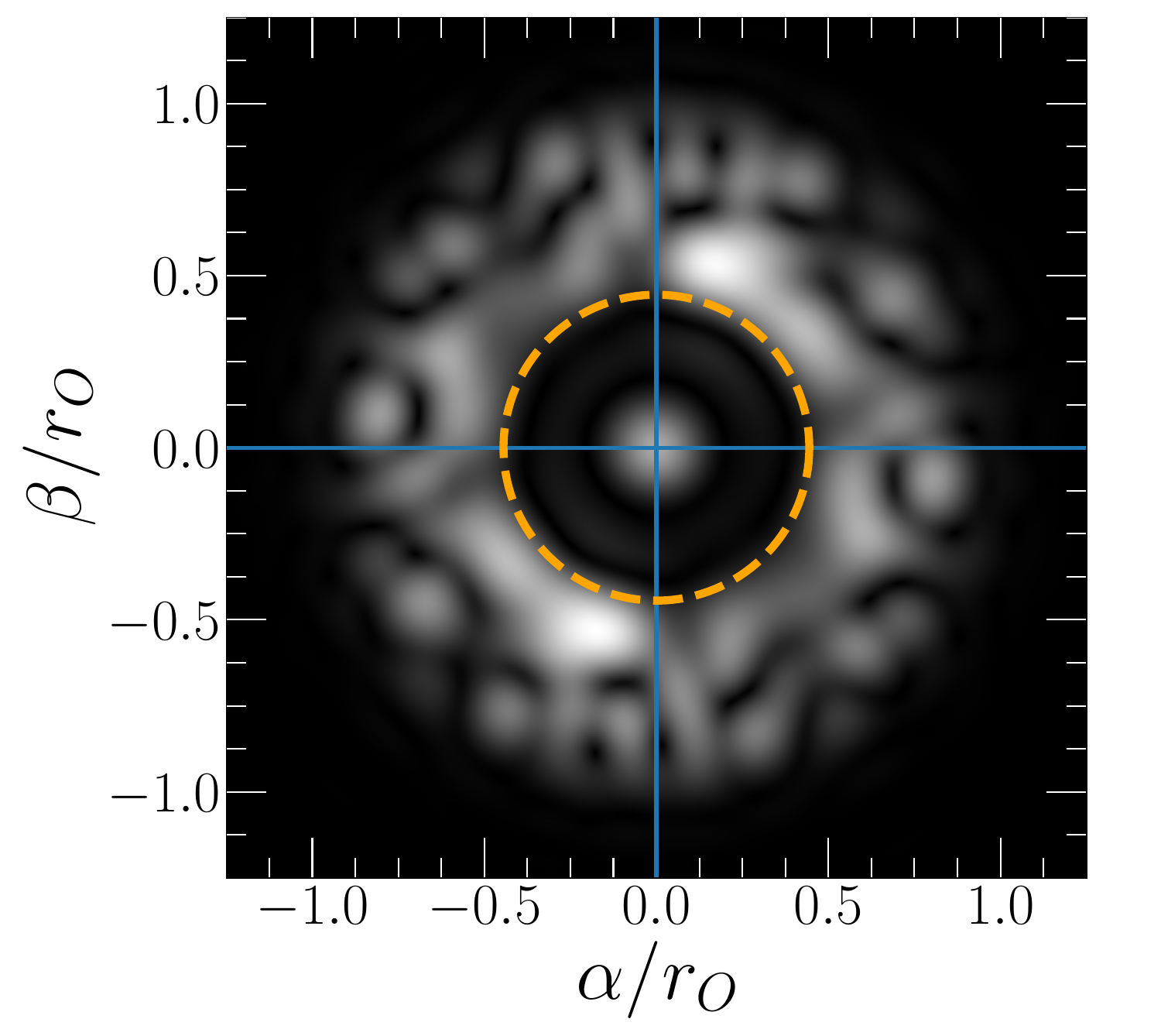}\label{Results:ShadowCompare:Fig:MainFig2:7-8}}
\caption{Similar results as those shown in \cref{Results:ShadowCompare:Fig:MainFig1}, but with fixed $a = 0.99M$ and varying $\theta_O$ are shown. 
The equatorial case for this parameter set can be found in \cref{Results:ShadowCompare:Fig:MainFig1:0.99}. 
The ray-optical shadow and the wave-optical shadow are compatible.
The emerging circle in the center of the shadow turns out to be a Poisson spot, which shrinks as the frequency increases.}
\label{Results:ShadowCompare:Fig:MainFig2}
\end{figure*}

Despite the variation of $a$, changing the polar position $\theta_O$ of the observer also has a crucial effect on the morphology of the characteristic Kerr shadow. 
While the deformation of the shadow is maximal in the equatorial plane, the shadow becomes increasingly circular at the poles.
For an observer located at $\theta_O \in \{0, \pi\}$, the photon region also degenerates to a photon sphere, leading to a circular shadow as in the case $a = 0$. 
Although the photon region became spherical, it should be emphasized that in this case the scattering results are not the same as in the SdS case. 
\cref{Results:ShadowCompare:Fig:MainFig2} shows a variation of the polar coordinate in $\pi/6$ steps from the equatorial plane to $\theta_O = \pi$. 
The hemisphere of the point sources co-moves in such a way that the antipodal alignment is maintained. 
All other parameters are fixed, and for the Kerr parameter an extreme choice $a = 0.99 M$ is considered. 
The equatorial case can be found in \cref{Results:ShadowCompare:Fig:MainFig1:0.99}. 
The results show that the wave shadow is again consistent with the theoretical prediction of \cref{Results:ShadowCompare:Eq:GrenzebachCelestAng}. 
However, in contrast to the equatorial variation of $a$, a significant spot appears in the center of the figure. 
Such spots are called Poisson's spots and are caused by constructive interference, which is more pronounced in the variation of angular position. 
Increasing the frequency shrinks the spot and it would theoretically disappear in the high-frequency limit, thus, not being observable in ray optics.

\section{Conclusion and Outlook}
In this work, the exact wave-optical imaging of point sources in the KdS spacetime is discussed.
The solutions of separated radial and angular Teukolsky equations are given in terms of solutions to Heun's equation. 
The main problem in wave scattering is the normalization of angular solutions. 
The solution of the angular Teukolsky equation and the derivation of the corresponding eigenvalues lead to the so-called Heun functions $Hf$, which possess an orthogonality relation that allows for normalization.
Omitting the normalization constant prevents the set of solutions from being an orthonormal basis, which is mandatory for the expansion of arbitrary square integrable functions. 

Comparison of wave-optical images shows that the SdS case is a limit of the KdS case for $a\to 0$.
Omitting the normalization constant for $a \neq 0$ results in unusable images that neither agree with the results of the ray-optical approaches nor yield the SdS limit.
A key difference between wave-optical and ray-optical results (e.g., lens maps, see \cite{Bohn2015,Frost2021}, or numerical ray tracing of GRMHD models \cite{EHTC2019}), are observable amplitude distributions caused by interference.

Our results show expected properties, e.g., primary and secondary images appearing in non-antipodal alignments as well as Einstein rings for apparent antipodal cases, which agree with results from ray optics.
Frame-dragging and image splitting, which are a consequence of rotating compact bodies, are also reproduced. 
The nature of \cref{WavOpt:Green:Eq:GreenFun} allows for a superposition of multiple point sources, which is necessary to construct BH shadows, since all wave paths to the observer must cover the entire photon region of the black hole, causing a shadow as seen in \cref{Results:ShadowCompare:Fig:MainFig1,Results:ShadowCompare:Fig:MainFig2}. 
The variation of $a$ and $\theta$ shows wave-optical shadow regions in agreement with the results derived from ray tracing approaches.

The computational efficiency of current implementations of the Heun function limits its applicability in our investigation.
In principle, the method can be used to investigate arbitrary frequencies.
However, in order to obtain results in a reasonable time, we limit our evaluations to $M\omega \leqslant 20$.
Otherwise, the computation time increases non-linearly because \cref{WavOpt:Green:Eq:GreenFun} has a higher cut-off $L$ (see \cref{Appendix:Convergence}). 
Becker \cite{Becker1997} suggests in his derivation of the normalization constant the use of already computed expressions, so that they are not evaluated unnecessarily multiple times in each step. 
Promising results of an alternative implementation of the solutions of Heun's equations \cite{Birkandan2021a} suggest an improvement in computational efficiency of several orders of magnitude, greatly reducing the effort required to compute the exact equations. 

Until now, only low frequencies have been used in the scalar case $s=0$. 
Therefore, the derivation of a high-frequency approximation in terms of solutions to Heun's equation may be of interest. 
This will reduce the computational cost and allow higher-frequency results to be computed in a reasonable amount of time, while serving as a bridge between the low frequencies studied in this work and the ray-optical computations of previous works.
This allows wave-optical investigations of the effects of $a$ and $\Lambda$, e.g. as in \cite{Nambu2016}.

In addition to different frequency regimes, other bosonic perturbations are also of interest, in particular $s=1$ electromagnetic fields and gravitational fields for $s=2$.
The employed NP formalism and our results allow for reconstruction of, e.g., the full Faraday tensor.
Scalar fields have been used as a simple model and to approximate other perturbations \cite{Andersson2000}. 
However, polarization degrees of freedom, e.g., in the low-frequency scattering of gravitational waves, may be worth further analysis for the study of effects around a black hole.
Assuming that the observer is sufficiently far away from the block hole, the construction of a simple coordinate plane as the observer plane results in apparent images that match the ray-optical results for a ZAMO observer. 
This is reflected in the shift of the coordinate origin of the image plane along the $X_I$ axis in \cref{Results:ShadowCompare:Eq:BardeenAlpha}. 
The question naturally arises as to why the results here are related to those seen by a ZAMO observer.
A fully satisfactory answer cannot be given yet, but will be the subject of future work. 
In addition, it is of interest to construct the observer plane in such a way that arbitrary velocities of an observer can be included, as done in \cite{Grenzebach2016}. 
This will, of course, not only result in a different wave-optical shadow, but will also alter scattered images.
In the case of de Sitter-spacetimes, the study of observers comoving with the expansion presents an interesting case in its own.

Besides this conceptual discussion of wave-optics via black hole scattering, the generalization of the background spacetime will be of interest in future work, e.g., including more free parameters of Plebanski-Demianski spacetimes, wormhole spacetimes, or in alternative gravitational theories.

\section{Acknowledgements}
The authors would like to thank Volker Perlick, Torben Frost and Oleg Tsupko for fruitful discussions and hints during the preparation of the manuscript. 
The research is funded by the Deutsche Forschungsgemeinschaft (DFG, German Research Foundation) – Project-ID 434617780 – SFB 1464 and
funded by the Deutsche Forschungsgemeinschaft (DFG, German Research Foundation) under Germany’s Excellence Strategy – EXC-2123 QuantumFrontiers – 390837967, and through the Research Training Group 1620 ``Models of Gravity''.

\section{Appendix}
\appendix
\section{On the TME}
The discussion of linear perturbations of a background metric is intuitively done by considering the decomposition $g_{\mu\nu} = \eta_{\mu\nu} + h_{\mu\nu}$.
It is used to study, e.g., the quasinormal modes of black holes \cite{Zerilli1970}. 
Another approach to the perturbation problem was taken by Teukolsky \cite{Teukolsky_1973} via the Newman-Penrose (NP) formalism. 
The advantage of perturbing the metric in the NP formalism is the extension to fields with an arbitrary spin.
The 12 NP scalars form the basis of the formalism, and their derivation is performed in terms of a null tetrad system. 
The choice of the two real null tetrads $l$ and $n$, representing radial in- and outgoing null rays viewed from an asymptotic region, and a complementary complex null tetrad $m$ is substantial. 
They fulfil
\begin{subequations}
\begin{align}
l_a n^a &= -1, \\
m_a \bar{m}^a &= 1,
\label{TME:KT:Eq:NonZero}
\end{align}
\end{subequations}
which are the only non-zero contractions.
In Petrov type-D metrics, some degrees of freedom vanish, leading to $\kappa = \sigma = \lambda = \nu = 0$. 
However, a complex null rotation is not yet fixed. 
Performing such a rotation with $\epsilon = 0$, also fixes the last degree of freedom, leading to the so-called Kinnersley tetrads \cite{Kinnersley_1969,Teukolsky_1973}. 
For the KdS metric, expressed in PB functions, this yields
\begin{subequations}
\begin{align}
    l^\mu &= \frac{1}{\Delta_r} \left[(\Sigma + a \chi) \partial_t + \Delta_r \partial_r + a \partial_\phi \right], \\
    n^\mu &= \frac{1}{2 \Sigma} \left[(\Sigma + a \chi) \partial_t - \Delta_r \partial_r + a \partial_\phi \right], \\
    m^\mu &= -\frac{i \rho^*}{\sqrt{2}} \left[\frac{\chi \csc\theta}{\sqrt{\Delta_\theta}} \partial_t - i \sqrt{\Delta_\theta} \partial_\theta + \frac{\csc\theta}{\sqrt{\Delta_\theta}} \partial_\phi \right].
\end{align}
\end{subequations}
The remaining non-zero coefficients are
\begin{subequations}
\begin{align}
    \rho &= -\frac{1}{r - i a \cos\theta} \\
    \tau &= -i \frac{a \sin\theta \sqrt{\Delta_\theta}}{\sqrt{2} \Sigma}\\
    \pi &= i \rho^2 \frac{a \sin\theta \sqrt{\Delta_\theta}}{\sqrt{2}}\\
    \mu &= \frac{\rho \Delta_r}{2 \Sigma}\\
    \gamma &= \frac{2 \Delta_r \rho + \Delta_r'}{4 \Sigma} \\
    \beta &= -\rho^* \frac{2 \cot\theta \Delta_\theta + \Delta'_\theta}{4 \sqrt{2} \sqrt{\Delta_\theta}}\\
    \alpha &= \rho \frac{2 \Delta_\theta\left(\cot\theta + 2 i \rho \sin\theta\right) + \Delta'_\theta}{4 \sqrt{2} \sqrt{\Delta_\theta}}
\end{align}
\end{subequations}
From these, 5 complex so-called Weyl scalars $\Psi_i$ can be composed. 
Type-D spacetimes eliminate certain scalars ($\Psi_1 = \Psi_3 = \Psi_4 = \Psi_5 = 0$).  
The remaining non-zero Weyl scalar is
\begin{align}
\Psi_2 = M \rho^3 \, .
\end{align}
In contrast to linear perturbation in the metric, all spin coefficients, Kinnersley tetrads and Weyl scalars are linearly perturbed. 
From these perturbations and related symmetries, the differential equations for different spin weights are derived in terms of the NP formalism \cite{Bini_2003}, where for scalar perturbations ($s = 0$)
\begin{align}
    [&D\Delta + \Delta D - \delta^* \delta - \delta \delta^* + (-\gamma - \gamma^* + \mu + \mu^*) D \notag \\
    & + (\epsilon + \epsilon^* - \rho^* - \rho) \Delta + (-\beta^* - \pi + \alpha + \tau^*) \delta \\
    & + (-\pi^* + \tau - \beta + \alpha^*) \delta^*] {}_s\Psi_{lm} = 0 \, . \notag
\end{align}
The resulting differential equation for the scalar case coincides with the Klein-Gordon equation in de Sitter spacetimes $\left(\Box - \frac{R}{6}\right) \Phi = 0$. 
For positive spin-weights $\left(s \in \left\{\frac{1}{2}, 1, 2\right\}\right)$ the Teukolsky master equation follows,
\begin{align}
    \{&[D - \rho^* + \epsilon^* + \epsilon - 2 s (\rho + \epsilon)] \left(\Delta + \mu - 2 s \gamma\right) \notag \\
    & - [\delta + \pi^* - \alpha^* + \beta - 2 s (\tau + \beta)] (\delta^* + \pi - 2 s \alpha) \\
    & - 2 (s - 1) (s - 1/2) \psi_2\} {}_s\Psi_{lm} = 0 \notag
\end{align}
and for negative spin-weights $\left(s \in \left\{-\frac{1}{2}, -1, -2\right\}\right)$ we obtain
\begin{align}
    \{&[\Delta - \gamma^* + \mu^* - \gamma - 2 s (\gamma + \mu)] (D - \rho - 2 s \epsilon) \notag \\
    & - [\delta^* - \tau^* + \beta^* - \alpha - 2 s (\alpha + \pi)] (\delta - \tau - 2 s \beta) \\
    & - 2 (s + 1)(s + 1/2) \psi_2\} {}_s\Psi_{lm} = 0 \, , \notag
\end{align}
where $D = l_\mu \partial^\mu$, $\Delta = n_\mu \partial^\mu$, and $\delta = m_\mu \partial^\mu$ are directional derivatives of the NP formalism. 
Another noteworthy extension proceeds to supersymmetric spin fields $s = \pm 3/2$, which is approached via the Geroch-Held-Penrose formalism \cite{Geroch1973}. 

The solution ${}_s\Psi_{lm}$ of the differential equation yields scalars of the Newman-Penrose or Geroch-Held-Penrose formalism according to \cref{TME:Tab:NPGHPsol}.
\begin{table}
\centering
\begin{tabular}{c||c|c|c|c|c|c|c|c|c}
$s$ & 0 & $\frac{1}{2}$ & $-\frac{1}{2}$ & 1 & -1 & $\frac{3}{2}$ & -$\frac{3}{2}$ & 2 & -2 \\
\hline
${}_s\Psi_{lm}$ & $\Phi$ & $\chi_0$ & $\rho^{-1} \chi_1$ & $\phi_0$ & $\rho^{-2} \phi_2$ & $\Omega_0$ & $\rho^{-3} \Omega_3$ & $\psi_0^B$ & $\rho^{-4} \psi_2^B$
\end{tabular}
\caption{Solution of ${}_s\Psi_{lm}$ for different spin-weights $s$. The respective expressions are scalars from Newman-Penrose or rather Geroch-Heldt-Penrose formalism \cite{Teukolsky_1973}.}
\label{TME:Tab:NPGHPsol}
\end{table}
Finally, for the KdS metric using the NP identity \cite{Chandrasekhar1998}
\begin{align}
    D\mu - \delta\pi = &(\rho^* \mu + \sigma \lambda) + \pi \pi^* - (\epsilon + \epsilon^*) \mu - (\alpha^* - \beta)\pi \notag \\
    & - \nu \kappa + \psi_2 + 2 \Lambda^\text{NP} 
\end{align}
where $\Lambda^\text{NP} = \frac{R}{24}$ and $R = 4 \Lambda$, the final form of the TME shown in \cref{TME:Eq:TME} follows.

\section{Coordinate system of the image plane}
\label{Appendix:WavOpt:Coord}
The simplification of the diffraction integral to a Fourier transformation, discussed as an approximation in the Kirchhoff-Fresnel theory, gives access to the image plane. 
In the image plane the coordinates
\begin{subequations}
\label{Appendix:WavOpt:Coord:Eq:DetPlaneCoord}
\begin{align}
    X_I &= r_I \sin \phi_I \, , 
\label{Appendix:WavOpt:Coord:Eq:DetPlaneCoordX} \\
    Y_I &= r_I \cos \phi_I \, ,
\label{Appendix:WavOpt:Coord:Eq:DetPlaneCoordY}
\end{align}
\end{subequations}
are used.
The image mapped onto the image plane reveals information about the apparent angular position of the observed object. 
Therefore, it is of great interest to derive relations between the angular position and coordinates in the image plane. 
\begin{figure}
\centering
\includegraphics[width=\linewidth]{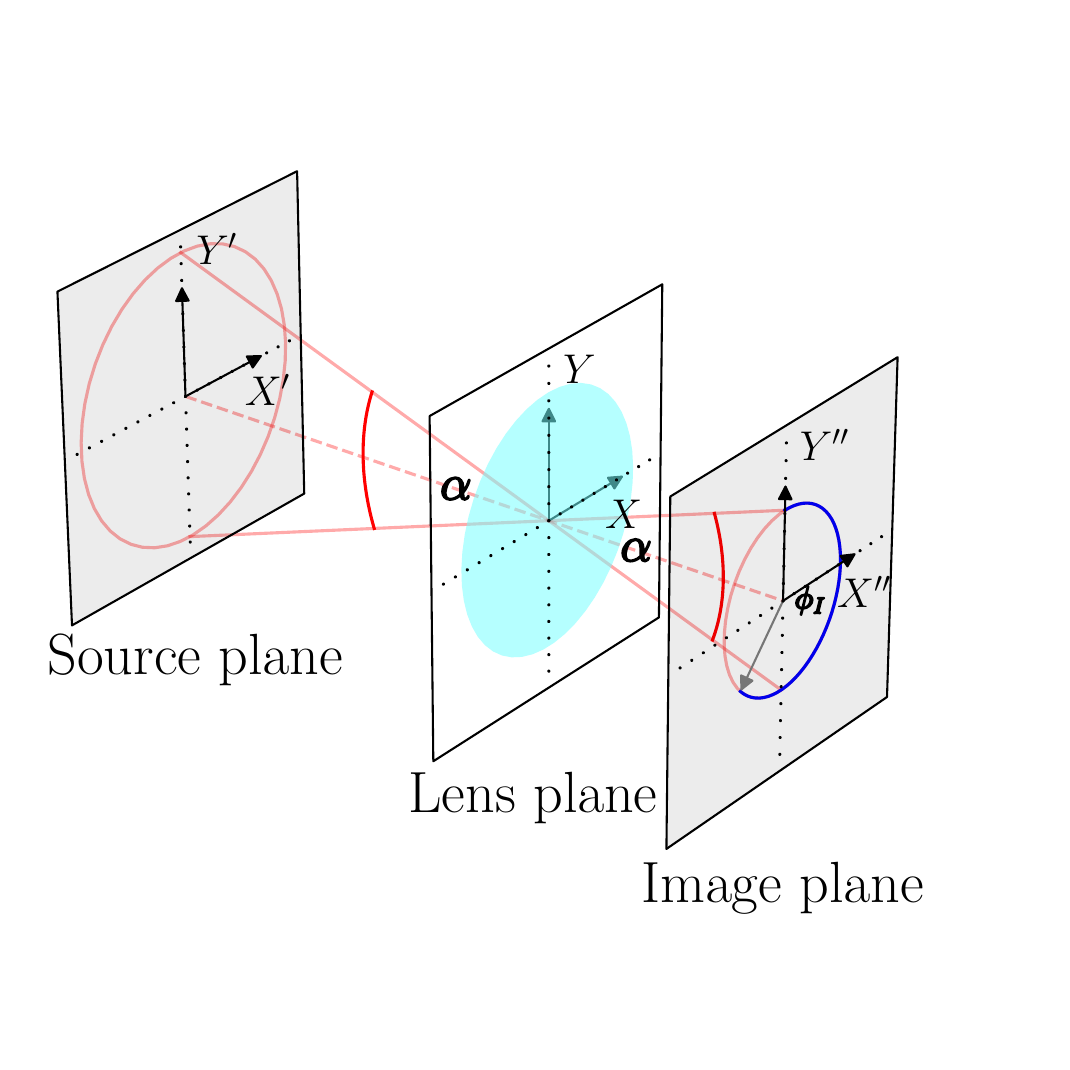}
\caption{Extension to \cref{WavOpt:FourierOpt:Fig:Scheme} to display the viewing angles $\alpha$ and $\phi_I$ in the case of Kirchhoff-Fresnel diffraction planes. 
The projection on the image plane connects the coordinates $X_I$, $Y_I$ with the viewing angles.}
\label{Appendix:WavOpt:Fig:OpeningAngles}
\end{figure}
Returning to the simple concept of a convex lens imaging an object on the image plane placed at a distance $f$ from the lens, as seen in \cref{Appendix:WavOpt:Fig:OpeningAngles,WavOpt:FourierOpt:Fig:Scheme}, the opening angle $\alpha$ defines the apparent angular size both from the object to the lens and from the lens to the detector. 
It can be derived by simple trigonometric relations and gives access to the radial distance $r_I$ of an imaged object from the center of the coordinate system,
\begin{align}
    r_I &= f \tan\vartheta \, , \label{Appendix:WavOpt:Coord:Eq:OpeningAngle}
\end{align}
where the half angle is $\vartheta = \frac{\alpha}{2}$. 
For points near the optical axis, the radius is approximately $r_I \approx f \theta$. 
Thus, the focal-length normalized image plane coordinates are
\begin{subequations}
\label{Appendix:WavOpt:Coord:Eq:ImagePlane}
\begin{align}
    \frac{X_I}{f} &= \tan\vartheta \sin\phi_I \approx \vartheta \sin\phi_I \, , 
\label{Appendix:WavOpt:Coord:Eq:ImagePlaneXI} \\
    \frac{Y_I}{f} &= \tan\vartheta \cos\phi_I \approx \vartheta \cos\phi_I \, . 
\label{Appendix:WavOpt:Coord:Eq:ImagePlaneYI}
\end{align}
\end{subequations}

Considering large distances results in points near the center of the projection. 
Consequently, an approximation with respect to $\vartheta$ up to the first order yields
\begin{subequations}
\begin{align}
    x(r_p)& = -2 \tan\left(\frac{1}{2} \vartheta(r_p)\right) \sin\psi(r_p) \approx -\vartheta(r_p) \sin\psi(r_p) \, , \\
    y(r_p) &= -2 \tan\left(\frac{1}{2} \vartheta(r_p)\right) \cos\psi(r_p) \approx -\vartheta(r_p) \cos\psi(r_p) \, .
\end{align}
\end{subequations}

Stereographic projection is usually defined without a minus sign, as well as with $x(r_p)$ defined by $\cos\psi(r_p)$ and $y(r_p)$ defined using $\sin\psi(r_p)$. 
This is due to the convention used by Grenzebach et al.\ \cite{Grenzebach2016}\footnote{See in particular the defining equation Eq. (4.4) and the complementary Fig. 4.2.} and is respected in \cref{Appendix:WavOpt:Coord:Eq:ImagePlane}. 
Considering the projection onto the celestial sphere, again for small $\vartheta$ (in the authors' notation), gives
\begin{subequations}
\begin{align}
    \tilde{x} &= \sin\psi(r_p) \sin\vartheta(r_p) \approx \vartheta(r_p) \sin\psi(r_p) \, , \\
    \tilde{y} &= \cos\psi(r_p) \sin\vartheta(r_p) \approx \vartheta(r_p) \cos\psi(r_p) \, , \\
    \tilde{z} &= \cos\vartheta(r_p) \approx 1 \, ,
\end{align}
\end{subequations}
where $\tilde{x}, \tilde{y}, \tilde{z}$ are the projections of the celestial coordinates on the celestial sphere, which becomes a plane for small viewing angles $\vartheta(r_p)$. 
This leads to agreement of all projections up to the first order of $\vartheta$. 
Therefore, it is reasonable to assume that the center of the observer plane is located at the observer's momentary coordinates.

Since the Fourier integral is evaluated discretely, $X_I,Y_I$ must be expressed accordingly in the context of a discrete Fourier transformation. 
Assuming that the Green's function is evaluated in the observer plane with the respective sample distances $X_T,Y_T$, and comparing the definitions of the continuous Fourier integral and the discrete Fourier transformation
\begin{subequations}
\label{Appendix:WavOpt:Coord:Eq:FT}
\begin{align}
    F(\omega) &= \int f(t) e^{-i \tilde{\omega} t} dt \, , 
\label{Appendix:WavOpt:Coord:Eq:CTFT} \\
    F(\omega_n) &= \sum\limits_{k=1}^{N} f(t_k) e^{-i \tilde{\omega}_n t_k}\, ,
\label{Appendix:WavOpt:Coord:Eq:DTFT}
\end{align}
\end{subequations}
a coefficient comparison of the continuous case \cref{WavOpt:FourierOpt:Eq:FourierDiffInt,Appendix:WavOpt:Coord:Eq:FT} yields the spatial frequencies
\begin{subequations}
\label{Appendix:WavOpt:Coord:Eq:CTFTwDef}
\begin{align}
    \tilde{\omega}_{X_I,n} = \frac{\omega}{f} X_{I,n} \, , 
\label{Appendix:WavOpt:Coord:Eq:CTFTwDefXI} \\
    \tilde{\omega}_{Y_I,n} = \frac{\omega}{f} Y_{I,n} \, .
\label{Appendix:WavOpt:Coord:Eq:CTFTwDefYI}
\end{align}
\end{subequations}
The discrete frequencies and time bins of the discrete case are related to the sample distance $T$ of the processed input by
\begin{subequations}
\begin{align}
    \tilde{\omega}_n &= 2 \pi \frac{n}{N T}, &n = 1, ..., N 
\label{Appendix:WavOpt:Coord:Eq:DFTwtDef} \, , \\
    t_k &= k T, &k = 1, ..., N \, , 
\end{align}
\end{subequations}
where $N$ is the number of samples and $n$ is the sample index. 
By equating \cref{Appendix:WavOpt:Coord:Eq:CTFTwDef} with \labelcref{Appendix:WavOpt:Coord:Eq:DFTwtDef}, $X_I,Y_I$ for sample $n$ are
\begin{subequations}
\label{Appendix:WavOpt:Coord:Eq:ImPlane}
\begin{align}
    \frac{X_{I,n}}{f} &= \frac{2 \pi}{\omega} \frac{n}{N X_T} \, , 
\label{Appendix:WavOpt:Coord:Eq:ImPlaneX} \\
    \frac{Y_{I,n}}{f} &= \frac{2 \pi}{\omega} \frac{n}{N Y_T} \, ,
\label{Appendix:WavOpt:Coord:Eq:ImPlaneY}
\end{align}
\end{subequations}
with $n \in \left\{-N/2 - 1, ..., -1, 0, 1, ..., N/2\right\}$ for even $N$ and $n \in \left\{-(N-1)/2, ..., -1, 0, 1, ..., (N-1)/2\right\}$ for odd $N$.

This gives access to the $X_I,Y_I$ coordinates depending on the sample index $n$ and thus to the opening angles, directions of observed structures, and also to the comparison of the result with other known results. 
Note that the focal length $f$ of a convex lens appears on the left-hand side, leading to a focal length-normalized coordinate. 

\section{observer plane and arbitrary rotations around the black hole}
\label{Appendix:WavOpt:PlaneRot}
The evaluation of the scattered wave and its interference is observed in the observer plane, which in principle is a coordinate-constructed plane. 
The construction of the observer plane starts with a quasi-Cartesian coordinate system built on Boyer-Lindquist coordinates,
\begin{align}
    \vec{r}_O = \begin{pmatrix} X \\ Y \\ Z \end{pmatrix}.
\end{align}
where in the non-rotated case $Z = r_O$ and $X, Y \in \left[-d, d\right]$, where $d$ is the aperture size. 
To describe the two-dimensional coordinates of the observer plane for arbitrary inclinations $\theta_O$, a rotation is performed in terms of the Boyer-Lindquist coordinates of this coordinate plane.
$\vec{r}$ is rotated around the $(x, z)$-plane by $\theta$. 
Thus, the center of the inclined plane will always have its normal pointing towards the black hole. 
Note that its center at $X = 0$, $Y = 0$ is located in Boyer-Lindquist coordinates at $\vec{r}_O = (r_O, \theta_O, \phi_O)$. 
This yields
\begin{align}
    \vec{r}\,' &= \vec{r} ~\text{Rot}_y(\theta) \notag \\
    &= 
\begin{pmatrix}
X \cos \theta + r_O \sin \theta \\ 
Y \\ 
r_O \cos \theta - X \sin \theta 
\end{pmatrix} \label{Appendix:WavOpt:PlaneRot:Eq:RotatedPlaneCoord} \\
&\overset{!}{=} r'_O 
\begin{pmatrix}
\sin \theta_O \cos \phi_O \\
\sin \theta_O \sin \phi_O \\
\cos \theta_O 
\end{pmatrix} \notag \, .
\end{align}
$\theta_O$ can be derived from \cref{Appendix:WavOpt:PlaneRot:Eq:RotatedPlaneCoord}:
\begin{align}
\cos \theta_O = \frac{1}{r'_O} \left(r_O \cos \theta + X \sin \theta \right)
\label{Appendix:WavOpt:PlaneRot:Eq:thetaO}
\end{align}
\cref{Appendix:WavOpt:PlaneRot:Eq:thetaO} now describes $\theta_O$ inserted in the evaluation of the Green's function \cref{WavOpt:Green:Eq:GreenFun} in terms of the observer planes coordinates $(X,Y)$.

\section{Windowing of the observer plane}
\label{Appendix:Tukey}
\begin{figure}
\centering
\subfloat[$\text{Im}\left(G(\vec{r}, \vec{r}_s, l_\text{max}\right)$]{\includegraphics[width=0.5\linewidth]{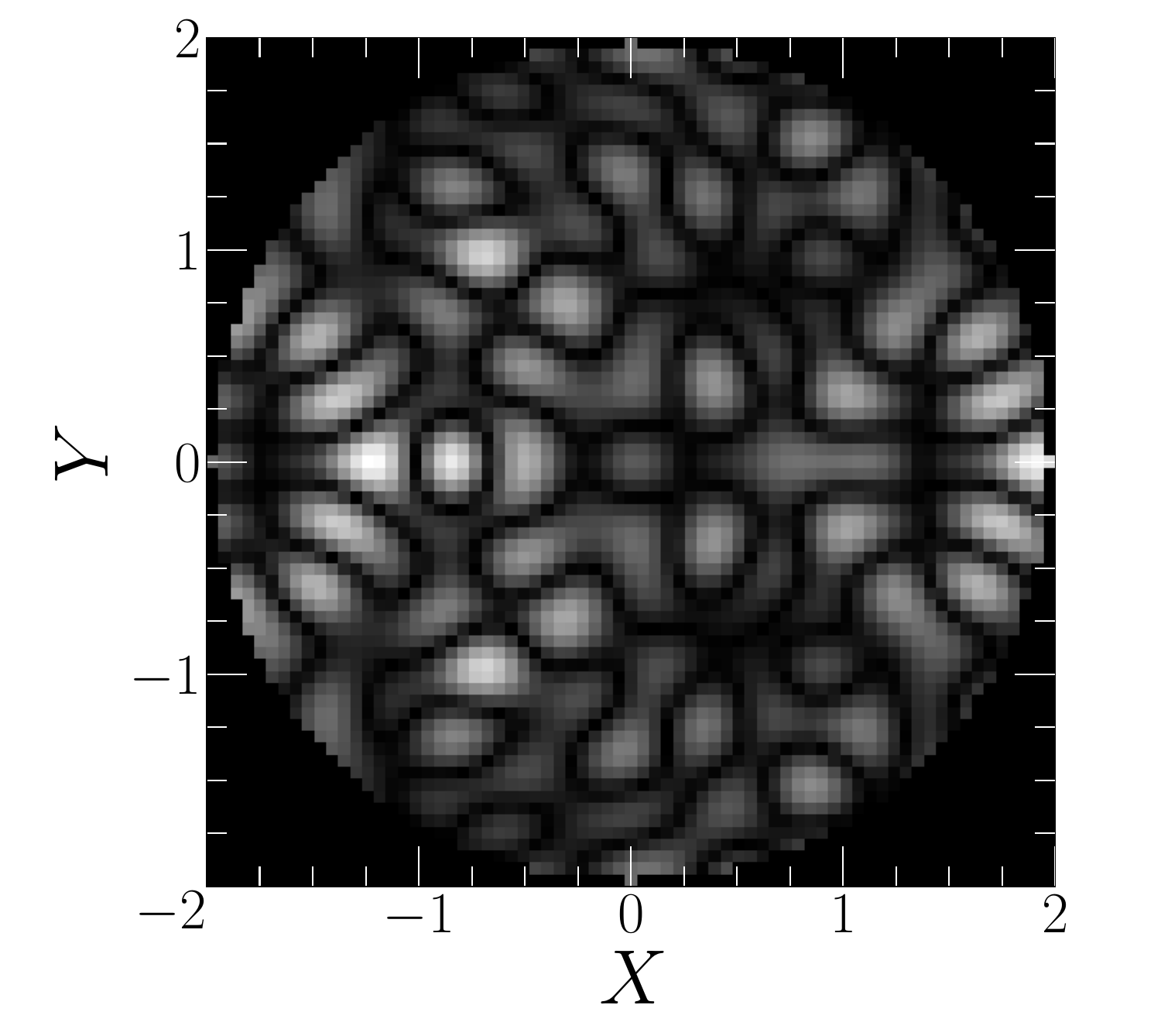}\label{Appendix:Tukey:Fig:TukeyFilter:IM}} \hfill
\subfloat[$\text{Im}\left(G(\vec{r}, \vec{r}_s, l_\text{max}\right)$ filtered with a Tukey filter according to \cref{Appendix:Tukey:Eq:TukeyFiltering}.]{\includegraphics[width=0.5\linewidth]{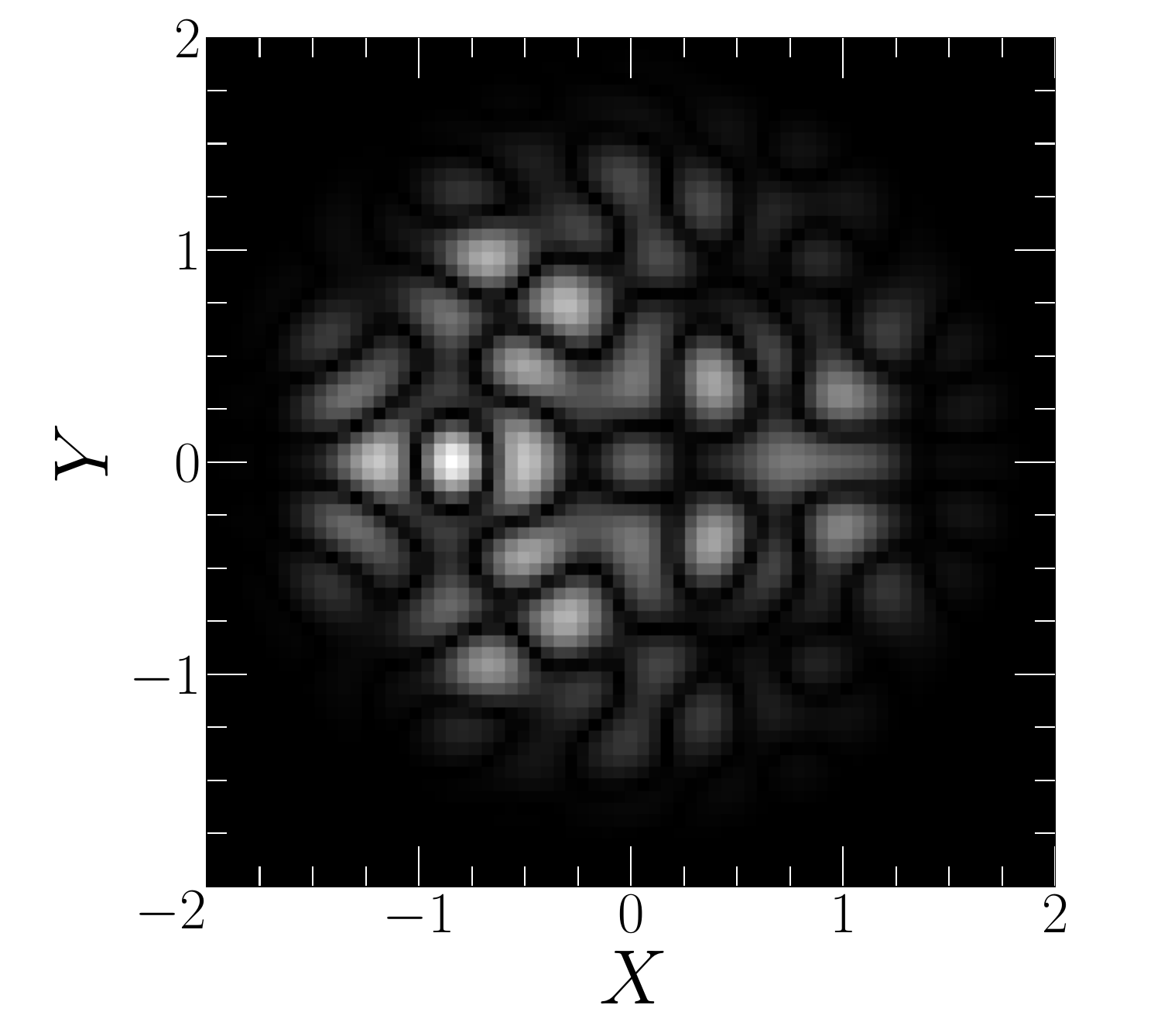}\label{Appendix:Tukey:Fig:TukeyFilter:IMTukey}}
\caption{Demonstration of the Tukey filter.
The apparent effect of the application can be summarized as a smoothing of the edges to zero. 
The result seen here is the observer plane of \cref{Results:ShadowCompare:Fig:MainFig1:0.99}, where a superposition of multiple sources is applied; see \cref{Results:ShadowCompare}. 
observer planes of single sources can be seen for SdS in \cref{Results:SdS:Fig:Mwvar:IMMw4,Results:SdS:Fig:Mwvar:IMMw12,Results:SdS:Fig:Mwvar:IMMw18} and for KdS in \cref{Results:KdS:Fig:aVar:IMa0.90,Results:KdS:Fig:aVar:IMa0.60,Results:KdS:Fig:aVar:IMa0.00}.}
\label{Appendix:Tukey:Fig:TukeyFilter}
\end{figure}

The complex result of \cref{WavOpt:Green:Eq:GreenFun} is filtered with a Tukey filter (also called a tapered cosine window), as used in \cite{Nambu2022}. 
It reduces aliasing effects of the Fourier transform by the sharp edge of the observer plane. 
This step is technically inspired and should be considered as a kind of post-processing of the data, which does not introduce any physical effect. 
\cref{Appendix:Tukey:Fig:TukeyFilter} shows an example of the imaginary part of the evaluated Green's function \cref{WavOpt:Green:Eq:GreenFun} at the observer plane, which coincides with the lens plane of \cref{WavOpt:FourierOpt:Fig:Scheme}. 
In \cref{Appendix:Tukey:Fig:FFTNoTukeyFilter} results in the image planes for two particular cases are shown, exemplifying results without an applied Tukey filter. 
The most prominent effect of omitting the Tukey filter is a more prominent emergence of aliasing effects and a more outstanding Poisson spot in the center of the Einstein ring.
Note that for \cref{Appendix:Tukey:Fig:FFTNoTukeyFilter:KdS} the source has been shifted to $\phi_s = 0.94 \pi$ such that it \enquote{appears} antipodal to the observer, compensating the frame-dragging.
By applying the Tukey filter to the results of the lens plane, Poisson spots are not emerging, e.g. as in Refs. \cite{Kanai2013,Nambu2016,Zeng2023,Zhang2021}.
These would shrink in size as the frequency increases.
It should be emphasized that the filter is a technical and not a physical step in the calculation. 
The filter function modifies the impinging scattering at the observer plane by
\begin{align}
    \Phi(X, Y) = w\left(\frac{1}{2} \left(\frac{\sqrt{X^2 + Y^2}}{d} + 1\right), \frac{2}{3}\right) \tilde{\Phi}(X, Y).
\label{Appendix:Tukey:Eq:Windowing}
\end{align}
\begin{figure}
\centering
\subfloat[$a = 0.99 M$, $\phi_s = 0.94 \pi$]{\includegraphics[width=0.5\linewidth]{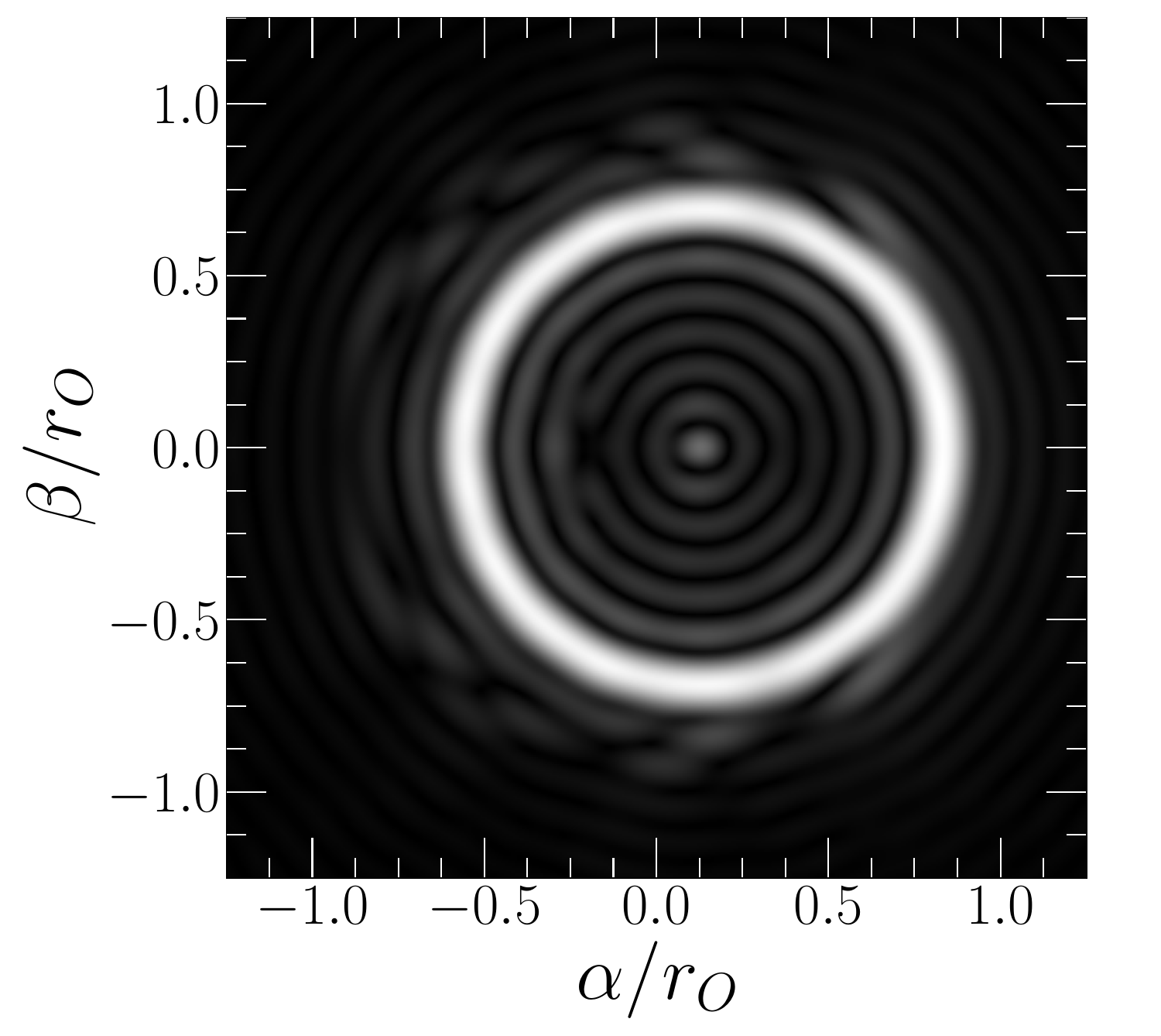}\label{Appendix:Tukey:Fig:FFTNoTukeyFilter:KdS}} \hfill
\subfloat[$a = 0.00 M$, $\phi_s = 1.00 \pi$]{\includegraphics[width=0.5\linewidth]{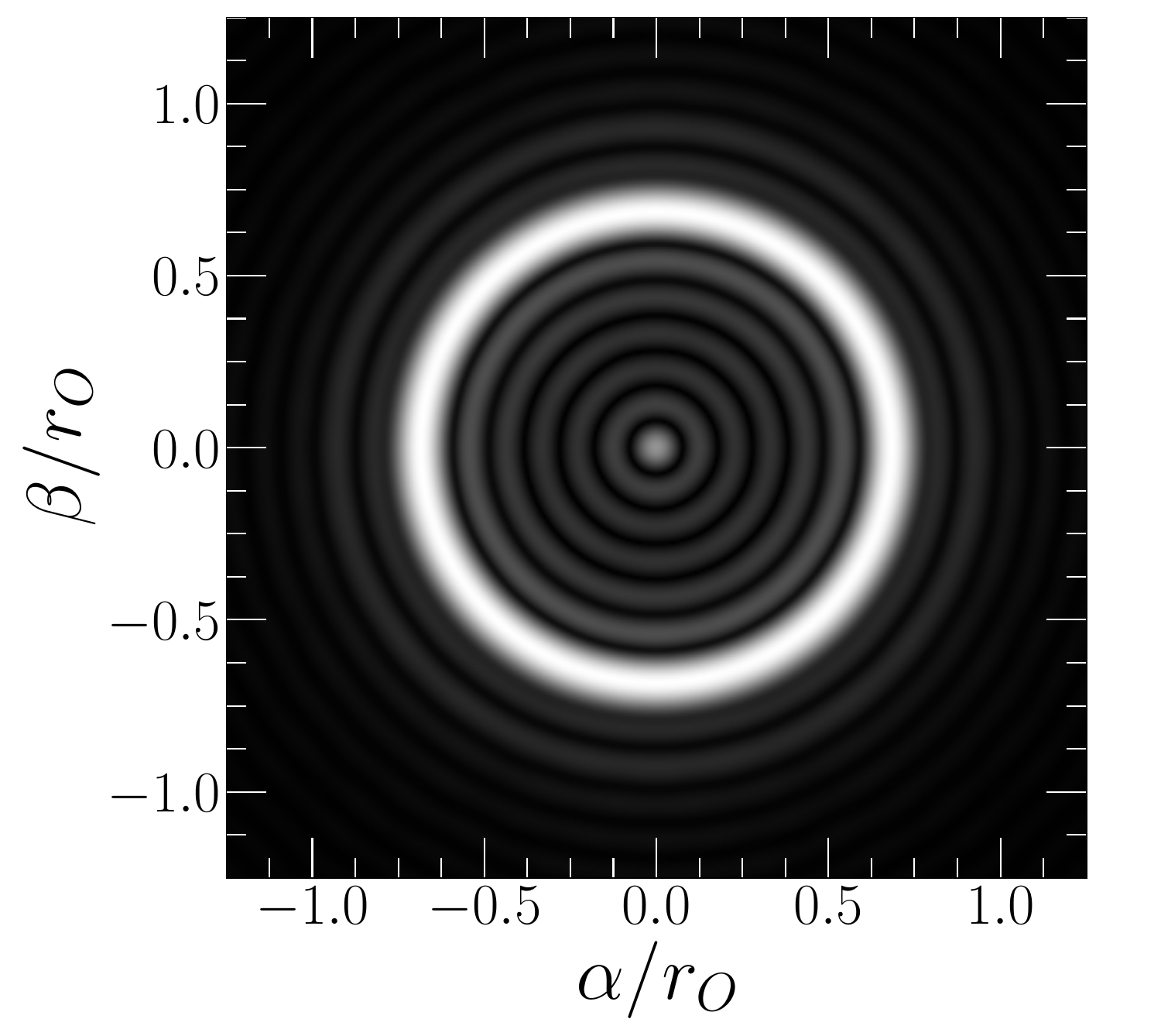}\label{Appendix:Tukey:Fig:FFTNoTukeyFilter:SdS}}
\caption{
Demonstration of the results of the image plane $\left|\mathcal{F}\left(G(\vec{r}, \vec{r}_s, l_\text{max})\right)\right|^2$ without Tukey filtering applied. 
Shown above are two examples for $M\omega = 15$, $\Lambda M^2 = 10^{-3}$, $r_O = 10 M$, $r_s = 20 M$, $\theta_O = \frac{\pi}{2}$, $\theta_s = \frac{\pi}{2}$, and choices of $a$ and $\phi_s$ in the respective captions. 
The most prominent observation is the presence of aliasing effects and a more outstanding Poisson spot.
}
\label{Appendix:Tukey:Fig:FFTNoTukeyFilter}
\end{figure}
The choice of the first argument is motivated by a radial symmetric weighting and a shift to the center of $X = 0$, $Y = 0$. 
The Tukey filter function $w(x, p)$ is defined as \cite{Bloomfield2000}
\begin{align}
w(x, p) = \begin{cases}
\frac{1}{2} \left(1 - \cos\left(\frac{2 \pi x}{p}\right)\right), &0 \leqslant x < \frac{p}{2} \\
1, &\frac{p}{2} \leqslant x < 1 - \frac{p}{2} \\
\frac{1}{2} \left(1 - \cos\left(\frac{2 \pi (1 - x)}{p}\right)\right), &1 - \frac{p}{2} \leqslant x \leqslant 1 \\
0, & x < 0 \wedge x > 1 
\end{cases}.
\label{Appendix:Tukey:Eq:TukeyFiltering}
\end{align}
The $p$-parameter controls how sharply the edges are smoothed, where $p = 0$ corresponds to a rectangular filter and $p = 1$ corresponds to a Hann filter. 
$p = \frac{2}{3}$ turns out to be an appropriate choice so that the edges are smoothed without losing too much of the original information.

\section{Notes on convergence}
\label{Appendix:Convergence}
\begin{figure}
\centering
\includegraphics[width=\linewidth]{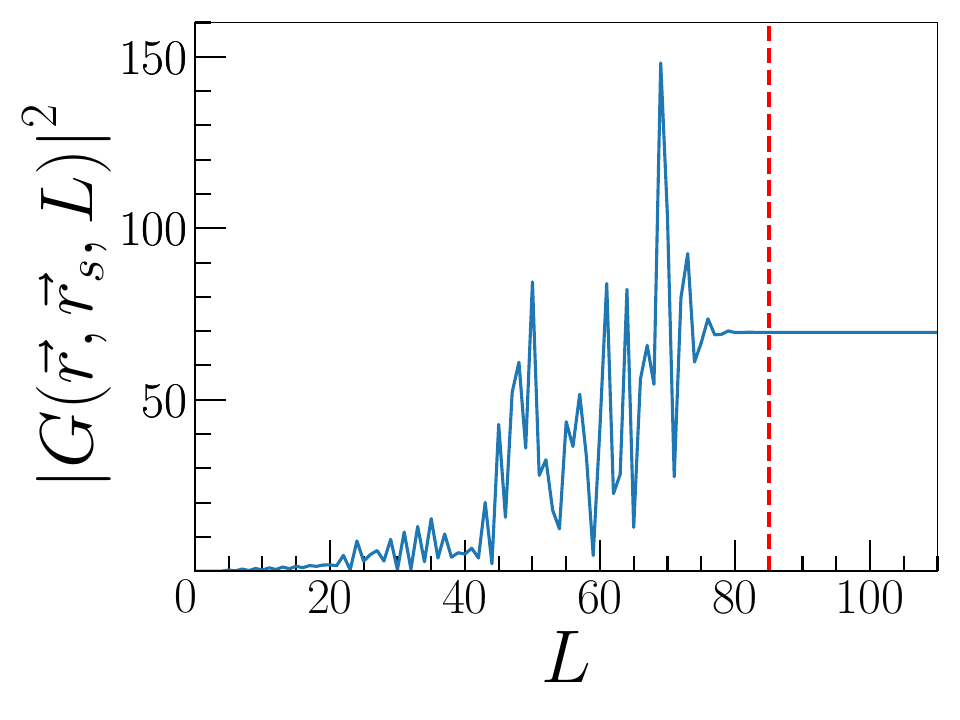}
\caption{Convergence behaviour of the Green's function $G(\vec{r}, \vec{r}_s, L)$ over $L$. 
The following parameters are used: $a=0.99M$, $\Lambda M^2 = 10^{-3}$, $M\omega = 9$, $r = 10M$, $r_s = 6M$, $\theta = \pi / 2$, $\theta_s = \pi / 2$, $\phi = \pi$, $\phi_s = 0$. 
The vertical red dashed line indicates where $C(L) \leqslant 10^{-10}$.}
\label{Appendix:Convergence:Fig:Convergence}
\end{figure}

\cref{WavOpt:Green:Eq:GreenFun} is a sum of infinite terms. 
However, in the evaluation, the sum runs to a finite $L = l_\text{max}$, at which point the sum converges sufficiently. 
Of course, one naturally has to consider that errors are introduced this way\footnote{These errors can be \enquote{smoothed} by an Ansatz of Handler and Matzner \cite{Handler1980,Glampedakis2001}}. 
The evaluation of an appropriate choice of $l_\text{max}$ is done heuristically here. 
\cref{Appendix:Convergence:Fig:Convergence} shows the convergence of Green's functions for a given set of parameters. 
For a certain $L = l_\text{max}$ the sum converges sufficiently. 
Sufficient is defined here as an $L$ for which
\begin{align}
C(L) = \left|G(\vec{r}, \vec{r}_s, L-1) - G(\vec{r}, \vec{r}_s, L)\right| \overset{!}{\leqslant} 10^{-10}.
\label{Appendix:Convergence:Eq:lmaxDetermine}
\end{align}
The value for the lower limit is chosen by considering sufficient small contributions for higher partial sums. 
One parameter that significantly varies the $l_\text{max}$ derived by \cref{Appendix:Convergence:Eq:lmaxDetermine} is the frequency $\omega$ because finer structures of the scattering take longer for the Green's function to converge. 
The choice of the radial coordinates $r_O$, $r_s$ also has a significant influence which should not be neglected. 
A complete analytical formula for the inequality \cref{Appendix:Convergence:Eq:lmaxDetermine} cannot be given yet, since the problem is highly non-trivial. 
For now, the heuristic condition is sufficient in the calculations. 

%

\end{document}